\documentclass[manuscript,screen]{acmart}
\usepackage{multirow}
\usepackage{booktabs}
\usepackage{array}
\usepackage{url}
\usepackage{orcidlink}
\usepackage{algorithm}
\usepackage{algpseudocode}
\usepackage{graphicx}
\usepackage{float}
\usepackage{pgfplots}
\usepackage{booktabs}  
\usepackage{graphicx}  
\usepackage{multirow}   
\usepackage{graphicx}
\usepackage{booktabs}
\usepackage{multirow}
\usepackage{float} 
\usepackage{placeins}
\usepackage{wrapfig}
\usepackage{ragged2e}
\usepackage{pgfplots}
\usetikzlibrary{positioning}
\usepackage{caption}
\usepackage{lscape}
\usepackage{placeins}
\usepackage{tikz} 
\usepackage{booktabs}
\usepackage{float} 
\usepackage{placeins} 
\usetikzlibrary{chains,shapes.arrows,fit}
\usepackage{forest}
\usepackage{graphicx}
\usepackage{enumitem}
\usepackage{forest}
\usepackage{placeins}
\usepackage[justification=centering]{caption}

\usepackage{amsmath}
\usepackage{fancybox}
\usepackage[utf8]{inputenc}
\usepackage{longtable}
\usepackage{tikz}
\usepackage{booktabs}
\usetikzlibrary{trees}

\usetikzlibrary{angles,quotes,patterns}
\AtBeginDocument{%
  }
    
\usepackage{graphicx}
\graphicspath{ {./images/} }

\acmYear{2025}
\acmDOI{XXXXXXX.XXXXXXX}

\setcopyright{none}  
\renewcommand\footnotetextcopyrightpermission[1]{}  
\authorsaddresses{}

\acmISBN{978-1-4503-XXXX-X/18/06}

\definecolor{arrowcolor}{RGB}{201,216,232}
\definecolor{circlecolor}{RGB}{79,129,189}
\colorlet{textcolor}{white}
\colorlet{bordercolor}{white}

\pgfdeclarelayer{background}
\pgfsetlayers{background,main}
\pgfplotsset{compat=1.18}
\newcounter{task}

\newlength\taskwidth
\newlength\taskvsep

\def\taskpos{}
\def\taskanchor{}

\newcommand\task[1]{%
  {\parbox[t]{\taskwidth}{\scriptsize\Centering#1}}}

\tikzset{
inner/.style={
  on chain,
  circle,
  inner sep=4pt,
  fill=circlecolor,
  line width=1.5pt,
  draw=bordercolor,
  text width=1.2em,
  align=center,
  text height=1.25ex,
  text depth=0ex
},
on grid
}

\newcommand\Task[2][]{%
\node[inner xsep=0pt] (c1) {\phantom{A}};
\stepcounter{task}
\ifodd\thetask\relax
  \renewcommand\taskpos{\taskvsep}\renewcommand\taskanchor{south}
\else
  \renewcommand\taskpos{-\taskvsep}\renewcommand\taskanchor{north}
\fi
\node[inner,font=\footnotesize\sffamily\color{textcolor}]    
  (c\the\numexpr\value{task}+1\relax) {#1};
\node[anchor=\taskanchor,yshift=\taskpos] 
  at (c\the\numexpr\value{task}+1\relax) {\task{#2}};
}

\newcommand\drawarrow{
\ifnum\thetask=0\relax
  \node[on chain] (c1) {}; 
\fi
\node[on chain] (f) {};
\begin{pgfonlayer}{background}
\node[
  inner sep=10pt,
  single arrow,
  single arrow head extend=0.8cm,
  draw=none,
  fill=arrowcolor,
  fit= (c1) (f)
] (arrow) {};
\fill[white] 
  (arrow.before tail) -- (c1|-arrow.west) -- (arrow.after tail) -- cycle;
\end{pgfonlayer}
}

\begin{document}

\title[]{Understand your Users, An Ensemble Learning Framework for Natural Noise Filtering in Recommender Systems}

\author{CLARITA HAWAT \orcidlink{0009-0008-4992-3996}}
\email{clarita.hawat@st.ul.edu.lb}

\author{WISSAM AL JURDI \orcidlink{0000-0001-9497-0515}}
\email{wissam.al_jurdi@univ-fcomte.fr}

\author{JACQUES BOU ABDO \orcidlink{0000-0002-3482-9154}}
\email{ouabdjs@ucmail.uc.edu}

\author{Jacques Demerjian \orcidlink{0000-0001-9798-8390}}
\email{jacques.demerjian@ul.edu.lb}

\author{Abdallah Makhoul \orcidlink{0000-0003-0485-097X}}
\email{abdallah.makhoul@univ-fcomte.fr}

\renewcommand{\shortauthors}{HAWAT et al.}

\ccsdesc[500]{Information systems~Recommender Systems}
\keywords{Recommender systems, offline evaluation, Ensemble Learning, Group Validation, Serendipity, Noise Signature, Multi-Layer Noise Management Framework}



\begin{abstract}
The exponential growth of web content is a major key to the success for Recommender Systems. This paper addresses the challenge of defining noise, which is inherently related to variability in human preferences and behaviours. In classifying changes in user tendencies, we distinguish three kinds of phenomena: external factors that directly influence users' sentiment, serendipity causing unexpected preference, and incidental interaction perceived as noise. To overcome these problems, we present a new framework that identifies noisy ratings. In this context, the proposed framework is modular, consisting of three layers:  known natural noise algorithms for item classification, an Ensemble learning model for refined evaluation of the items and signature based noise identification. We further advocate the metrics that quantitatively assess serendipity and group validation, offering higher robustness in recommendation accuracy. Our approach aims to provide a cleaner training dataset that would inherently improve user satisfaction and engagement with Recommender Systems. 
\end{abstract}

\maketitle

\FloatBarrier
\section{Introduction}

In an era where users are bombarded with endless choices, the pressure on Recommender Systems (RS) to cut through the noise and to deliver meaningful suggestions is growing. The main objective is to enhance user experience, usually personalized recommendations that boost engagement, simplify decision-making, and ultimately drive revenue. These systems are becoming more relevant and needed in Human Computer Interaction (HCI), where feedback mechanisms enhance interaction effectiveness, focused on in many research papers \cite{schafer2007collaborative, ekstrand2011collaborative, cui2020personalized, li2021research}. As the market continues to grow — over 600 million products on Amazon alone — and competitors continue to rise, users face 'choice fatigue'. This fatigue diminishes their ability to distinguish between relevant and irrelevant options. In this context, effective recommendations are crucial to maintaining relevance, reducing cognitive overload, enhancing user engagement, and building their trust in these systems.
\\ 
\\
Two major approaches have evolved for Recommender Systems over time: the first is collaborative filtering which relies on past behaviors and preferences, by recommending items enjoyed by like-minded users. The second approach is Content-based filtering, which relies on item properties, and matches them to the user's preferences. When both methods are combined, we get a Hybrid Recommender System. Finaly, the Knowledge-Based Recommender System relies on little to no historical data when suggesting items, instead it relies on its "knowledge" such as rules, constraints or specific requirements to recommend items.
Its roots date back to the sixties when systems like "Selective Dissemination of Information" \cite{hensley1963selective} were designed to distribute newly arrived information by matching it to users' interests stored in profiles. 
During the digital age, content-based techniques were utilized in vast domains such as personalized recommendations of interesting web pages \cite{pazzani1996syskill}.
\\
Between 1991 and 1992, collaborative filtering was initially user-based, meaning the algorithm identified users with similar interests (such as purchase patterns) and then recommended items those users discovered but the target user had not yet encountered, \cite{jaronski1990combining}, \cite{belkin1992information} and \cite{goldberg1992using}. In 1994, the GroupLens system \cite{resnick1994grouplens} introduced machine learning to predict whether a user would like specific unseen items. The core assumption is that to be able to detect if two users share similar tastes, we can check if they rate k items similarly, then they will most probably rate different items in a similar manner. The Different approaches differ in their definition of “ratings", their definition of the number of shared items (k), and their definition of “similarity”.
\\
\\
The 90s marked a transformative period for technology across various fields. In particular, Recommender Systems began to lay the groundwork for the personalized content seen today in major corporations like Amazon, Netflix, and YouTube \cite{jannach2021recommender}. From the humble beginnings of RSs in the 1960s, to today's sophisticated machine learning algorithms, Recommender Systems have transformed the way we consume information. Yet, amidst this evolution, these systems are also prone to errors due to human involvement. These errors, referred to as 'noise', occur when user ratings do not accurately reflect their true preferences or intentions.
\\
In 2006, noise was categorized into \textit{Natural} and \textit{Malicious} by  O'Mahony et al. \cite{o2006detecting}. The first arises from unintentional inconsistencies in user ratings, while the second results from intentional false ratings aimed at manipulating the system. Filtering out such noise is important for maintaining the accuracy and credibility of recommendations. Unlike malicious Noise, natural noise did not receive the focus of researchers in the past years. 
The problem in research working on improving RS, is only focusing on improving the accuracy, whereas the correct realistic way to identify the efficiency of RS, is by its ability to make personalized recommendations \cite{Critique_WJ_JBA_JD}.
\\ 
\\
In addition to existing noise filtering mechanisms, \cite{al2022strategic} introduced a signature-based algorithm to detect predefined patterns of noise. In cybersecurity, a signature refers to a specific pattern or set of characteristics that signals unusual or suspicious activity within a network. For example, repeated failed login attempts within a short period of time, are considered malicious. This type of activity is tagged as a 'signature' or a 'pattern' used to identify potential attacks in cybersecurity. Similarly, in Recommender Systems a noise signature is a pattern used to identify noise, such as malicious data manipulation or common errors that distort user feedback. If not detected early, these activities can negatively impact the learning process of the ML model. By leveraging these signatures, the system can more effectively filter out noise, improving the accuracy and reliability of user-generated ratings. The signature proposed in \cite{al2022strategic} was designed to detect obfuscation, a phenomenon where users intentionally manipulate their profiles to confuse the system. This concept parallels 'black-box obfuscation' described by \cite{goldwasser2007best}, where an obfuscated program hides any deterministic information about its original input. In a similar way, users can alter their profiles to mask their true behaviors. For example, a user is classified as 'opted-out' if more than 50\% of their activity on the last day contains noise, making it difficult for the system to generate accurate recommendations. This 'opt-out' was classified as obfuscation phenomenon by the authors.
\\ 
\\
While noise management is essential in RS to ensure the quality of user feedback, not every outlier in a dataset impacts it negatively. Some outliers, while unexpected can actually enrich user experience. This important phenomenon in RSs is known as Serendipity. Unlike noise, which introduces errors through inaccurate recommendations, serendipity works on enhancing the user experience. Often confused with noise, it refers to an unplanned yet meaningful discovery as mentioned in Smets et al. \cite{smets2022serendipity}. According to El Badran et al. \cite{el2019serendipity}, serendipity represents a calculated risk, where the system deliberately offers recommendations that might seem inaccurate. However, this risk, can lead to unsatisfactory results if the user dislikes the recommendation. It is typically measured by three factors: unexpectedness, utility, and quality. They also argue that the goal of Recommender Systems should be to filter out noise while preserving serendipity.
\\
\\ 
Furthermore, a thorough evaluation of RS' performance is challenging to achieve, with attempts dating back to 2004 \cite{herlocker2004evaluating}. These challenges arise from the multiple dimensions that need to be addressed to create an adequate and effective evaluation framework. \cite{zangerle2022evaluating} explained that research on evaluating Recommender Systems typically differentiates two methods: \textit{system-centric} and \textit{user-centric} evaluation. System-centric metrics refer to the predictive accuracy of the system, such as MAE, RMSE, Recall and Precision. While user-centric evaluation, focuses on assessing the system from the users' perspective, emphasizing how quality and experience are perceived when interacting with RS. It typically includes key aspects such as user engagement, as well as the serendipity and novelty of the recommendations.
While the system-centric metrics are commonly used to quantify the accuracy of Recommender Systems, however, as \cite{Critique_WJ_JBA_JD} pointed out, algorithms can show varying performance on different datasets, in other words, a better RMSE for example, does not necessarily translate to more personalized recommendations. An evaluation technique that can bypass these issues, and  measure the improvements/reduction in RS performance was needed. To address this, \cite{Group_Validation_in_Recommender_Systems} proposed a modular accuracy slicing evaluation framework called '\textit{Group Validation}'. The framework is based on '\textit{Simpson's Paradox}', a statistical phenomenon where a trend or direction observed within a group of data, can disappear or change direction when the groups are combined. Similarly, 
the authors identifed that the analysis conducted on a smaller group of users can yield different results compared to the analysis of the overall population \cite{Cohen2023}. This framework was designed to provide a better, more detailed evaluation of Recommender Systems performance. A thorough evaluation essentially needs to address all relevant aspects, as it should consider both user and system metrics. This is important because recommendations that perform well on one side, may not satisfy the other side \cite{zangerle2022evaluating, konstan2012recommender, mcnee2006making}.
\\ 
\\
Up until now, noise filtering algorithms have been tied to specific rules used to detect noisy items. These algorithms often neglects several factors present in the dataset. One common issue is their sensitivity to unbalanced datasets, that leads to actual activities getting mislabeled due to poor representation. As shown in \cite{mienye2021performance} ML algorithms often struggle on imbalanced datasets. Or Gupta et al. \cite{gupta2019dealing} studied noise filtering algorithms in general systems, and showed that Ensemble-based methods are better at identifying noise items, than other techniques (single model techniques). Ensemble learning (EL) has proven to be accurate to enhance the adaptability of Recommender Systems with unbalanced datasets by integrating many models to achieve a more accurate and well-rounded recommendation strategy. Ensemble learning might offer a comprehensive solution to the constraints set by the diversity of users and items in RS by reducing overfitting, enhancng genralizaton, handling imbalanced data and improving detection accuracy \cite{sagi2018Ensemble}.
\\ 
\\
Numerous promising noise management algorithms and frameworks have been proposed, where each framework intrinsically assumes a different definition of noise. What remains missing is the unifying structure that integrates their distinct noise definitions together ensuring positive user feedback.
The growing need for a unified framework that addresses noise management, while preserving serendipity, is supported by an evaluation system capable of identifying performance improvement or degradation is evident.
\\
Building on the issues and needs mentioned above, we propose a modular framework that consists of three different layers that work together to improve noise filtering in Recommender Systems. The first layer utilizes many noise filtering algorithms to eliminate identifiable noise. In this layer, noise filtering algorithms form a decision board, to vote on items' classification as noisy. The second layer is an Ensemble model that combines several models to handle residual items that the first layer did not arrive to a definitive answer on. Finally, the third layer introduces noise signatures serving as an effective guard against known noise patterns.
Second, and to balance system metrics and user metrics, we propose a 2D metric that includes both \textit{Group Validation} metrics and \textit{serendipity} scores. That will keep our system at a high precision and relevance for recommendation results while encouraging serendipity in users' interest. All of the layers and metrics combined provide a comprehensive framework that fits the noise management while fostering user satisfaction and trust.
\\ 
\\
The following sections will be divided into: section \ref{sec:relatedworks} which will talk about related works and background, section \ref{sec:framework} to discuss the proposed framework, section \ref{sec:env_var} to show environment parameters used across the paper, \ref{sec:chosenalgo} to explain the selected algorithms for layer one and two, and last but not least, \ref{sec:testing_analysis} to test the proposed framework and an analysis based on the results while we will finalize in section \ref{sec:conclusion}, with the conclusion.

\FloatBarrier
\section{Related Work and Background}
\label{sec:relatedworks}
In this section, we provide an overview of related work relevant to the proposed framework. The discussion is organized into four subsections: section \ref{sec:nfreview} addresses natural noise filtering algorithms and their categories; section \ref{sec:elreview} focuses on Ensemble learning algorithms; we move forward to shed some light on the noise signature that will be included in our framework in section \ref{sec:signature}, and section \ref{sec:metricsused} examines known accuracy metrics.

\subsection{Natural Noise Selection Review}
\label{sec:nfreview}
As explained in the introduction, natural noise in Recommender Systems did not yet receive the full focus of researchers. To establish a foundation for the first layer of our framework, we considered two recent papers, that surveyed noise filtering algorithms \cite{Critique_WJ_JBA_JD} and \cite{Sampling_and_noise_filtering_methods_A_literature_review}.

\cite{Critique_WJ_JBA_JD} took into consideration 27 papers between 2006 and 2019, the authors explained that researchers took two different paths while they were studying the magic barrier, and introduced two new terminologies: logic barrier, and accuracy barrier. The \textit{logic barrier} refers to the challenges and limitations in the system' data and design according to \cite{herlocker2004evaluating}. The authors in the study \cite{herlocker2004evaluating} started speculating the \textit{'magic barrier'} (MB), they reasoned that Recommender Systems fail to get more accurate after a certain point. They explained that this threshold was set due to inconsistencies in user profiles and the authors in \cite{Critique_WJ_JBA_JD} explained that the studies following this path, did not work on algorithm enhancement, and stuck to evaluation methods of RSs \cite{Critique_WJ_JBA_JD}. They hypothesized that \textit{accuracy barrier} is the threshold where the system's accuracy cannot be improved due to the presence of \textbf{noise}. 

The authors in \cite{Critique_WJ_JBA_JD} grouped natural noise management into three categories
\begin{itemize}
    \item Magic barrier: Studies guided by the accuracy barrier and the logic barrier in noise management (NM).
    \item Classical natural Noise Management (NNM): Techniques that identify natural noise (NN), to identify noise, then remove it or correct it.
    \item Preference-dependent Natural Noise Management (NNM): Techniques using data beyond explicit ratings like movie reviews, director information ...
\end{itemize}

On the other hand, \cite{Sampling_and_noise_filtering_methods_A_literature_review} a review on noise management in Recommender Systems, considered 50 papers approx. for both malicious and natural noise. The authors, classified noise into crisp and fuzzy methods for the first layer:
\begin{enumerate}
\item \textbf{Crisp}:
Crisp algorithms are rigid and rely on exact data. They rely on binary decision-making models, where things are either true or false, without any room for ambiguity.
\item \textbf{Fuzzy}:
Fuzzy algorithms typically handle ambiguity and uncertainty in ratings and NNM. Usually, user preferences are not binary and ratings sometimes overlap between categories.
\end{enumerate}

They also subgrouped the crisp method into: Re-Rating and Ranking, Classification and Clustering, Magic Barrier, Outliers Detection and Global Information:

\begin{itemize}
    \item Re-Rating and Ranking: this category includes algorithms that require users to re-rate items they had rated before by following specific segmentations.
    \item Classification and Clustering: Classification techniques in noise filtering use predefined categories to label and filter out irrelevant data, while clustering group similar data points together. Algorithms such as decision trees and hierarchical clustering were mentioned.
    \item Magic Barrier: this category grouped algorithms that take into consideration the improvement of the accuracy barrier to classify items.
    \item Outliers Detection:  are statistical techniques used to identify observations that significantly deviate from the majority of the dataset.
    \item Global Information: Algorithms that utilise user and item preferences to detect noise.
\end{itemize}

\subsubsection{\textbf{Categories:}} 

Considering the survey in these two papers \cite{Critique_WJ_JBA_JD, Sampling_and_noise_filtering_methods_A_literature_review}, and given that layer 1 of our proposed framework incorporates distinct noise filtering algorithms, each handling noise differently, we established the following categories on a 2D scale. We first grouped them by fuzzy and crisp, then on the second axis, the algorithms were divided by how they consider and utilize data to make their prediction: 'User Centered Filtering', 'Accuracy Enhancement Ability' and 'User-Item Centered Filtering'. This serves as the basis for selecting the algorithms to include in our framework.
\newline
\textbf{Accuracy Enhancement Ability} 
\newline
This category refers to the capability of noise filtering algorithms to enhance the accuracy by effectively identifying and mitigating noise in the data. It involves techniques and methods to handle noise effectively ensuring high prediction accuracy. This relates to the magic/accuracy barrier explained above. Table \ref{tab:AccuracyPredictionPapers} summarizes all the algorithms in the crisp Accuracy Enhancement Ability category, ordered by citation count and then year, both in descending order.

\begin{table}[ht]
\centering
\caption{Summary of accuracy enhancement ability papers}
\begin{tabular}{p{6cm} p{2cm} p{2cm}}
\toprule
\textbf{Crisp Algorithm} & \textbf{Citation \#} & \textbf{Publication Year} \\
\midrule
\cite{amatriain2009like} & 328 & 2009 \\
\cite{amatriain2009rate} & 230 & 2009 \\
\cite{o2006detecting} & 177 & 2006 \\
\cite{10.1007/978-3-642-31454-4_20} & 90 & 2012 \\
\cite{said2018coherence} & 39 & 2018 \\
\cite{bellogin2014magic} & 39 & 2014 \\
\cite{saia2016semantic} & 29 & 2016 \\

\bottomrule
\end{tabular}

\label{tab:AccuracyPredictionPapers}
\end{table}

\FloatBarrier

\textbf{User Centered}\\
The studies in this category focus on user preferences and behaviors. Algorithms in this category will use behavior analysis and contextual filtering and analyze the noise in user data by looking at their preferences and anomalies in user interactions, ratings. For example, if a user consistently rates items highly, but suddently starts giving low ratings, the algorithm will consider this sudden change in tendencies as noise, using the user's historical data.
Table \ref{tab:UserCenteredPapers} lists the research papers that fit this category, ordered by citation count and then year, both in descending order.

\begin{table}[ht]
\centering
\caption{Summary of User Centered Papers}
\begin{tabular}{ p{3cm} p{3cm} p{2cm} p{2cm}}
\toprule

\textbf \textbf{Crisp} & \textbf{Citation \#} & \textbf{Publication Year} \\
\midrule
\cite{pham2013preference} & 59 & 2013 \\
 \cite{yu2016novel} & 10 & 2016 \\
 \cite{pham2013integrating}& 3  & 2012 \\
\bottomrule
\end{tabular}

\label{tab:UserCenteredPapers}
\end{table}

\textbf{User-Item Centered} \\
These algorithms aim to identify and filter out noisy ratings by examining patterns and relationships within user and item groups. The process typically involves creating profiles for users and items and then filtering out noise based on inconsistencies against user and item group.
For instance, if a user known for his high tendency in positive ratings gives an unusually low score to an action movie expected to be successful, the algorithm will flag this rating as noisy.
\\ 
\begin{enumerate}

\item Starting with the \textbf{fuzzy subcategory}, 
\cite{yera2016fuzzy} worked on a flexible process based on fuzzy tools that has been introduced to handle natural noise, making the identification and correction of noisy ratings in Recommender Systems more adaptable. An extensive case study demonstrates that this fuzzy method outperforms previous approaches, they showed that managing natural noise with fuzzy techniques yields better accuracy in Recommender Systems compared to rigid methods, this approach was mentioned in multiple studies \cite{martinez2016managing, yera2020natural}. 
\item As for the Crisp algorithms:
\label{it:useritemcrispnf1}
the noise detection method in \cite{toledo2015correcting, martinez2016managing, toledo2013managing, dixit2019proposed, bag2019noise} is based on the classification of users, items and ratings. These classes are: 
\begin{itemize}
    \item For users:
     critical, average and benevolent -- based on the ratings they give.
     \item For items:
     strongly, averagely and weakly preferred -- based on the ratings they get.
\end{itemize}
The authors explain that for a rating $R_ui$ if the user u and item i are classified into a similar category, then if the rating  $R_ui$ doesn't belong to that same category, the rating is noisy. They mention noise correction after noise identification following a collaborative Recommender Systems.
Table \ref{tab:userItemAlgorithm} summarizes all the algorithms in the crisp User-Item category, ordered by citation count and then year, both in descending order.
\end{enumerate}
\FloatBarrier

\begin{table}[h!]
\caption{Summary of User-Item Centered Papers}
\centering
\begin{tabular}{|c|c|c|c|}
\hline
\textbf{Fuzzy} & {Crisp} & \textbf{Citation \#} & \textbf{Publication Year} \\
\hline
 & \cite{toledo2015correcting} & 121 & 2015 \\
\hline
\cite{yera2016fuzzy} &  & 75 & 2016 \\
\hline
 & \cite{castro2017empirical} & 66 & 2017 \\
\hline
 & \cite{bag2019noise} & 56 & 2019 \\
 \hline
 & \cite{li2013noisy} & 56 & 2013 \\
 \hline
 & \cite{al2018Serendipity} & 45 & 2018 \\
 \hline
\cite{castro2018fuzzy} & & 44 & 2018 \\
\hline
\cite{saia2014semantic} & & 23 & 2014 \\
\hline
 & \cite{martinez2016managing} & 17 & 2016 \\
\hline
\cite{wang2021effective} &  & 15 & 2021 \\
\hline
 & \cite{toledo2013managing} & 9 & 2014 \\
\hline
 & \cite{latha2015ranking} & 6 & 2015 \\
\hline
 & \cite{dixit2019proposed} & 4 & 2019 \\
\hline
\end{tabular}
\label{tab:userItemAlgorithm}
\end{table}

Now that we've laid the foundation for noise filtering algorithm selection in Layer 1, we will now proceed to define the same for the subsequent layers. These brief reviews will be utilized in algorithm selection in section \ref{sec:chosenalgo}.

\subsection{Ensemble Learning Review}
\label{sec:elreview}
Given that the noise filtering algorithms in Layer 1 will have distinct noise identification processes and may produce varying results, and as considering the imbalanced noisy data, given the ability of Ensemble Learning to mitigate these issues, it forms the layer 2 of the proposed framework. As we established the foundation for the first layer, we will now explore recent studies and methodologies on various Ensemble learning techniques to lay the foundation for the second layer.
\\
\\
Ensemble methods overcome three problems that Machine Learning (ML) usually encounter: statistical problems, computational problems and representation problems \cite{dietterich2002Ensemble}. Also while working with ML algorithms on high-dimensional or imbalanced datasets, learning can be particularly challenging, especially when considering multiple factors in decision making \cite{mienye2022survey}. 
EL is a concept in ML where many models, mostly known as base learners - are combined and their predictions averaged out to produce a final prediction. This can be helpful in reducing errors and enhancing accuracy.
\\
We reviewed different Ensemble learning algorithms by analyzing surveys and research papers. We used \cite{mienye2022survey}, a 2022 IEEE survey on Ensemble Learning, as well as \cite{de2021reliable} 2021 Elsevier article, \cite{dong2020survey} 2020 Springer survey, \cite{liu2008isolation} 2019 IEEE article. 
We surveyed different categories for Ensemble learning, and we'll briefly introduce them and list the papers for each one:

\begin{enumerate}
    \item \textbf{\textit{Boosting}} is a ML technique able to convert "weak learners into a strong classifier" \cite{AWODELE_2023}. The study on boosting started in 1990 in \cite{schapire1990strength} which had a good impact on ML and statistics \cite{mienye2022survey}. 
    \textbf{weak learners} and \textbf{strong learners} are terms used In EL Algorithms, to describe the performance and capabilities of the models that collaborate together to form the Ensemble.
    Papers around stacking are highlighted in table \ref{graph:sup_boost_el}.

\begin{table}[htbp!]
\centering
\caption{Ensemble methods: Boosting}
\label{graph:sup_boost_el}
\begin{tabular}{ p{8cm} p{2cm} p{2cm}}
\hline
\textbf{Boosting} & \textbf{Year} & \textbf{Citations} \\ \hline
\cite{prokhorenkova2018catboost} & 2022 & 5319 \\ \hline
\cite{ke2017lightgbm} & 2017 & 17346 \\ \hline
\cite{chen2016xgboost} & 2016 & 54258 \\ \hline
\cite{wu2011some} & 2011 & 47 \\ \hline
 \cite{friedman2001greedy} & 2001 & 33416 \\ \hline
\end{tabular}

\end{table}

    \item \textbf{Bagging}
    \cite{breiman1996bagging} first developed the bootstrap aggregating also called bagging to improve the classification of ML models. This idea behind it is that it creates multiple versions of a predictor using different random subsets, results of each model are averaged to produce the results.
    The information on the algorithms are shown in Table \ref{graph:sup_bagging}.

    \begin{table}[htbp!]
    \centering
    \caption{Ensemble methods: Bagging}
    \begin{tabular}{ p{8cm} p{2cm} p{2cm} }
    \hline
    \textbf{Bagging} & \textbf{Year} & \textbf{Citations} \\ \hline
    \cite{altman2017Ensemblev1} & 2017 & 289 \\ \hline
    \cite{ho1995random} & 1995 & 11038 \\ \hline
    \end{tabular}
    \label{graph:sup_bagging}
    \end{table}
\vspace{2.5em}


\FloatBarrier
    \item \textbf{Stacking}
    In 1992, \cite{wolpert1992stacked} introduced the \textbf{\textit{Stacking}} as an EL framework that trains multiple machine learning algorithms, called base learners, and then it combines their predictions. Stacking is utilized when different task aspects are captured by different models. A separate model called meta-learner, is utilized to make the final decision based on the predictions of the base learners. 
    The difference between Stacking and the other ensemble methods is that it employs various ML algorithms, each learning differently from the dataset. As shown in Table \ref{graph:sup_stack_el}, various Ensemble methods and their corresponding citation counts are presented.


\end{enumerate}  
\vspace{-0.0cm}
\begin{table}[htbp!]
\centering
\caption{Ensemble methods: Stacking}
\begin{tabular}{ p{8cm} p{2cm} p{2cm}}
\hline
\textbf{Stacking} & \textbf{Year} & \textbf{Citations} \\ \hline
\cite{liang2021stacking} & 2022 & 67 \\ \hline
\cite{jain2021systematic} & 2021 & 338 \\ \hline
\cite{chatzimparmpas2021empirical} & 2021 & 24 \\ \hline
\cite{ribeiro2020Ensemble} & 2020 & 539 \\ \hline
\cite{ouyang2018multi} & 2018 & 68 \\ \hline
\end{tabular}

\label{graph:sup_stack_el}
\end{table}
\FloatBarrier

\cite{mienye2022survey} explained that 
"bagging reduces variance without increasing the bias", while boosting reduces bias, whereas Stacking reduces both bias and variance by combining the strengths of multiple models.

\FloatBarrier
\justify
\textbf{(2) Semi-Supervised Algorithms:}
\mbox{}\\[0.5em]
The semi-supervised method is usually used when labeled data is scarce. Additionally, EL is employed to boost generalization, addressing issues associated with the limited adaptability of individual learning methods. \cite{dong2020survey} explained that it "focus on augmenting the training set and utilizing them" \cite{dong2020survey}. The authors also mentioned through extensive results that semi-supervised Ensemble methods perform better than other Ensemble methods on classification jobs, where labeled data is insufficient. Papers adressing this category is listed in figure \ref{graph:semisup_el}.


\begin{figure}[ht!]
    \centering
    \caption{Semi-Supervised Ensemble Learning Publications per Year}
        \label{graph:semisup_el}
    \begin{minipage}{0.7\textwidth}
        \centering
        \begin{tikzpicture}
        \begin{axis}[
            ybar,
            width=\textwidth,
            symbolic x coords={2006, 2007, 2012, 2013, 2014, 2015, 2016, 2017, 2018, 2021},
            height=8cm,
            xtick=data,
            ylabel={Publications},
            xlabel={Year},
            nodes near coords,
            ymin=0,
            ymax=7,
            x tick label style={rotate=30, anchor=east, font=\scriptsize},
            bar width=10pt,
            enlargelimits=true,
            grid=major,
            grid style=dashed,
            xlabel style={yshift=-2pt},
        ]
        \addplot coordinates {(2006,1) (2007,1) (2012,3) (2013,3) (2014,1) (2015,2) (2016,1) (2017,4) (2018,3) (2021,1)};

        \node at (axis cs:2006,5.5) {\cite{yan2006subspace}};
        \node at (axis cs:2007,5.5) {\cite{wang2007Ensemble}};
        \node[align=center] at (axis cs:2012,5.5) {\cite{zhang2012semi} \\ \cite{woo2012semi} \\ \cite{iqbal2012semi}};
        \node[align=center] at (axis cs:2013,5.5) {\cite{mahmood2013semi} \\ \cite{zhang2013exploiting} \\ \cite{chen2013convergence}};
        \node[align=center] at (axis cs:2014,5.5) {\cite{fersini2014sentiment}};
        \node[align=center] at (axis cs:2015,5.5) {\cite{yu2015incremental} \\ \cite{yang2015parallel}};
        \node at (axis cs:2016,5.2) {\cite{xiao2016semi}};
        \node[align=center] at (axis cs:2017,6) {\cite{lu2017hyperspectral} \\ \cite{soares2017cluster} \\ \cite{yu2017adaptive} \\ \cite{alves2017social}};
        \node[align=center] at (axis cs:2018,5.5) {\cite{wei2018combined} \\ \cite{yu2018semi} \\ \cite{yu2018multiobjective}};
        \node at (axis cs:2021,5.5) {\cite{de2021reliable}};
        
        \end{axis}
        \end{tikzpicture}
        
    \end{minipage}
\end{figure}


\textbf{(3) Unsupervised Algorithms:}
\mbox{}\\[0.5em]
And the last category, is unsupervised algorithms which center around Clustering Ensemble methods. They operate by creating a series of clustering partition through various clustering algorithms, then they merge these partitions together to produce a unified consensus solution.
\cite{dong2020survey} noted that Ensemble Clustering methods generally outperform single clustering approaches in stability, accuracy and robustness, as they leverage all the information available to each clustering member. Methods like nearest neighbor, Euclidean median, Gaussian-mixture models, graph-based consensus clustering scheme and many more were mentioned in this topic. 
While outlier models and classification models appear different, they share theoretical underpinnings in the bias-variance trade-off \cite{rayana2016sequential}. Traditional outlier detection algorithms, such as density-based and distance-based methods, struggle to detect all types of outliers across diverse datasets. Ensemble methods aim to overcome this by combining the strengths of multiple base detectors. 
Initially most outlier Ensemble methods focused on parallel frameworks, primarily using Bagging-inspired approaches to reduce variance by aggregating results from independent base detectors. Then it was later developed to utilize more sophisticated Ensemble approach, that utilizes sequential approaches instead of parallel ones.Papers that reference unsupervised Ensemble learning are included in figure \ref{graph:merged_pub}.
With this, we conclude the EL layer's quick survet; in the following subsection, we will be discussing the Signature layer. Layer 3 consists of noise signatures, the order of this layer can vary based on the signature chosen by the user.
\begin{landscape}
\begin{figure}[h!]
    \centering
            \caption{Unsupervised Ensemble Learning Algorithms publications}
        \label{graph:merged_pub}
    \begin{minipage}{\textwidth}
        \centering
        \begin{tikzpicture}
        \begin{axis}[
            ybar,
            width=\textwidth,
            symbolic x coords={1995, 2001, 2002, 2003, 2004, 2005, 2006, 2007, 2008, 2009, 2010, 2012, 2013, 2014, 2015, 2016, 2017, 2019, 2021},
            width=20cm,
            height=10cm,
            xtick=data,
            ylabel={Publications},
            xlabel={Year},
            nodes near coords,
            ymin=0,
            bar width=3pt,
            ymax=7,
            x tick label style={rotate=30, anchor=east, font=\scriptsize},  
            enlargelimits=upper,
            bar width=10pt,
            enlargelimits=true,
            grid=major,
            grid=none, 
            grid style=dashed,
            xlabel style={yshift=-2pt},
        ]
        \addplot coordinates {
            (1995,1) (2001,1) (2002,3) (2003,2) (2004,3) (2005,2) (2006,2) (2007,2) (2008,4) (2009,3) 
            (2010,2) (2012,5) (2013,1) (2014,5) (2015,2) (2016,2) (2017,2) (2019,1) (2021,1)
        };

        \node[align=center] at (axis cs:1995,2.5) {\footnotesize \cite{ho1995random}};
        \node[align=center] at (axis cs:2001,2.5) {\cite{friedman2001greedy}};
        \node[align=center] at (axis cs:2002,5.5) {\footnotesize \cite{dash2002feature} \\ \cite{ mitra2002unsupervised} \\ \cite{strehl2002cluster}};
        \node[align=center] at (axis cs:2003,5.5) { \footnotesize\\ \cite{fern2003random} \\ \cite{kuncheva2003measures}};
        \node[align=center] at (axis cs:2004,5.5) {\footnotesize \cite{topchy2004analysis} \\ \cite{frigui2004unsupervised} \\ \cite{dy2004feature}};
        \node[align=center] at (axis cs:2005,3.5) {\cite{topchy2005clustering} \\ \cite{huang2005automated}};
        \node[align=center] at (axis cs:2006,3.5) {\cite{shi2006unsupervised} \\ \cite{kuncheva2006evaluation}};
        \node[align=center] at (axis cs:2007,3.5) {\cite{jing2007entropy} \\ \cite{yu2007graph}};
        \node[align=center] at (axis cs:2008,5.5) {\cite{liu2008isolation} \\ \cite{hong2008unsupervised} \\ \cite{li2008localized} \\ \cite{hong2008consensus}};
        \node[align=center] at (axis cs:2009,5.5) {\cite{azimi2009adaptive} \\ \cite{hore2009scalable} \\ \cite{grozavu2009variable}};
        \node[align=center] at (axis cs:2010,3.5) {\cite{wang2010tree} \\ \cite{lourencco2010scalability}};
        \node[align=center] at (axis cs:2012,6.5) {\cite{zhang2012generalized} \\ \cite{yu2012sc3} \\ \cite{wu2012efficient} \\ \cite{yu2012hybrid}  \cite{gullo2012multiobjective}};
        \node[align=center] at (axis cs:2013,2.5) {\cite{yu2013hybrid}};
        \node[align=center] at (axis cs:2014,6.8) {\cite{franek2014Ensemble}  \cite{yu2014hybrid} \\ \cite{yu2014adaptive} \\ \cite{yang2014exploring} \\ \cite{yu2014probabilistic}};
        \node[align=center] at (axis cs:2015,4) {\cite{yu2015adaptive} \\ \cite{elghazel2015unsupervised}};
        \node[align=center] at (axis cs:2016,4.5) {\cite{chen2016xgboost} \\
        \cite{yu2016distribution}};
        \node[align=center] at (axis cs:2017,3.5) {\cite{ke2017lightgbm} };
        \node[align=center] at (axis cs:2019,2.5) {\cite{hariri2019extended}};
        \node[align=center] at (axis cs:2021,2.5) {\cite{de2021reliable}};
        
        \end{axis}
        \end{tikzpicture}

    \end{minipage}
\end{figure}
\end{landscape}

\subsection{Noise Signature}
\label{sec:signature}
A \textbf{\textit{signature}} is a unique mark or sign made by an individual to indicate their identity, it can be a pattern of malicious activity, a set of known IPs of attackers or a pattern of traffic originating from these IPs. A \textbf{\textit{noise signature}} is explained by the unique pattern of noise produced by a specific source.  
In \textit{Recommender Systems}, a \textit{noise signature} refers to the unique pattern of noise present in the data used by the system. This noise can significantly impact the performance and accuracy of recommendations.
If certain information is being introduced in a consistent manner, whether they are true or not, they will affect the system's understanding of a user, potentially resulting in irrelevant recommended content, ultimately deteriorating the user experience.
\\ 
A powerful RS is one that doesn't compromise on privacy while ensuring good accuracy and personalized recommendations. Many users loose trust in a system that tracks their information, \cite{wang2024enhancing} studied the concept of recommendation acceptance intention (RAI) where they disccus the importance of users accepting the RS. User's privacy is an issue that should be reflected on, \cite{al2022strategic} identified obfuscation as a new type of noise and called the noise itself 'Opt-out', a method used by users to leave untrusted systems where privacy is lacking. The authors called it user-intended type of noise, where the behavior of the user is affecting the validity of item feedback. They defined an \textbf{\textit{Opt-out attack}} to be used by a person as a method of obfuscating the user's true personal profile. 
They explained that while online platforms continuously assure their users about personal privacy, most of the times this protection is exaggerated and masked deep in privacy policies, which 91\% of people in the US consent to without reading (\href{https://www.businessinsider.com/deloitte-study-91-percent-agree-terms-of-service-without-reading-2017-11}{Deloitte survey 2017}).

\subsection{Metrics Used:}
\label{sec:metricsused}
As explained in the introduction, a well balanced system is one that takes into consideration both system and user metrics to meet the needs of its users. In this section, we will be diving into the metrics we chose to evaluate the effectiveness of the system.
The proposed system metric bases its decisions on two metrics, one of each type. We will be introducing each one in the following subsections.
\subsubsection{ Group Validation}
The first metric introducedin the proposed system is Groud Validation; it is a modular system metric introduced in \cite{Group_Validation_in_Recommender_Systems}. It works by slicing the users present in the dataset into different clusters, and then measuring the chosen accuracy metric, cluster by cluster, this way the accuracy is better tracked across different groups.
Studies have shown that evaluating the system in smaller slices of the datasets can provide more nuanced insights \cite{ovaisi2022rgrecsys, chung2019automated, chung2018slice, chung2019slice}. One of the papers studied the accuracy on 2 groups divided by gender, the system was better for male users than for female users. These studies show that evaluating the accuracy on smaller groups will give us more insights on how the system is working and where it might be lacking, it also presented a need to have good information on the user before making these slices, these methods could not be applied in systems, where the dataset lacks information. Group Validation took these gaps into consideration, and implemented a clustering method that works on most common datasets: rating-based datasets, they utilized the k-means algorithm to form these slices, k-means groups data in n clusters where their variance is equal. 
\\
The proposed framework is modular in nature, in other words, other clustering techniques can be utilised as per the system's needs. To effectively choose the number of clusters a system should be forming, they followed the elbow method. The main goal behind Group Validation is to be able to see how many clusters were affected negatively by invalid user behavior.
\\
As Group Validation is a modular framework, the accuracy metric that would be measured in groups(clusters) is relative to the user. We utilized multiple system-centric accuracy metrics: \textit{nDCG}, \textit{Precision}, \textit{Recall} and \textit{F1-Score}. It falls between both system-centered and user-centered metrics. 

The second metric taken into consideration is called Serendipity which will be explained in the following subsection.
\subsubsection{Serendipity}
\mbox{}\\[0.5em]
It has been investigated in multiple papers, such as \cite{badran2019adaptive} \cite{ge2010beyond} \cite{al2023exploring} \cite{kotkov2018investigating} \cite{al2018Serendipity} that \textbf{Serendipity} in RS, is important to enhance user satisfaction. Serendipity is defined as the "happy coincidence" of finding unexpected relevant items. 

\cite{kotkov2016survey} explained that at any moment, there is an intersection between what users perceive as unexpected, familiar, relevant, novel and serendipitous. A lot of methods were proposed to measure the Serendipity in a system, for example: \cite{onuma2009tangent} proposed the TANGENT where items are recommended mainly from user's group, but from others as well.



Another paper that addresses Serendipity is \cite{nakatsuji2010classical} where it developed a recommendation algorithm that focuses on novelty aspect. The authors enhanced the kNN method by switching the similarity method with a relatedness method, so instead of forming a neighborhood of similar users, it forms one of related users. It enhances the novelty of recommendations compared to classical kNN.
Similarly \cite{said2013user}'s goal was also to improve novelty where they proposed "k-furthest neighbor (kFN) recommendation algorithm".  kFN is meant to manage the bias towards popular items that kNN struggles with, by selecting items that different users dislike, it forms neighborhood of users different to a target user. Even while this algorithm is recommending more novel items, it was questionable whether these items are really serendipitous or not.
More Recent studies on Serendipity aim to improve it.
\\ 
\cite{badran2019adaptive} expressed that Serendipity lies in the intersection between relevance, novelty and unexpectedness. \cite{yan2020Serendipity} calculated Serendipity of item \textit{i} using the following equation:
\begin{equation}
\text{Serendipity}(i) = \text{unexpectedness}(i) \times \text{relevance}(i)
\label{eq:Serendipity}
\end{equation}

\cite{al2023exploring} chose to use one of the many ways to calculate unexpectedness, by calculating the cosine similarity between a user's recommended items tagged with (I), and historical connections (H). As for the global Serendipity, it is calculated by averaging all users (U) and all recommended items (I).

\begin{equation}
\left\{
\begin{array}{l}
\text{unexpectedness}(i) = \text{unexpectedness}(I, H) = \frac{1}{|I|} \sum_{i \in I} \sum_{h \in H} \cos(i, h) \\
\text{relevance}(i)=
\begin{cases} 
1 & \text{if the item was interacted with} \\
0 & \text{if not}
\end{cases}
\end{array}
\right.
\label{eq:breakdown}
\end{equation}

\cite{al2023exploring} took into consideration the above equation \ref{eq:Serendipity} to measure Serendipity in the system. Their method uses "social network theory" \cite{AWODELE_2023}, leveraging weak ties to improve recommendation performance. By utilizing weak-linked groups for item recommendations, their approach showed an improvement in the unexpectedness and surprise for users.

\section{Design and Architecture}
\label{sec:framework}

The previous sections define the environment of each layer aiming at laying the foundation, before diving into the specifics of the proposed systems.
We propose two systems: the first is a meta-heuristic framework, that aims at identifying noise hwile ensuring the disctinct noise definitions are taken into consideration; the second is a 2d evaluation system, mixing both system and user metrics to ensure user satisfaction.
In this section, we will provide an overview of the framework's design, inspiration and functionality in Section \ref{subsec:nds-proposal}, along with the evaluation system in Section \ref{subsec:evaluation_proposed}.

\subsection{Noise detection system}
\label{subsec:nds-proposal}
\begin{figure}[h!]
    \centering
    \caption{Secure Network Architecture}
    \includegraphics[width=\textwidth]{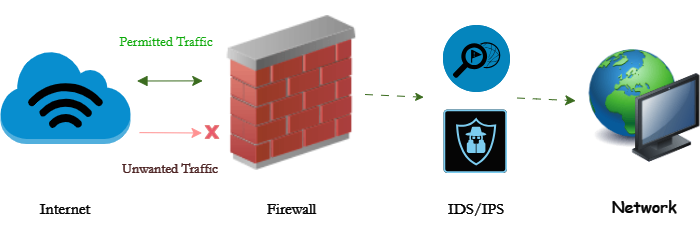}
    
    \label{fig:Secure_Network_Architecture}
\end{figure}

Noise isn’t limited to Recommender Systems; it can be found across various systems both on and off the internet, and it can be defined in numerous ways. Noise refers to any element that isn’t reflective of the truth. While its ‘signature’ can change from system to system, the core concept remains the same. Noise can manifest in different forms, as stress cluttering the perspective on life, as rumors tarnishing someone’s reputation, as a malicious transaction undermining a bank account's integrity, or as a cyberattack compromising a network.
All these use cases can be categorized as noise attacks, as they affect the correct understanding of a specific profile. When we examine how various systems mitigate introduced anomalies, various techniques can be found. Yet, the core idea behind identifying it does not necessarily need to change from an environment to another, as the underlying problem remains the same, only its symptoms vary.
\\ 
Noise management has been widely studied cybersecurity. Given the sensitivity of the topic, cybersecurity has been a major focus for many researchers. Our proposed framework draws an analogy to Intrusion Detection Systems (IDS) used in cybersecurity. Just as an IDS detects and an IPS filters out malicious activities (or 'noise') that threaten the integrity of a network, by using both signature and anomaly based detections, our framework identifies and removes 'noisy' data that can distort recommendations.
\\ 
To ensure the protection of private networks, firewalls, Intrusion Prevention Systems (IPS) and Intrusion Detection Systems (IDS) \cite{al2022risk} are typically used (Figure \ref{fig:Secure_Network_Architecture} illustrates the architecture of a secure network).
\FloatBarrier
\cite{gouda2007structured} compared a firewall to a security guard positioned at the entrance between a \textit{private network} and the \textit{Internet}. The firewall checks every incoming and outgoing packet ensuring that only safe and authorized data passes through.

In such systems, the firewall comes as a first layer protection followed by an IPS and/or an IDS to make sure the data entering the network is valid and threat free. \cite{karatas2018neural} defined an Intrusion Detection System as a software developed to identify attacks that could harm a network or system.
While \cite{khraisat2019survey} explained that failing to prevent the intrusions on a system can degrade the security level, such as integrity, availability and data confidentiality. IDS are classified into Signature-based and Anomaly-based. The first searches for known patterns resembling known attacks, while the second builds a normal behavior model and flags anomalies. The steps of SIDS and AIDS are illustrated in the
Figure \ref{fig:sids} and Figure \ref{fig:aids}.
\cite{bai2003intrusion} highlighted that IDS which only utilize one of these methods will be limited. IDS typically use both AIDS and SIDS synchronously.  Such IDS can be called a Hybrid IDS \cite{khraisat2019survey}. The authors also mentioned that utilizing ensemble models instead of a single algorithm demonstrates better predictive performance.
Hybrid Ensemble IDSs have been studied in \cite{khraisat2020hybrid, mhawi2022advanced, naz2022Ensemble, thockchom2023novel, hajj2023cross, psathas2024hedl}.

\begin{figure}[h!]
    \centering
    
    \begin{minipage}{0.4\textwidth}
        \centering
        \caption{SIDS: Signature-Based IDS}
        \label{fig:sids}
        \begin{tikzpicture}[node distance=1cm, every node/.style={rectangle, draw, fill=blue!20}]
        
        \node (packets1) {Network Packets};
        \node (pattern) [above=of packets1] {Pattern Matching};
        \node (signature) [above=of pattern] {Signature DB};
        \node (alert1) [above=of signature] {Trigger Alerts};

        \draw[->] (packets1) -- (pattern);
        \draw[->] (pattern) -- (signature);
        \draw[->] (signature) -- (alert1);
        
        \end{tikzpicture}
        
    \end{minipage}
    \hfill
    \begin{minipage}{0.4\textwidth}
        \centering
        \caption{AIDS: Anomaly-Based IDS}
        \label{fig:aids}
        \begin{tikzpicture}[node distance=1cm, every node/.style={rectangle, draw, fill=green!20}]
        
        \node (packets2) {Network Packets};
        \node (profile) [above=of packets2] {Build Normal Profile};
        \node (anomalies) [above=of profile] {Detect Anomalies};
        \node (alert2) [above=of anomalies] {Trigger Alerts};

        \draw[->] (packets2) -- (profile);
        \draw[->] (profile) -- (anomalies);
        \draw[->] (anomalies) -- (alert2);
        
        \end{tikzpicture}
        
    \end{minipage}

\end{figure}

Our proposed Framework is designed to be modular, allowing users to customize each layer based on their requirements. Just like how an IPS works to detect and filter out the intrusions, our proposed framework will operate similarly. It will use both anomaly detection methods, as well as signature-based methods to detect noise in the system. The overall structure of the framework, along with the specific steps involved, is shown in Figure \ref{fig:FrameworkArchitecture}.

It is classified into three distinct layers, each serving a specific function in the noise filtering process:
\begin{enumerate}

    \item Layer 1: Natural Noise Algorithms:
Multiple Noise Filtering (NF) algorithms from different categories are applied to identify and filter out noisy data. They will act as a decision board. Items are categorized into three groups:
    \begin{enumerate}
        \item
         \textbf{Noisy items}: Classified as noisy by every member of the decision board.
        \item
         \textbf{Clean Items:} Classified as non noisy by every member of the decision board.
        
        \item \textbf{Uncertain Items:} Where the NF algorithms do not reach full consensus.
    \end{enumerate}

\item Layer 2: Ensemble Learning Algorithm:
        Ensemble learning model is utilized to classify the uncertain items that are passed down from the previous step.

\item Layer 3: Signature Algorithms:
        Noise is identified using known noise patterns. This layer can be positioned as the first step, before initial noise removal, if the patterns are independent of noise classification, ensuring all detected noisy items are filtered out. Alternatively, it can be placed later if a classified dataset is needed.
\item Removal of noisy items from the database.
\end{enumerate}
\par
The proposed framework is modular in design; the chosen algorithms used in each layer as well as the selection process will be explained in detail in Section \ref{sec:chosenalgo}.
Recommender Systems should be judged based on their ability to meet individual user needs and purposes \cite{mcnee2006being}. We will dive into the evaluation system's details in the next subsection.
 \\

\begin{minipage}{\textwidth}
\captionof{figure}{The architecture of the proposed framework}
    \label{fig:FrameworkArchitecture}
\includegraphics[width=\textwidth]{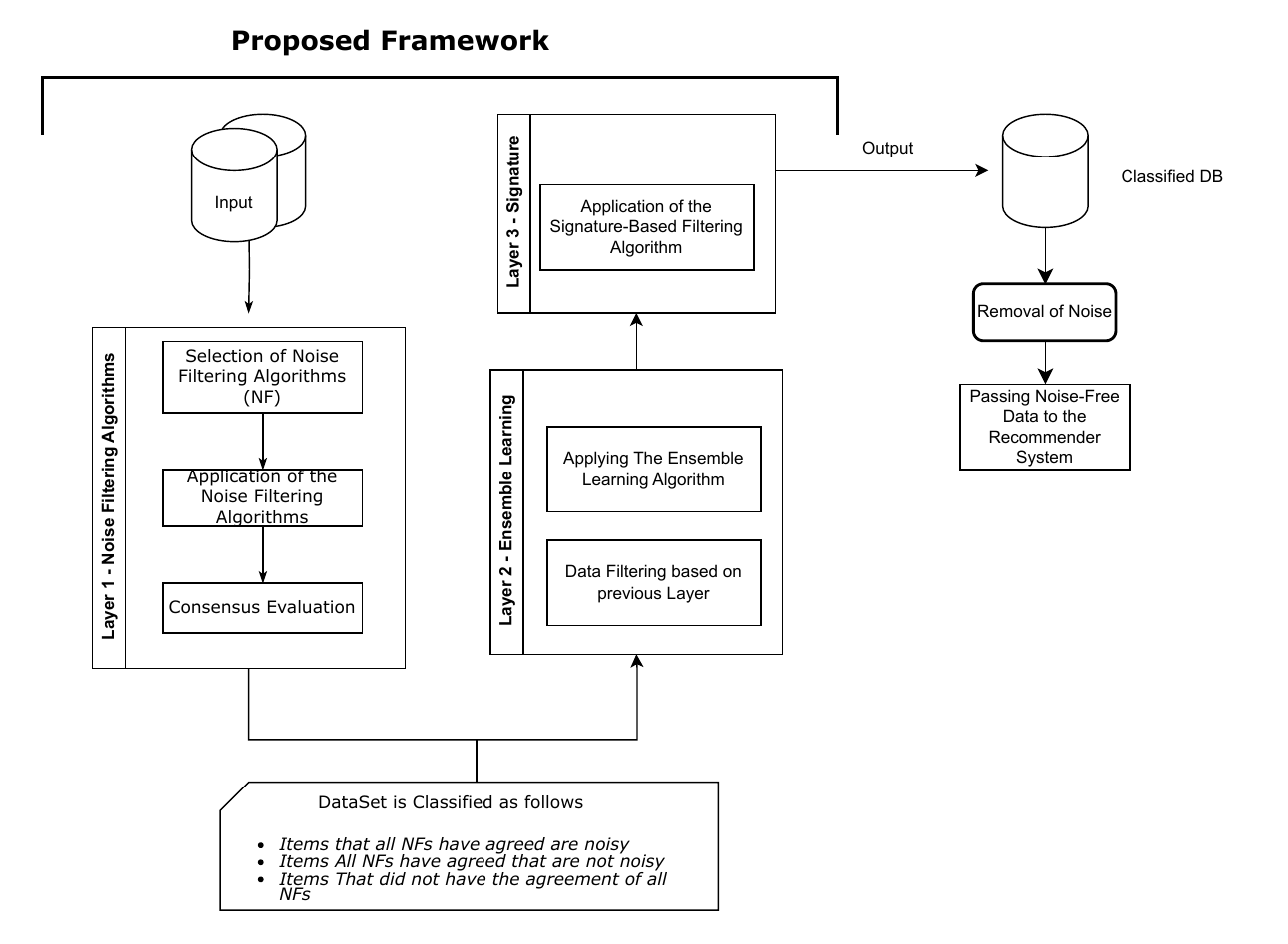}
\end{minipage}
\vspace*{-\baselineskip}
\subsection{Evaluation System}
\label{subsec:evaluation_proposed}
As previously mentioned, we propose a 2d evaluation system, taking into consideration both system and user metrics. The idea behind the proposed evaluation system is to look into the improvement in the system, comparing the state before and after a noise filtering algorithm is applied.
To study the impact of the action done by an algorithm, we can group the items into 4 sections based on their location on the orthogonal axis.
We will be using Group Validation to measure the accuracy highlighting the impact on groups of users, which showed that performance issues often are hidden from traditional metrics. We will measure user Serendipity to ensure a pleasant user experience.
Picture an XOY axis, the Y axis will display the accuracy improvement for a certain user in their group (new accuracy value - old accuracy value), while the X axis will display the Serendipity improvement for the user (new Serendipity value - old Serendipity value). This graph will be divided into four sections, each section represents a different 'quadrant'.

\begin{tabular}{@{}p{0.5\textwidth}@{} p{0.4\textwidth}@{}}
    \begin{minipage}{\linewidth}
    In theory, we propose that we take the four quadrants to evaluate the impact by rating in Table \ref{tab:quadrants_repartition}: 
        \begin{itemize}
            \item Both Group Validation and Serendipity have increased. In this case, the items removed are noisy, as their effects are negative on the system. This corresponds to Quadrant I in Figure \ref{fig:quadrants}.
           
            \item Both Group Validation and Serendipity have decreased. In this case, the items removed are not noisy, as their effects are positive on the system. This corresponds to Quadrant III in Figure {fig:quadrants}.
            
            \item If Group Validation and Serendipity do not deviate in the same way (one increases and the other decreases), a threshold value, alpha, is proposed to divide the noisy from actual values. This corresponds to Quadrants II and IV in Figure {fig:quadrants}.
            Alpha, is the threshold that controls how much influence the accuracy has on the classification of the item (data point) over the Serendipity. It is entirely up to the user to decide on the alpha based on what their requirements.
        \end{itemize}
    \end{minipage} &
    \begin{minipage}{\linewidth}
        \centering
         
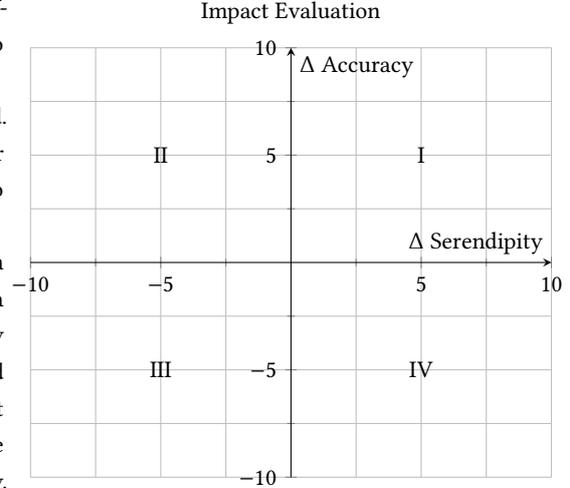
\captionof{figure}{Quadrants for Impact Evaluation}
        \label{fig:quadrants}
        \begin{tikzpicture}
            \begin{axis}[
                axis lines=middle,
                xlabel={\(\Delta\) Serendipity},
                ylabel={\(\Delta\) Accuracy},
                title={Impact Evaluation},
                grid=both,
                minor tick num=1,
                xmin=-10, xmax=10,
                ymin=-10, ymax=10,
                ]
                \node at (axis cs:5,5) {I};
                \node at (axis cs:-5,5) {II};
                \node at (axis cs:-5,-5) {III};
                \node at (axis cs:5,-5) {IV};
            \end{axis}
        \end{tikzpicture}
        \vspace{-0.5em}
       
    \end{minipage}
\end{tabular}
\begin{table}[htbp!]
    \centering
    \caption{Evaluation of Impact by Quadrants}
    \label{tab:quadrants_repartition}
    \begin{tabular}{|c|c|c|c|}
        \hline
        \textbf{Item} & \textbf{Group Validation} & \textbf{Serendipity} & \textbf{Quadrant} \\
        \hline
        Item 1 & Increased & Increased & I \\
        \hline
        Item 2 & Decreased & Decreased & III \\
        \hline
        Item 3 & Increased & Decreased & IV \\
        \hline
        Item 4 & Decreased & Increased & II \\
        \hline
    \end{tabular}
    
\end{table}


    

Now that the structure of the two systems have been introduced and defined, we will proceed to detail the experiment conducted to test and utilize them.

\section{Environment Parameters}
\label{sec:env_var}
In the following section, we will cover the environment parameters used throughout this paper, including datasets, metrics, and Recommender Systems.
Based on MovieLens dataset, we randomly sampled Subset 1 to represent a large-scale scenario with 5M ratings, while Subset 2 (2.1M ratings) provides a moderate-scale environment for additional testing and validation.
Table \ref{tab:movielens_statistics} presents the details of the two subsets extracted from the MovieLens 25M dataset.
\begin{table}[h!]
\centering
\caption{MovieLens Dataset Statistics \\ \url{https://grouplens.org/datasets/movielens/}}
\begin{tabular}{lrrr}
\toprule
\textbf{Statistic} & \textbf{ML-25m-subset1} & \textbf{ML-25m-subset2} \\
\midrule
Number of Users & 25,873 & 5,624 \\
Number of Movies & 38,501 & 27,320 \\
Number of Ratings & 5,000,000 & 2,114,467 \\
Rating Scale & 0.5-5 & 0.5-5 \\
\bottomrule
\end{tabular}
\label{tab:movielens_statistics}
\end{table}
\FloatBarrier

To evaluate Group Validation metrics, we employed the method introduced in \cite{Group_Validation_in_Recommender_Systems} and to measure Serendipity, we utilized the methodology described in \cite{al2023exploring} ensuring a robust framework for our analysis. As for system accuracy metric, we utilzed nDCG, Recall, Precision and F1-Score. The algorithms were chosen because they are widely recognized and utilized in research on noise filtering algorithms. Their established use provides a reliable basis for comparison and validation of employed noise filtering algorithms.

\begin{figure}[H]
\centering
\caption{Proposed evaluation system's architecture}
\includegraphics[width=0.8\textwidth]{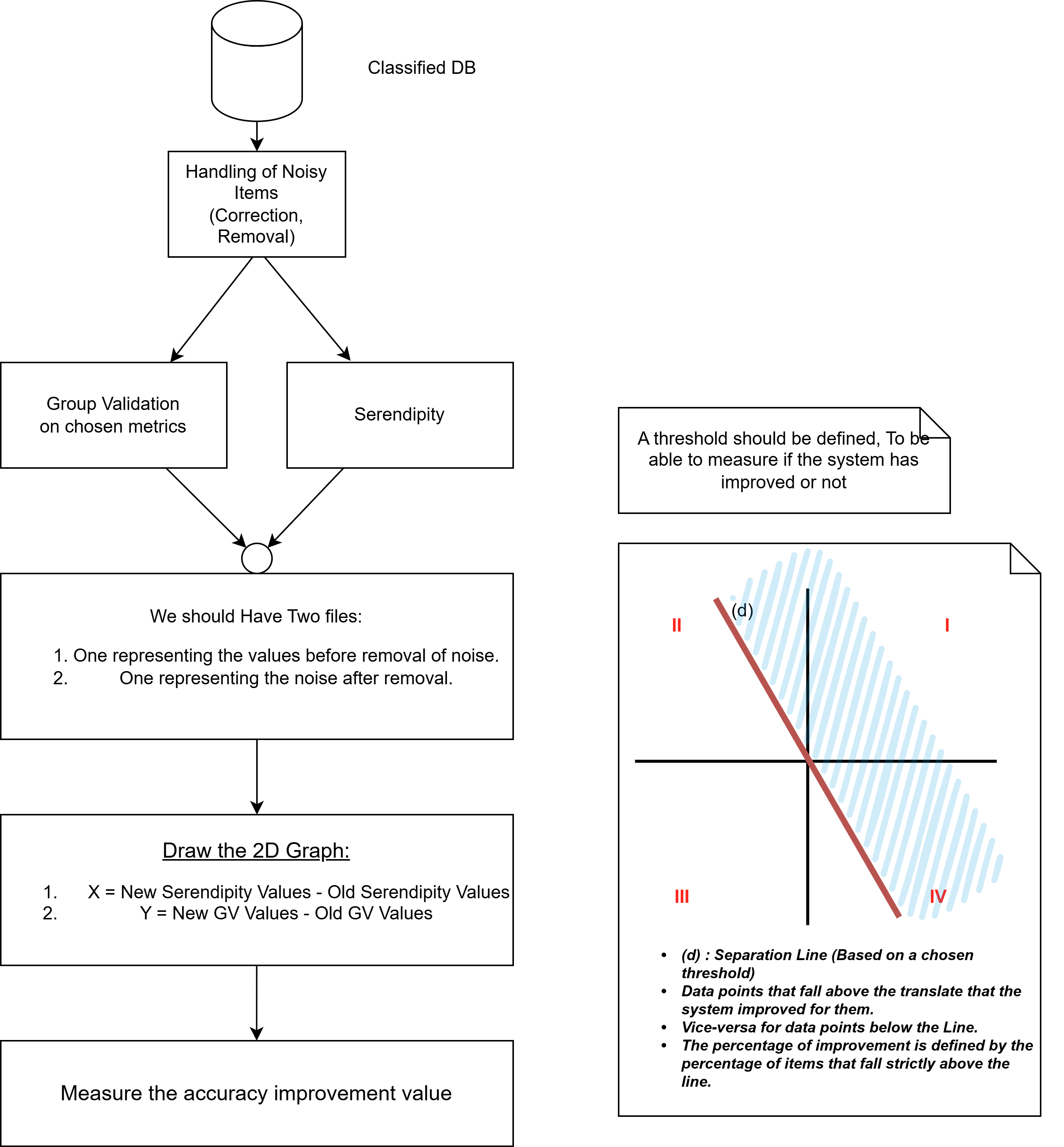}
\label{fig:2Daccuracy}
\end{figure}

In Recommender Systems, metrics like Recall, Precision, Normalized Discounted Cumulative Gain (nDCG) and F1-Score are essential for evaluating model performance. Precision measures the accuracy of the items recommended, indicating the proportion of relevant items among the recommended ones, while Recall assesses the model's ability to identify all relevant items, representing the proportion of relevant items retrieved from the total available \cite{Alkhalil2025}. When either the Precision or Recall are unevenly distributed, the F1-Score serves as a harmonic mean between them, its job is to produce a value that balances out these two metrics. Meanwhile, nDCG complements these metrics by focusing on the position of relevant items in the ranked list, giving higher importance to relevant items appearing earlier in the list. These metrics enable an efficient evaluation of RS, ensuring they not only provide accurate recommendations but also deliver them in a user-friendly manner.
In the subsequent sections, we will denote the system metrics that we utilized 
nDCG, F1, Recall, and Precision metrics to evaluate the noise filtering algorithms.
Figure \ref{fig:2Daccuracy} illustrates the step-by-step process by which the proposed evaluation system functions.
These metrics were utilized in the Group Validation evaluation method on the Y axis of the 2D evaluation system, alongside Serendipity on the X axis, forming four distinct evaluation systems. These 4 evaluation metrics are labeled as follows:
\begin{itemize}
    \item nDCG-Serendipity
    \item F1-Serendipity
    \item Recall-Serendipity
    \item Precision-Serendipity
\end{itemize}

To further refine the analysis, a threshold was proposed to determine whether the improvement was significant enough to be counted as positive improvement.
In this paper we will be using the following equation as the plane separating positive improvements from the negatives: 
\centering
\begin{equation}
 0.07 \cdot X + 0.17 \cdot Y = 0
\label{threshold_equation}
\end{equation}
\par
\justify
Where X represents the improvement in Serendipity, and Y denotes the improvement in the system metrics specified above. The threshold (the variable) is modular. In our experiment, we prioritized accuracy over Serendipity, while the system works on still identifying Serendipity, good Serendipity without good accuracy is only accepted where the accuracy falls within the chosen threshold.

All data points falling above this line will be counted positive, while all data points on and under this plane are considered negative.
This threshold assigns greater weight to accuracy compared to Serendipity, as reflected by the higher coefficient for Y. The choice of these weights emphasizes the importance of maintaining high accuracy, while ensuring adequate Serendipity improvements.

Critical group metric was introduced by \cite{Group_Validation_in_Recommender_Systems} as the group of users whose preferences and behaviors are particularly important for evaluating the performance and effectiveness of the system. In this experiment, we have use kmeans to cluster the data, with k=100.
These users are the most sensitive to the system, as their accuracy tends to fall below the average accuracy across all users. To determine the critical groups, we compared the cluster accuracy, based on nDCG, to the average cluster accuracy. Clusters with below-average accuracy were classified as critical groups, and the metric was expressed as the percentage of affected clusters. This approach was used in the framework evaluation and for comparisons with other prominent noise management systems.



\section{Experiment Setup - Algorithm Selection}
\label{sec:chosenalgo}

As discussed in Section \ref{sec:framework} of the proposed framework, each layer has its distinct role, contributing to a comprehensive and robust noise detection mechanism. The framework consists of three layers: known noise filtering algorithms, Ensemble Learning, and Signature Analysis. While we already selected the algorithm for layer 3 (Opt-out signature introduced by \cite{al2022strategic} and explained in Section \ref{sec:signature}), we still have to select the algorithms to be included in layer 1 and layer 2.
The selected algorithms are grouped as follows:
\begin{enumerate}
    \item Natural Noise Algorithms
    \item Ensemble Learning
\end{enumerate}

In this section, we will detail the criteria used to select the algorithms implemented in the proposed framework for both noise filtering and Ensemble learning layers. Key factors, such as publication year and citation counts, were considered to ensure that the chosen algorithms reflects those that have been recognized by their effectiveness in the field. For implementation details, please refer to the GitHub repository \cite{Hawat_Understand_your_Users}.

\subsection{Layer 1: Noise Filtering Algorithms Selection}
\label{subsec:NFSelection}
The selection process for the natural filtering algorithms is to decide which algorithms will be included in layer 1. The objective of layer 1 is to measure the consensus on the identified noise by the selected noise filtering algorithms. 
We prioritized well-known noise filtering algorithms, organizing them first by citation count and then by publication year in descending order. We eliminated algorithms that employ re-rating concepts  and those designed for group Recommender Systems.

Based on the mini review done in Section \ref{sec:relatedworks}, we identified the top listed algorithm for each group \ref{tab:AccuracyPredictionPapers} \ref{tab:userItemAlgorithm} \ref{tab:UserCenteredPapers}, Table \ref{tab:NoiseFilteringChoices} shows the chosen algorithms by category.

\begin{longtable}{|p{0.3\textwidth}|p{0.1\textwidth}|p{0.1\textwidth}|p{0.4\textwidth}|}
\caption{Chosen Noise Filtering Papers} \label{tab:NoiseFilteringChoices} \\
\hline
Category & Fuzzy & Crisp & Selection Notes \\
\hline
\endfirsthead
\multicolumn{4}{c}{{\tablename\ \thetable{} -- continued from previous page}} \\
\hline
Category & Fuzzy & Crisp & Selection Process \\
\hline
\endhead
\endfoot
\hline
\endlastfoot
User Centered Filtering & - & \cite{yu2016novel} & \cite{pham2013preference} was discarded because it relies on re-rating. \\
\hline
User-item Interactions Centered & \cite{yera2016fuzzy} & \cite{toledo2015correcting} & - \\
\hline
Accuracy Enhancement Ability & - & \cite{o2006detecting} & \cite{amatriain2009like} got discarded because it doesn't identify noisy items and quantifies noise across the DB, \cite{amatriain2009rate} was discarded because it relies on re-rating \\
\hline
\end{longtable}


We will start by providing an overview of each selected algorithm. This will be followed by applying these algorithms to a dataset, outcomes and patterns observed on the results will be analyzed and discussed by the end of this subsection.

\subsubsection{Crisp, User-item Interactions Centered (NF1)}

This algorithm gained recognition not only through its citation count but also due to its popularity and utility across multiple papers. It was highlighted in the overview above, specifically in Section \ref{it:useritemcrispnf1} and utilized in the following papers \cite{toledo2015correcting, 
 martinez2016managing, toledo2013managing, dixit2019proposed, bag2019noise}. This paper introduces a novel method for improving collaborative filtering Recommender Systems by addressing "natural noise". They begin their algorithm by classifying users into the following groups:
\begin{itemize}
    \item Benevolent: users who usually give high ratings.
    \item Average: users who usually give average ratings.
    \item Critical: users who usually give low ratings.
    \item Variable: users with variable preferences.
\end{itemize}

Following the same classification process, items were tagged as:
\begin{itemize}
    \item Strongly-preferred: items that are highly preferred.
    \item Averagely-preferred: items that are moderately preferred.
    \item Weakly-preferred: items that are not very preferred.
    \item Variably-preferred: items that do not belong to any of the other categories.
\end{itemize}
Based on these categories, as well as the specific rating categories $R_{u,i}$ as weak, mean and strong, the authors defined three particular sets for each user/item: $W_u$, $A_u$ and $S_u$ for users and $W_i$, $A_i$ and $S_i$ for items.

The authors then describe a set of homologous classes (Table \ref{tab:nf1_Homologous}) that defines the usual behavior for a rating $R_{u,i}$, if the user u and item i, fall into one of these combinations, then the rating $R_{u,i}$, must fall into the chosen rating class, else, the rating is then tagged as possible noisy.

\begin{table}[h]
    \centering
    \caption{Homologous classes.}
    \begin{tabular}{l c c c}
        \hline
        & \textbf{User class} & \textbf{Item class} & \textbf{Rating class} \\
        \hline
        Group 1 & \textit{Critical} & \textit{Weakly-preferred} & \textit{Weak} \\
        Group 2 & \textit{Average} & \textit{Averagely-preferred} & \textit{Average} \\
        Group 3 & \textit{Benevolent} & \textit{Strongly-preferred} & \textit{Strong} \\
        \hline
    \end{tabular}
    \label{tab:nf1_Homologous}
\end{table}

\FloatBarrier

\subsubsection{Crisp, User Centered Filtering (NF2)}

Moving to the second algorithm, the authors of \cite{yu2016novel} address the challenges posed by varying user data. Their approach classifies users based on the amount and quality of their ratings.
\\
\begin{minipage}{.9\textwidth}
\centering
\captionof{figure}{Classification of user groups by engagement level}
\begin{forest}
    for tree = {draw, 
                align=center, 
                anchor=north, 
                s sep=4mm,  
                l sep=4mm,  
                tier/.option=level, 
                edge label/.style = {node[midway, fill=white, inner sep=2pt, font=\footnotesize\itshape, anchor=center, text depth=0.3ex]{#1}} 
                } 
  [User Groups 
    [Heavy 
      [HEUG] 
      [HDUG] 
    ] 
    [Medium 
      [MEUG] 
      [MDUG] 
    ] 
    [Light 
      [LEUG] 
      [LDUG] 
    ] 
  ] 
\label{forest:usercenteredforest}
\end{forest}
\end{minipage}

\FloatBarrier

Users are classified into these categories based on two axes: quantity and quality. The quantity is defined by the number of ratings a user provides, classifying them into 'heavy', 'medium' or 'light' users. While the quality is measured using a metric called coherence. It measures the consistency of a user's rating across items with similar features, resulting in a deeper division of users, into 'easy' (high coherence, consistent) and difficult (low coherence, inconsistent).
The combination of these two factors results in six different groups as presented in the figure \ref{forest:usercenteredforest} above.  
After grouping the users, to identify noise, they used a metric called Rating Noisy Degree (RND). RND is calculated for each rating as the ratio of features with high relative deviation to the total number of features for that item. 

The authors propose different actions to be taken after identifying the noisy items, depending on the user group. This difference in treatment is based on the understanding that a single noise-handling strategy may not be suitable for all user types. They discuss that heavy users, easy or difficult, will have an abundance of ratings, so noise removal can be done. However light users, easy or difficult, and medium-difficult users, with moderate numbers of inconsistent ratings, can be corrected instead of removed. While medium easy users, are left with their inconsistencies as the profile doesn't have enough data as is, so no processing is recommended.

\subsubsection{Crisp, Accuracy Enhanced Centered (NF3)}
The third algorithm falls under the accuracy enhanced centered proposed by \cite{o2006detecting}. The authors propose a framework for removing noise addressing both natural and malicious noise. As the target of layer 1 is to identify natural noise, we will be focusing on the NN algorithm.
They propose that natural noise is detected by comparing actual user ratings to predictions generated by a recommendation algorithm. This process of comparison assesses the accuracy of the recommendations to identify noise. This accuracy is measured by using a normalized Mean Absolute Error (MAE), ratings that exceed the threshold are flagged as noise.
In other words, a rating $R_ui$ is classified as noise, if the consistency(Rui) is bigger than a certain threshold. And This consistency is measured by MAE.

\begin{equation}
c(G(T)_{u,v}) = \frac{r_{x,v} - p_{u,v}}{r_{max} - r_{min}}
\label{eq:normalizedMAE}
\end{equation}

They explain that certain prediction/recommendation algorithms are accurate to a certain point, this is where they set their threshold.
\begin{equation}
c(G,T)_{u,v} > th -> R_{u,v} \text{is Considered as Noise}
\end{equation}
As to identify the noise the authors explored different thresholds, including 0.01, 0.05 0.1 and higher values. They showed that at the lowest threshold (\textit{th = 0.01}) accuracy performance was the poorest. Although they don't specify the best threshold to use, they present that a threshold of 0.05 provides good balance for the datasets due to its performance in accuracy and coverage.

They've employed the k–nearest neighborhood scheme, with neighborhood size chosen by experiment in order to optimize predictive accuracy, the number of neighboors (k) is this experiment was set to 35 for the MovieLens dataset, while considering the Pearson correlation similarity metric. The Goal is to weight the influence of similar users or items based on their correlation, because weights derived from a small numbers of co–rated items are assigned lower significance.

The authors highlighted that to be able to make such predictions while making sure the system does not overfit, there is a need to assume that a training set T in the database exists, on which all predictions are based. The selection of the members of the training dataset, usually falls on the system administrator's responsibility, as they choose known trusted users.

After identifying the noisy items, the authors propose to remove the noisy items from the recommendation process to enhance the performance of RS by eliminating inconsistent ratings from the system.

Overall this algorithm's approach by using MAE and the threshold focuses on the accuracy logic path. It provides a straightforward method to identify noise by quantifying prediction accuracy.

\subsubsection{Fuzzy, User-item Interactions Centered (NF4) }

The final algorithm selected for the classic noise detection layer (layer 1) takes a different approach compared to the previous algorithms. This algorithm employs a fuzzy logic approach. Unlike a traditional crisp method relying on strict thresholds or prediciton accuracies, this algorithm developed in \cite{yera2016fuzzy} uses fuzzy sets to handle variability in user and item profiles.
The authors argue that existing noise filtering algorithms are too rigid, failing to adequately manage the uncertainty  and vagueness in user preferences.
It involves 3 major steps:
\begin{enumerate}
\item Fuzzy Profiling: To handle rating uncertainties, ratings are converted into fuzzy linguistic labels, creating fuzzy profiles for users, items, and ratings. The function in \ref{graph:f1_graph} shows \textit{f1} used to amplify these profiles, using a "computing with words" process.
\begin{center}
\begin{minipage}{.4\textwidth}
\captionof{figure}{Fuzzy transformation function}
\includegraphics[width=\linewidth]{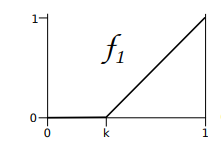}
\label{graph:f1_graph}
 \end{minipage}
\end{center}
 
\textit{}

\item Noise Detection: Using the created profiles, noisy ratings are identified. This involves an initial pre-filtering to remove unclear profiles, and the classification process that uses the distance between the rating profile and both the user and item profiles, using a minimum t-norm.
To get The eligible profiles, the Manhattan distance is utilized to calculate the distance between profiles, the authors used the following formula to calculate this distance:
\\
$d_{Manhattan}(x, y) = \sum_{k} |x_k - y_k|$ 
\\
where x and  \textit{y}  are the profiles being compared, k indexes the elements ("low," "medium," "high") within each profile, and $|x_k - y_k|$ represents the absolute difference between corresponding elements.
 \\
It has been demonstrated that the value  \(( d(p_{\text{Ru}}, p_{\text{Ri}}) )\) consistently falls within the range [0, 2]. Consequently, \((\delta_1)\) is set to 1 to identify profiles as similar if their dissimilarity is less than 50\% of the maximum possible distance, as indicated by Equation Above using the Manhattan distance.
The distance between the closest profiles of both the user and the item is evaluated against the rating profile, and this distance is then compared to the $\delta^2$ threshold.

\item Noise Correction: Ratings identified as noisy are then corrected using a combination of the original ratings and a new prediction made by a traditional Recommender Systems trained on the non-noisy ratings and while utilizing the noise degree.
\end{enumerate}
\subsection{Layer 1: Results and Analysis}
\label{subsubsec:layer1}

As we now have selected the noise filtering algorithms utilized in layer 1, we ran these algorithms against the two subsets from the movielens dataset.
Users with few ratings provide insufficient information, which can lead to noise and bias in the analysis. By focusing on more active users, the model can generate more accurate and robust predictions. A threshold of 50 was set arbitrarily to filter the dataset, users with less than 50 ratings, were removed from the dataset, and the removal of removing duplicate ratings, before actually passing by the algorithms. 
\begin{table}[h!]
\centering
\caption{Layer 1: Venn Diagrams for Different Datasets}
\renewcommand{\arraystretch}{1.5} 
\setlength{\tabcolsep}{10pt}     
\begin{tabular}{|c|c|}
\hline
\textbf{Dataset} & \textbf{Venn Diagram} \\
\hline
ML-25M-Subset (1) & 
\begin{minipage}{0.4\linewidth}
    \centering
    \includegraphics[width=\linewidth]{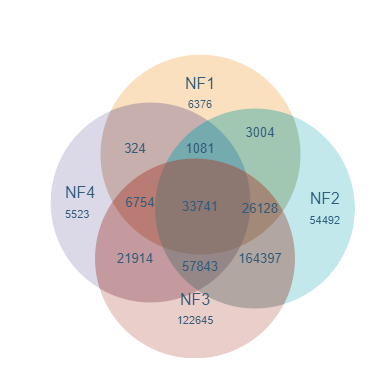}
    $\overline{NF1}$ $\overline{NF2}$ $\overline{NF3}$ $\overline{NF4}$ = 71035
    \vspace{5pt}
\end{minipage} \\
\hline
ML-25M-Subset (2) & 
\begin{minipage}{0.4\linewidth}
    \centering
    \includegraphics[width=\linewidth]{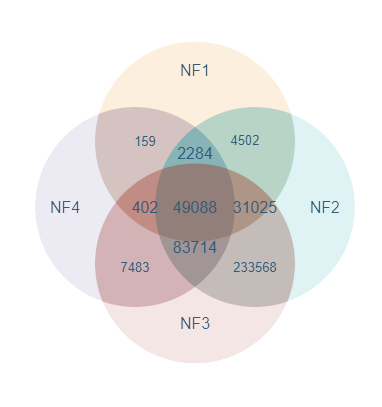}
    $\overline{NF1}$ $\overline{NF2}$ $\overline{NF3}$ $\overline{NF4}$ = 54755
    \vspace{5pt}
\end{minipage} \\
\hline
\end{tabular}
\label{tab:venn_diagrams}
\end{table}

\clearpage

In the identification process of NF3, it was emphasized that a training and a testing set are needed. To ensure that all algorithms are built on the same foundation, we began with NF3, split the dataset, and focused only on the testing dataset for all algorithms. This approach ensures that the benchmark among all four algorithms is based on the same solid foundation.
In order to identify the diversity in decisions taken by the algorithms, we utilized a Venn diagram to understand the overlap and intersection between noise management algorithms' classification, displayed in Table \ref{tab:venn_diagrams}.
The number displayed inside the Venn diagram, represent the number of overlapped noisy ratings between algorithms on noise classification. While the description of non noisy items, can be found at the top of the page showing the number of ratings that reached consensus on all noise filtering algorithms as non noisy rating.


The Venn diagrams in Table \ref{tab:venn_diagrams} offer a visual comparison of how each selected noise filtering classifies noise. The diagrams clearly illustrates notable variation in their results. We observe that NF2 and NF3 consistently classify the most data as noisy, followed by NF4, and NF1. This ranking pattern holds consistently across multiple datasets, indicating NF2 and NF3's heavy noisy classification tendencies, although we are still not sure whether this heavy rating is valid or not, we'll be able to measure this at the end of the testing section. 
 
Among the algorithms, NF1 is the quickest to deliver results. The diagrams also highlight the disparity in how each algorithm defines noise. NF1 and NF4 show moderate noise classification but lack consensus, suggesting they apply a different definition to noise.

While examining the overlap, only 17.958\% approx. of noise management on ML-25M-Subset(1) and 16.17\% approx. on ML-25M-Subset(2) gave consistent results by all algorithms. This variation is observed across both datasets, indicating that the variation in noise classification is not limited to a single case but is a recurring pattern across various scenarios. This \textit{non-consensus} emphasizes the need for a more unified approach in identifying noise. Since they all intend to capture unintentional natural noise, their results should align more closely to reflect this shared definition.



In summary, the analysis of noise classification across the algorithms reveals significant variations, with only partial consensus. This consistent pattern is observed across all datasets, which underscores the need for a more unified approach by an additional decision maker to finalize the consensus between them. To move from layer 1 to layer 2, the focus will be on ratings that did not reach consensus, as they need further assistance in classification. In the next section, we will explore the selection of the Ensemble learning algorithms that will form the foundation of layer 2.
\clearpage

\subsection{Layer 2: Selection of an Ensemble Learning Algorithm }  

The selection process for the Ensemble learning algorithm in layer 2 aims to identify the most suitable method to label the dataset remaining from layer 1. Our main focus was the newest Ensemble learning algorithms, so
we organized the algorithms first by publication year, followed by the number of citations in descending order. We excluded papers that focus on specific subjects, ensuring the algorithms align with our broader objectives. Applying these criteria on the reviewed Ensemble learning algorithm in \ref{sec:elreview}, resulted in the selection of five algorithms, one for each category:
\\ 
 \begin{minipage}[b]{0.9\textwidth}
\begin{longtable}{|p{0.2\textwidth}|p{0.2\textwidth}|p{0.2\textwidth}|p{0.2\textwidth}|p{0.2\textwidth}|}
\caption{Chosen Ensemble Learning Papers} \label{tab:EnsembleLearningChoices} \\
\hline
& Bagging & Stacking  & Boosting & Other \\
\hline
Supervised & \cite{altman2017Ensemblev1} (EL1)  & \cite{liang2021stacking} (EL2)&  \cite{prokhorenkova2018catboost} EL(3) &- \\
\hline
Semi-Supervised & \cite{de2021reliable} (EL4) & - & - & - \\
\hline
Unsupervised & - & - & - & \cite{hariri2019extended} (EL5) \\

\hline

\end{longtable}
\end{minipage}
\FloatBarrier
Similar to the process used for selecting algorithms for layer 1 in the previous section, we will begin by introducing the winning algorithm before proceeding to examine the final results and their analysis.
\subsubsection{Supervised, Bagging - EL1} 

The selected algorithm for the supervised bagging category is Random Forest. \cite{altman2017Ensemblev1} explained that the Random Forest model was built on the bagging algorithm while integrating variable selection methods. This algorithm was first introduced in \cite{ho1995random} and was later extended in \cite{breiman2001random}.
This algorithm works by applying the bagging technique to create multiple decision trees that work collaboratively to provide an answer to a specific question using boostrapped samples. It trains the Decision Trees from samples generated by the bagging technique generates.
The authors explain that Ensemble methods like bagging and random forest are used for solving a lot of issues that result from a weak or noisy dataset, such issues can be like underfitting and overfitting. 
In each bootstrap iteration, the usage of out-of-the-bag (OOB) sample with each bootstrap works similarly to using a test set, allowing the entire dataset to be used for model estimation and performance evaluation.
\par
\subsubsection{Supervised, Stacking - EL2} 

Stacking is generally considered a parallel Ensemble learning method. In Stacking, multiple base models (or learners) are trained independently on the same dataset. 
As the studies on Stacking found in the review were all dedicated for certain topics, we implemented a novel stacking approach that diverges from the methodologies typically discussed in the existing surveys. In this study, we developed two variations of the stacking algorithm to further evaluate its effectiveness. These variations were designed to explore different models, let's dive into their details:
\noindent
\subparagraph{Algorithm EL$2$:}
\begin{itemize}
       \item Base Learners:
    \begin{itemize}
        \item KNeighborsClassifier (KNN): This model captures local structures in the data, making it useful for identifying patterns based on proximity. Its simplicity and effectiveness in various scenarios add robustness to the Ensemble.
        \item DecisionTreeClassifier (CART): As a non-linear model, it can handle complex interactions and is interpretable, providing insights into feature importance.
        \item Support Vector Classifier (SVC): Known for its performance in high-dimensional spaces, SVC is effective for classification tasks, especially when data is not linearly separable.
        \item GaussianNB: This probabilistic model assumes independence among features and is effective for data with Gaussian distributions, adding a probabilistic perspective to the Ensemble.
    \end{itemize}
    \item Meta-Learner:
    \begin{itemize}
        \item LogisticRegression: Used as the meta-learner, Logistic Regression effectively combines the predictions from the base learners. Its sigmoid function allows for probabilistic interpretation, making it suitable for binary classification tasks.
    \end{itemize}
\end{itemize}
\par
\raggedright
\begin{minipage}{\linewidth}
\subparagraph{Algorithm EL$2.2$:}
\begin{itemize}
    \item Base Learners:
    \begin{itemize}
        \item RandomForestClassifier: This Ensemble method combines multiple decision trees to improve generalization and robustness. It reduces the risk of overfitting, enhancing predictive performance on unseen data.
        \item ExtraTreesClassifier: Similar to Random Forest, it utilizes random subsets of features and samples, providing greater randomness. This leads to increased diversity among the base learners, reducing variance.
        \item LogisticRegression: Including this model as a base learner allows for linear modeling of the data, complementing the non-linear base learners and providing a well-rounded approach to classification.
        \item DecisionTreeClassifier (CART): Provides the same benefits as in Algorithm 1, contributing to the overall robustness of the Ensemble.
        \item GaussianNB: Maintains its role as a probabilistic model, adding yet another layer of diversity in how predictions are generated.
    \end{itemize}
    \item Meta-Learner:
    \begin{itemize}
        \item LogisticRegression: Using Logistic Regression again as the meta-learner ensures consistency across the two algorithms while leveraging its capability to interpret combined predictions effectively.
    \end{itemize}
\end{itemize}
\end{minipage}
\par
\raggedright
\subsubsection{Supervised, Boosting - EL3}

The selected algorithm of this category, the Categorical Boosting algorithm, also known as CatBoost. This algorithm was introduced in \cite{prokhorenkova2018catboost} in 2017. 
As the name suggests, 'Cat'-'Boost' works on categorical data while it utilizes gradient boosting. 

 \cite{prokhorenkova2018catboost} conducted a comprehensive benchmark of various algorithms and CatBoost, evaluating their performance across a range of datasets using metrics such as log-loss and zero-one loss. They showed that CatBoost outperforms other boosting algorithms, such as XGBoost, LightGBM in terms of quality on various datasets (Amazon, Epsilor, Click Prediction and many more).
 \noindent
 \FloatBarrier

\subsubsection{Semi Supervised Ensemble Learning Algorithm - EL4}
Semi-Supervised (SS) algorithms in EL are recognized and studied, particularly for their ability to improve model performance when labeled data is scarce.
RESSEL Was introduced in \cite{de2021reliable} in 2021 as a wrapper method, it is used with known supervised learning algorithms. RESSEL, or Reliable Semi-Supervised Ensemble Learning, consists of: SS learning and EL, in a way that they enhance one another. It integrates bagging with self-training and incorporates an early-stopping mechanism following oob error.
We implemented two algorithms following the RESSEL method, and we utilized the models that were used in this paper to test the algorithm, we included 2 of them in the first algorithm and tagged this algorithm \#4.1, and utilized all the machine learning models used in \cite{de2021reliable} and tagged this algorithm \#4.2.

For \#4.1, we utilized the following classifiers:
\begin{itemize}[noitemsep, topsep=0pt]
    \item Random Decision Tree (RDT):
     \begin{itemize}
         \item Parameters: max depth=4, max features='sqrt'
     \end{itemize}
    \item Stochastic Gradient Descent (SGD) Classifier:
    \begin{itemize}
        \item Loss Function: log-loss
    \end{itemize}
\end{itemize}

For \#4.2, we adhered to the settings used in the referenced paper. The classifiers implemented are:
\begin{itemize}[noitemsep, topsep=0pt]
    \item Gaussian Naive Bayes (GNB)
    \item Support Vector Machine (SVM):
    \begin{itemize}
        \item Parameters: kernel='rbf', C=1.0, gamma='scale'
    \end{itemize}
    \item K-Nearest Neighbors (KNN):
    \begin{itemize}
        \item Parameters: number of neighbors=10
    \end{itemize}
    \item Random Decision Tree (RDT):
    \begin{itemize}
        \item Parameters: maxdepth=4, maxfeatures='sqrt'
    \end{itemize}
    \item Stochastic Gradient Descent (SGD) Classifier:
    \begin{itemize}
        \item Loss Function: logloss
    \end{itemize}
\end{itemize}

    

\FloatBarrier

The authors state that RESSEL demonstrates robustness to the variations in base classifier hyperparameters expanding the range of acceptable settings.
As we move forward to the next category of the Ensemble learning algorithms, we will examine an innovative algorithm that operates on the unlabeled data or in our case, the uncertain ratings only.

\subsubsection{UnSupervised Ensemble Learning Algorithm - EL5}
\justify
The Isolation Forest algorithm, first explained in \cite{liu2008isolation} in 2008, is a machine learning approach designed for anomaly detection, operating on the principle of isolating outliers within a dataset. It is rooted in the Decision Tree framework and posits that anomalies are rare observations different from the dataset, thus making them easier to identify. 
 \cite{hariri2019extended} introduces an enhanced splitting mechanism that improves anomaly detection in high-dimensional spaces. The authors propose that instead of selecting a random feature and value, EIF selects a random slope (normal vector) and intercept for the hyperplane at each branching point. The process begins by training the dataset, where multiple trees are constructed using the specified hyperplane splitting method. The path length is used to isolate the points, contributing to its anomaly score. Points isolated quickly(shorter path length) are more likely to be outliers.
In this study, we utilized the EIF provided by \href{https://docs.h2o.ai/h2o/latest-stable/h2o-docs/data-science/eif.html}{HE2.ai}, which follows the reference paper.

The following section will lead to the application of the algorithms, and the selection process of the algorithm, that will be included in layer 2 of our framework. 

\subsection{Layer 2 Algorithm Selection: Setup}
\justify
To train the datasets for the Ensemble learning, we used the data classified by the first Layer to be our labeled data through out the testing phase. In the following picture we will see the algorithms vs the improvement in accuracy. We also Note, That Training Features consist of the output from the previous layer. \\

\par
\makebox[\textwidth]{%
\begin{minipage}{0.6\textwidth}
    \centering
    
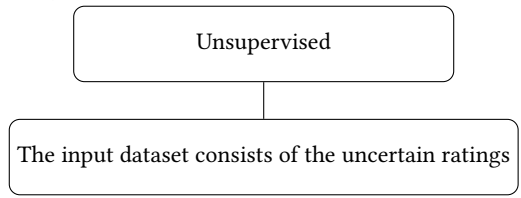
\captionof{figure}{Input format for the unsupervised algorithm}
    \begin{tikzpicture}[
        node distance=1.5cm,
        every node/.style={draw, rectangle, rounded corners, align=center, minimum width=5cm, minimum height=1cm}
    ]
        \node {Unsupervised}
            child { node {The input dataset consists of the uncertain ratings} };
    \end{tikzpicture}
\end{minipage}
}
\vspace{2em}

\par
\makebox[\textwidth]{%
\begin{minipage}{0.6\textwidth} 
    \centering
    
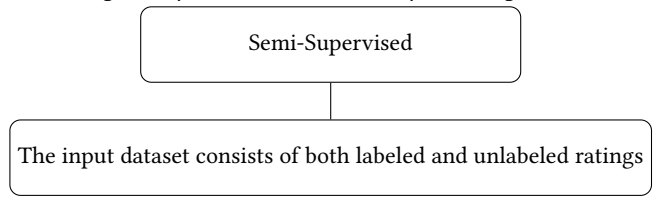
\captionof{figure}{Input format for the semi-supervised algorithm}
    \begin{tikzpicture}[
        node distance=1.5cm,
        every node/.style={draw, rectangle, rounded corners, align=center, minimum width=5cm, minimum height=1cm}
    ]
        \node {Semi-Supervised}
            child { node {The input dataset consists of both labeled and unlabeled ratings} };
    \end{tikzpicture}
\end{minipage}
}
\vspace{2em}

\par
\makebox[\textwidth]{%
\begin{minipage}{0.6\textwidth}
    \centering
    
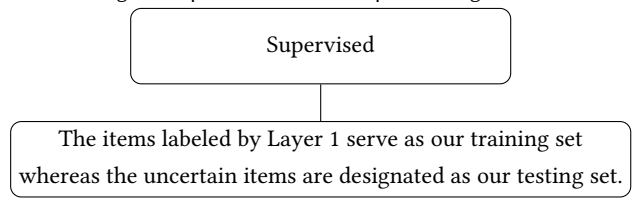
\captionof{figure}{Input format for the supervised algorithm}
    \begin{tikzpicture}[
        node distance=2.5cm,
        every node/.style={draw, rectangle, rounded corners, align=center, minimum width=5cm, minimum height=1cm}
    ]
        \node {Supervised}
            child { node {
            The items labeled by Layer 1 serve as our training set \\
            whereas the uncertain items are designated as our testing set.
            } };
    \end{tikzpicture}
\end{minipage}
}
\par

\justify
Given the demonstrated consistency in the output of layer 1 across the datasets, we selected one of the datasets for further experimentation. The selected dataset was the ml-25m-subset2 created randomly based on the MovieLens dataset as explained in the process phase on layer 1.
And to evaluate each algorithm's performance and be able to compare then against one another, we will employ the proposed 2d system metric, utilizing the LightGCN Recommender Systems across all accuracy assessments. LightGCN is a powerful linear recommendation algorithm that achieves strong performance despite its simplicity. Its architecture is well suited for collaborative filtering tasks making it effective for this experiment.

\subsection{Ensemble learning Selection}
\label{subsubsec:layer2}
So far, we have outlined the selection of the Ensemble learning algorithms. Next we will identify the most effective algorithm for implementation within this framework.
Layer 2's objective is enhancing accuracy by classifying the uncertain ratings left from layer 1 based on the proposed evaluation system.
In the previous section, we had 5 main categories that will be applied to the dataset, and compared against each other using the 2d evaluation system, to choose the best algorithm.

To ensure clarity, we will reintroduce the abbreviations used in this study:
\begin{enumerate}
    \item EL1 Bagging (Random Forest Classifier)
    EL2 \& EL2.2 (under the Stacking category) and EL3 (CatBoost), are the supervised algorithms.
    \item EL4 is the semi-supervised algorithm (RESSEL).
    \item EL5 is the unsupervised algorithm (Isolation Forest).

\end{enumerate}

The steps involved in our evaluation process are outlined below:
\begin{itemize}
    \item We will compare the performance of each algorithm, before and after it was employed. We utilize the algorithm performance at the output of layer 1, then with the results from the Ensemble algorithm.
    \item The four system metrics introduced in Section \ref{sec:env_var} will be utilized (2D)
    \item The percentage of improvement across these metrics will be quantified based on each improvement in rating.
\vspace{0.2cm}
\end{itemize}

\begin{center}
\begin{table}[H]
\centering
\caption{Improvement of EL Algorithms in Various Metrics}
\begin{tabular}{|l|c|c|c|c|}
\hline
\textbf{Algorithm} & \textbf{\% nDCG - Serendipity } & \textbf{\% F1 - Serendipity} & \textbf{\% Recall - Serendipity} & \textbf{\% Precision - Serendipity} \\ \hline
EL1   & 0.16 & 37.29 & 59.71 & 0 \\ \hline
EL2   & 0  & 2.9 & 23.1 & 0  \\ \hline
EL2.2   & 33.21 & 99.11 & 99.36 & 25.46 \\ \hline
EL3 & 12.12 & 94.11 & 96.38 & 17.57  \\ \hline
EL4.1 & 0.2  & 36.65 & 67.89 & 0   \\ \hline
EL4.2 & 0  & 5.97 & 28.38 & 0  \\ \hline
EL5   & 44.45  & 76.92 & 78.28 & 57.35 \\ \hline
\end{tabular}
\label{tab:improv_el}
\end{table}
\end{center}
\FloatBarrier
\noindent
\begin{table}
\caption{Rankings algorithms following the results}
\begin{tabular}{|l|c|c|c|c|}
\hline
\textbf{\% nDCG - Serendipity } & \textbf{\% F1 - Serendipity} & \textbf{\% Recall - Serendipity} & \textbf{\% Precision - Serendipity} \\ \hline
EL5 & EL2.2 & EL2.2 & EL5 \\ \hline
EL2.2 & EL3 & EL3 & EL2.2 \\ \hline
EL3 & EL5 & EL5 & EL3 \\ \hline
EL4.1 & EL1 & EL4.1 & EL4.1 \\ \hline
EL1 & EL4.1 & EL1 & EL1 \\ \hline
EL2 & EL4.2 & EL4.2 & EL4.2 \\ \hline
EL4.2 & EL2 & EL2 & EL2 \\ \hline
\end{tabular}
\label{tab_ranking}
\end{table} 

By examining the Table \ref{tab:graph_ELS} for data points distributed across systems we can see the improvement value as by axis, as well as in table \ref{tab:improv_el} we can see the percentage of ratings improved across the datasets after each algorithm identified and removed noisy ratings, and table \ref{tab_ranking} where we can see the algorithms ordered by the percentage of improvement across systems, to be able to identify the best algorithm.

By looking into the Table \ref{tab:improv_el}, we can see that the algorithms EL1, EL2, EL4.1 and EL4.2 have small to no improvements on nDCG-Serendipity and Precision-Serendipity, while EL3, EL4.2 and EL5 show better results on all systems. Lastly EL2.2 and EL3 made the best improvement on all systems. The graphs in \ref{tab:graph_ELS} interprets the results more clearly showing all data points under the red line (threshold), and shows the overall disparity of the data points in EL2.2, EL3 and EL5. The graph shows how data points are set following the 2 axes, the first one being the accuracy metric and the second one being serendipity. We can see that the algorithms showing worse results on EL1, EL2 and EL4.2 have smaller improvements for Serendipity. While for EL4.1, it showed promising levels of improvements for Serendipity.

\begin{table}[htbp!]
\centering
\caption{Performance Metrics for ML-25M-Subset(2) Dataset on Ensemble Learning Algorithms}
\label{tab:graph_ELS}
\resizebox{\textwidth}{!}{
    \begin{tabular}{@{}l@{}c@{}c@{}c@{}c@{}}
    \toprule
    \multirow{1}{*}{Implementation} & \multicolumn{4}{c}{ML-25M-subset} \\
    \cmidrule(lr){1-5}
      & \% nDCG - Serendipity & \% F1 - Serendipity  & \% Precision - Serendipity  & \% Recall - Serendipity\\
    \midrule
    EL1 & \includegraphics[width=3cm]{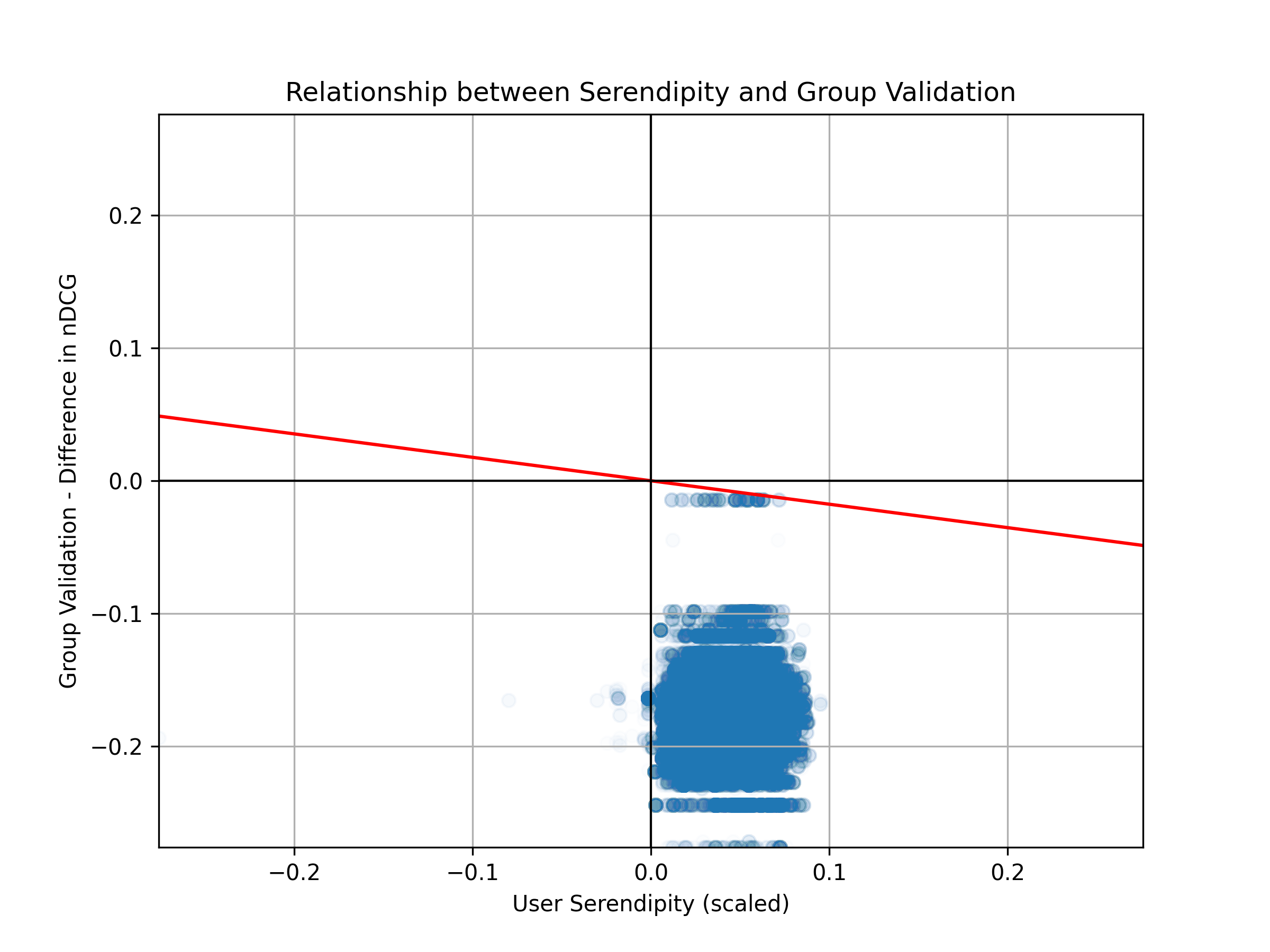} & \includegraphics[width=3cm]{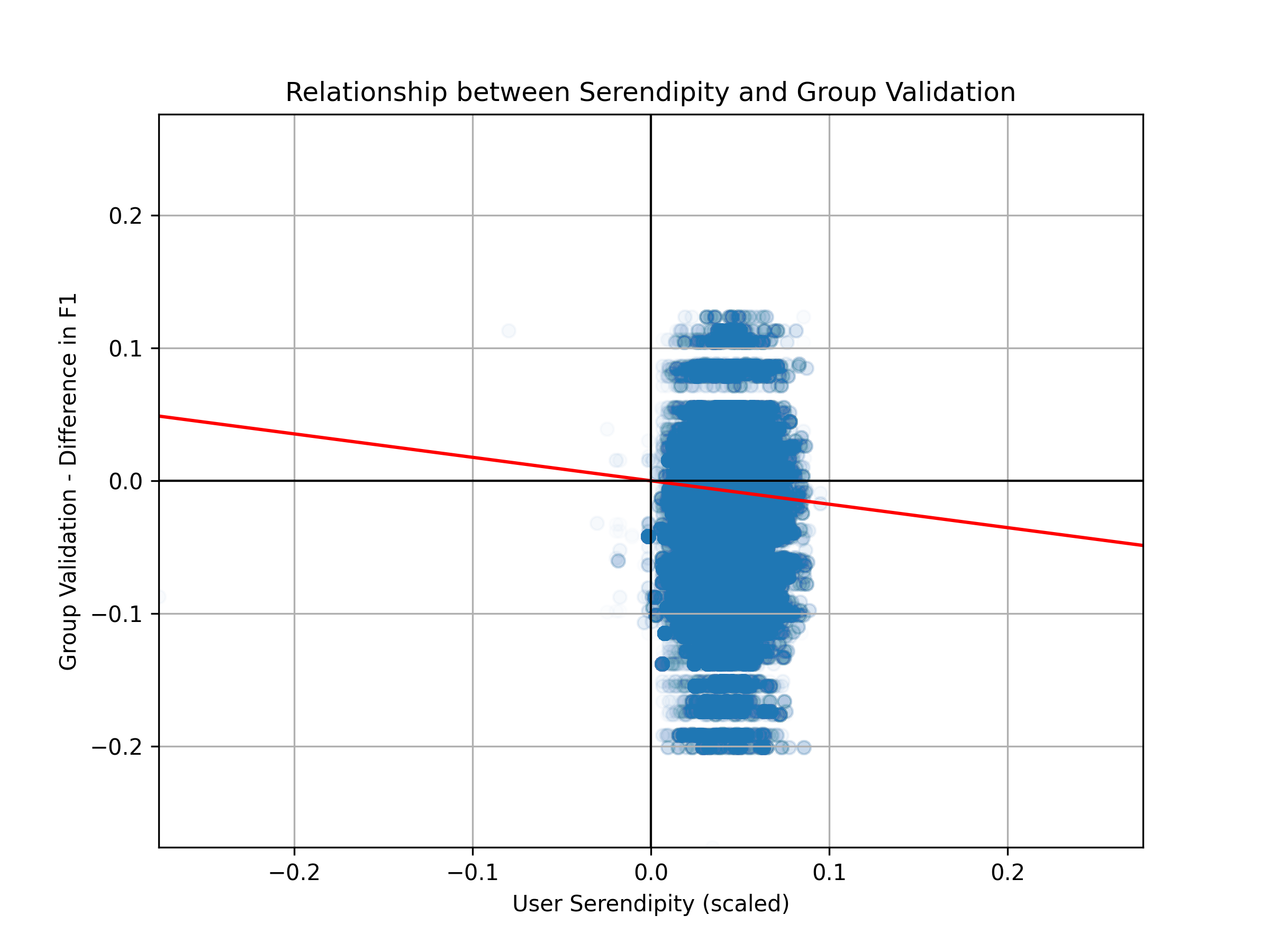}\ & \includegraphics[width=3cm]{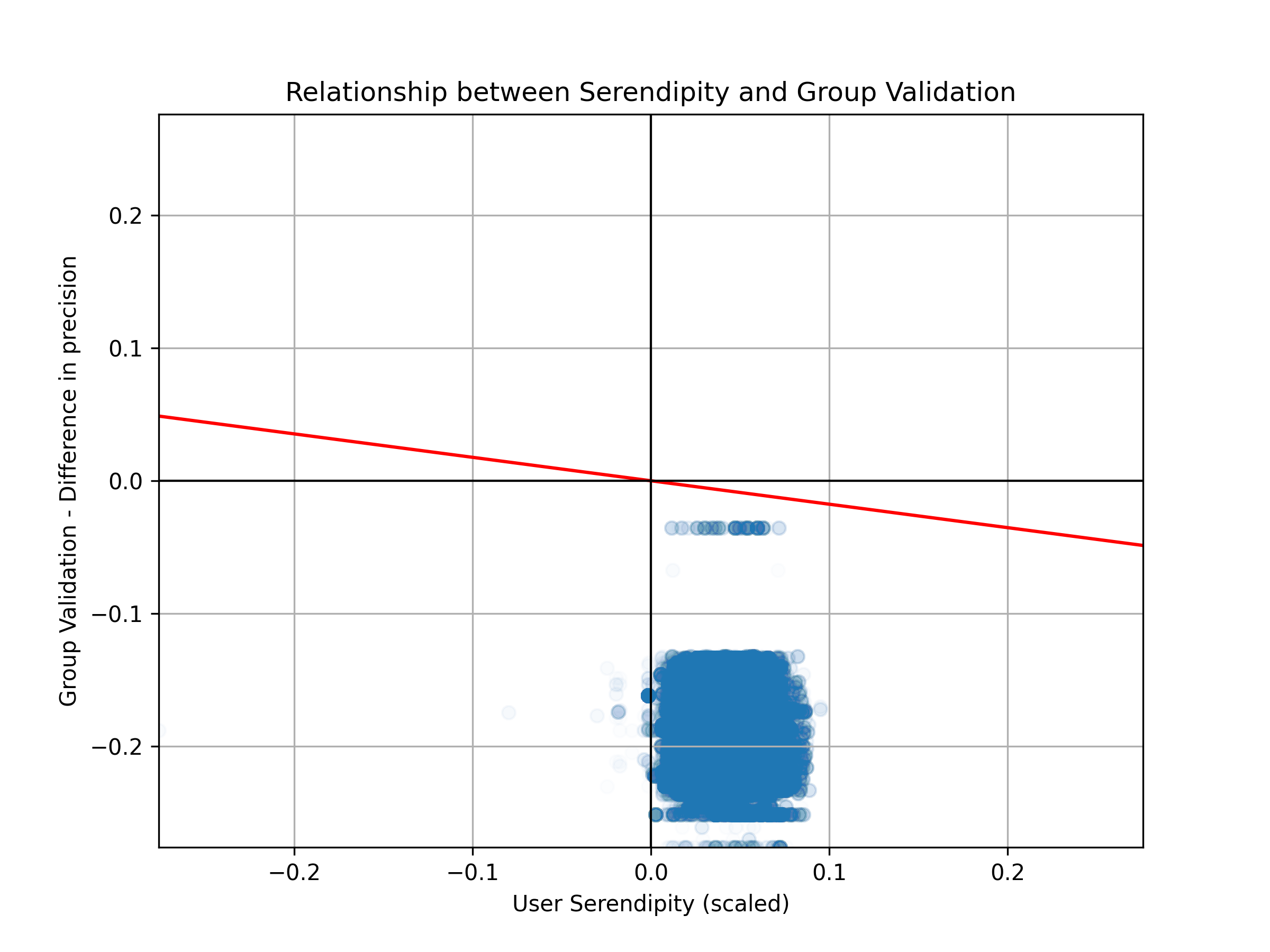} & \includegraphics[width=3cm]{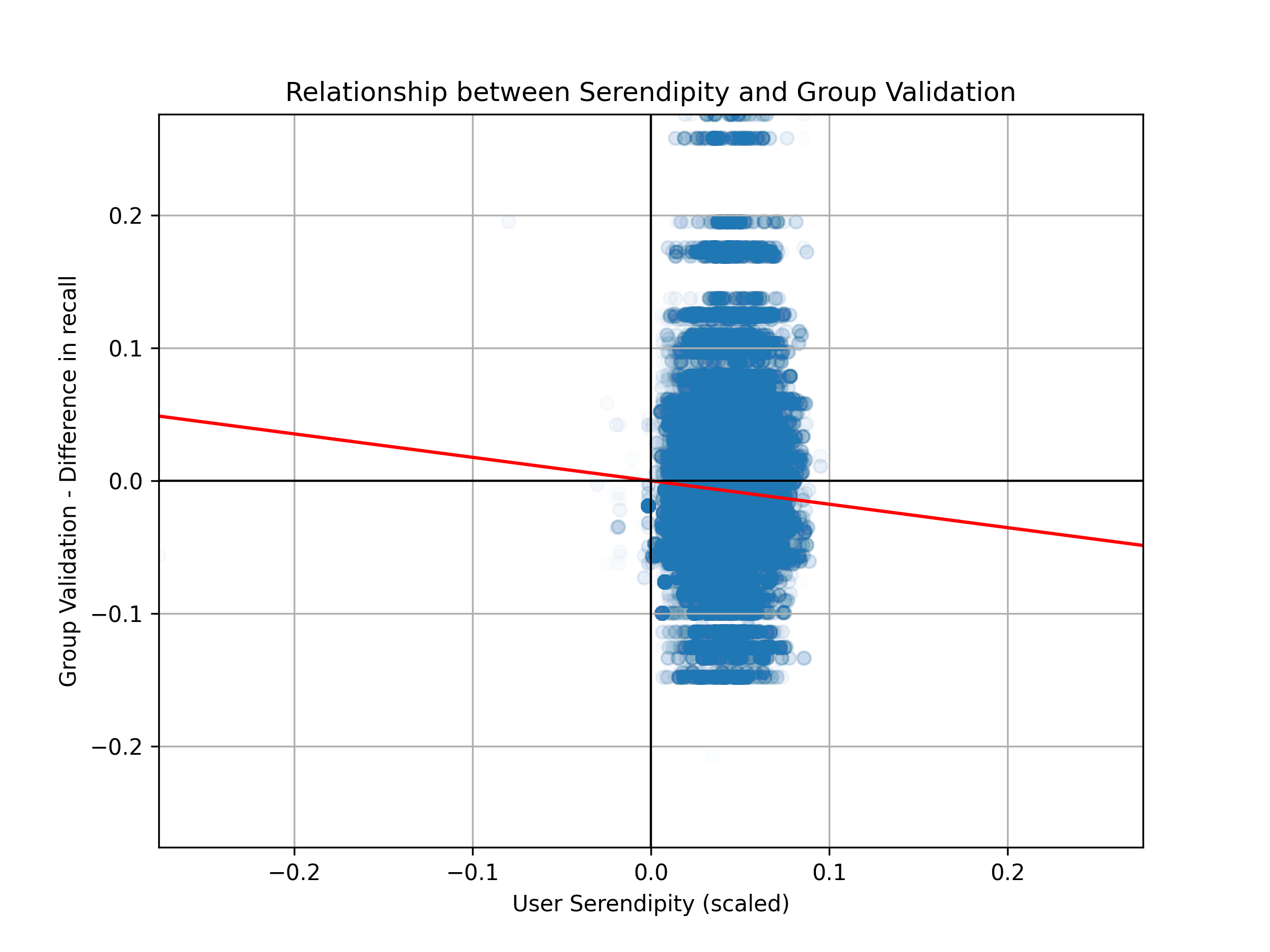} \\
    EL2 & \includegraphics[width=3cm]{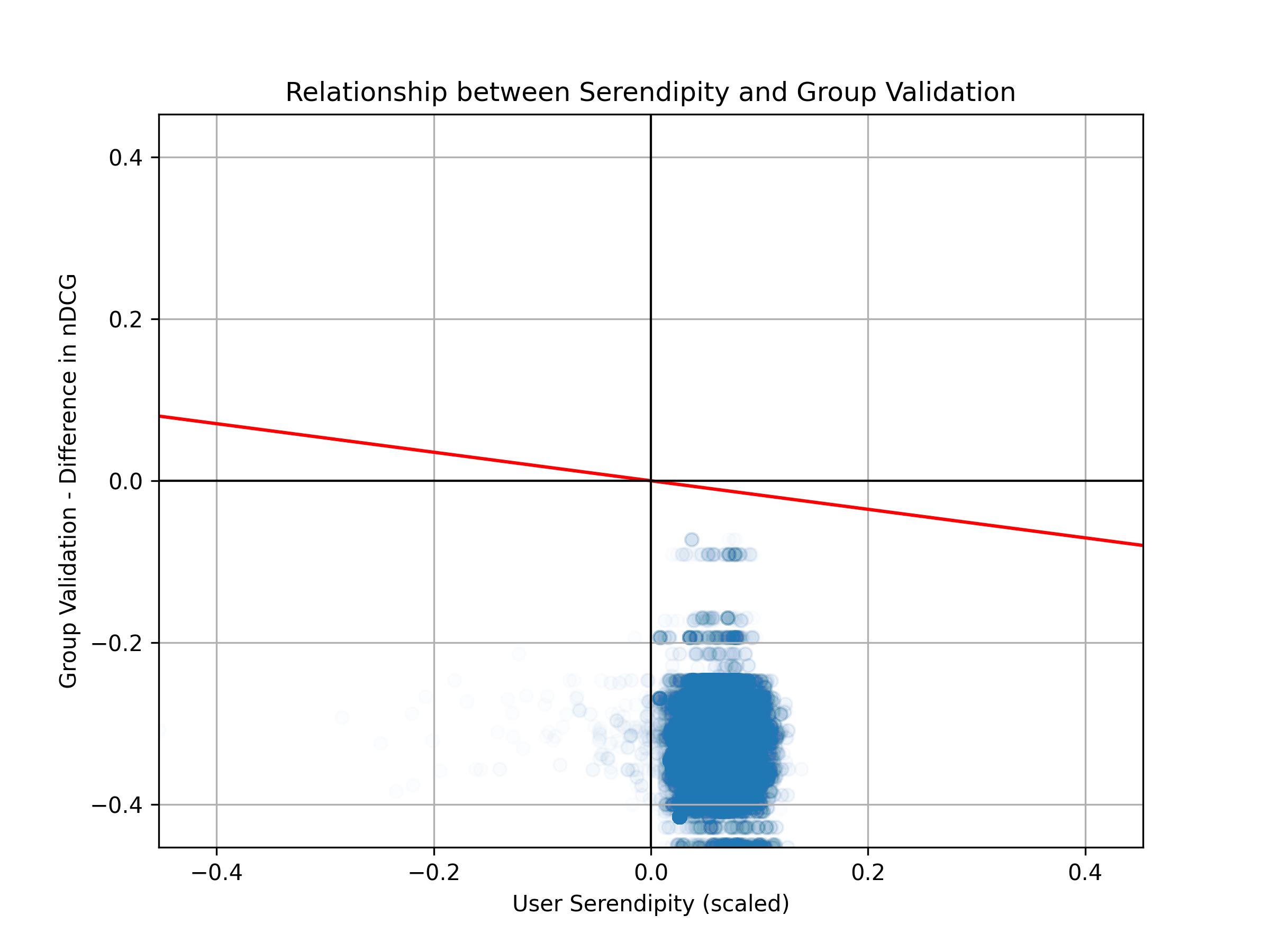} & \includegraphics[width=3cm]{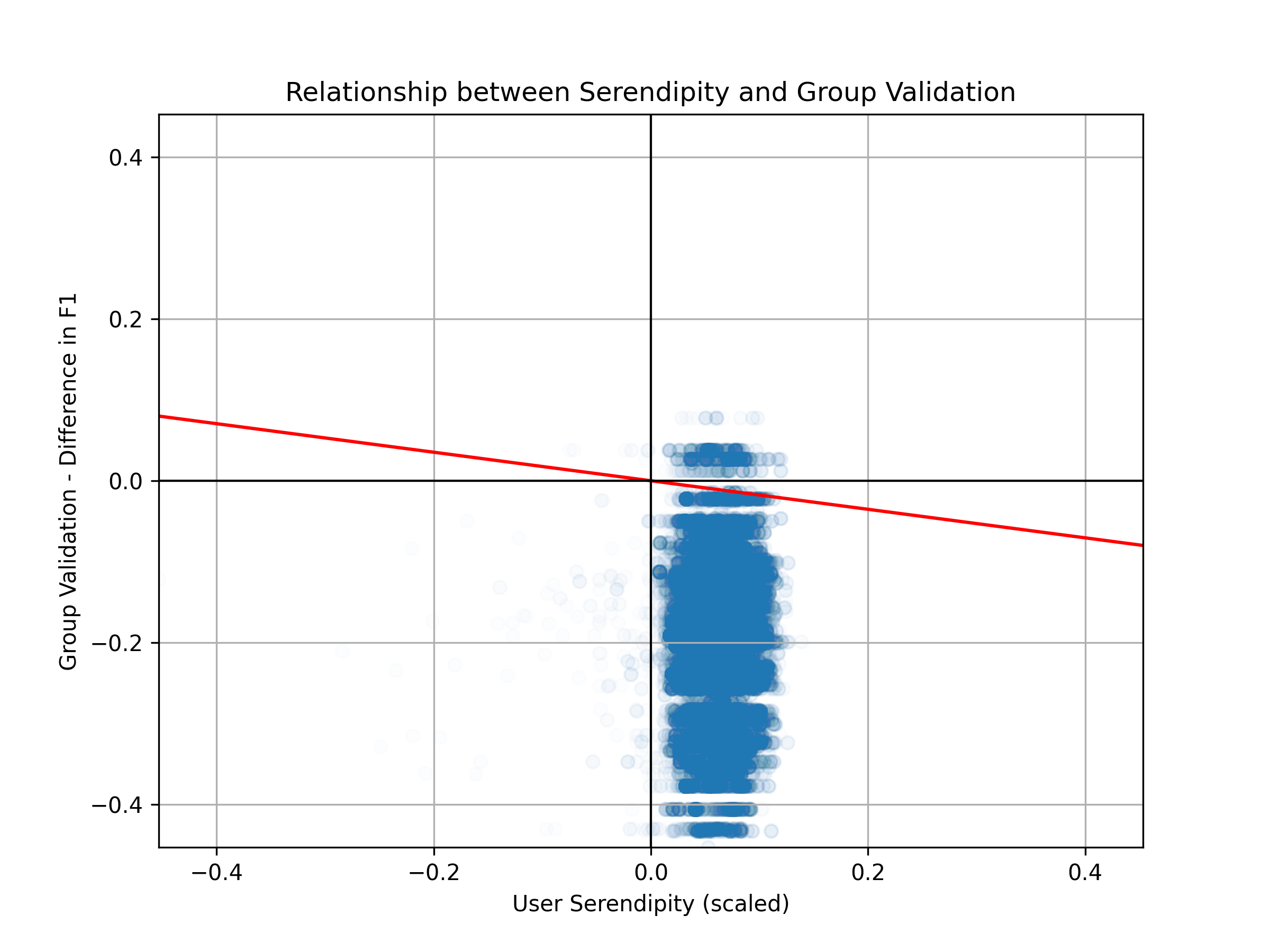}\ & \includegraphics[width=3cm]{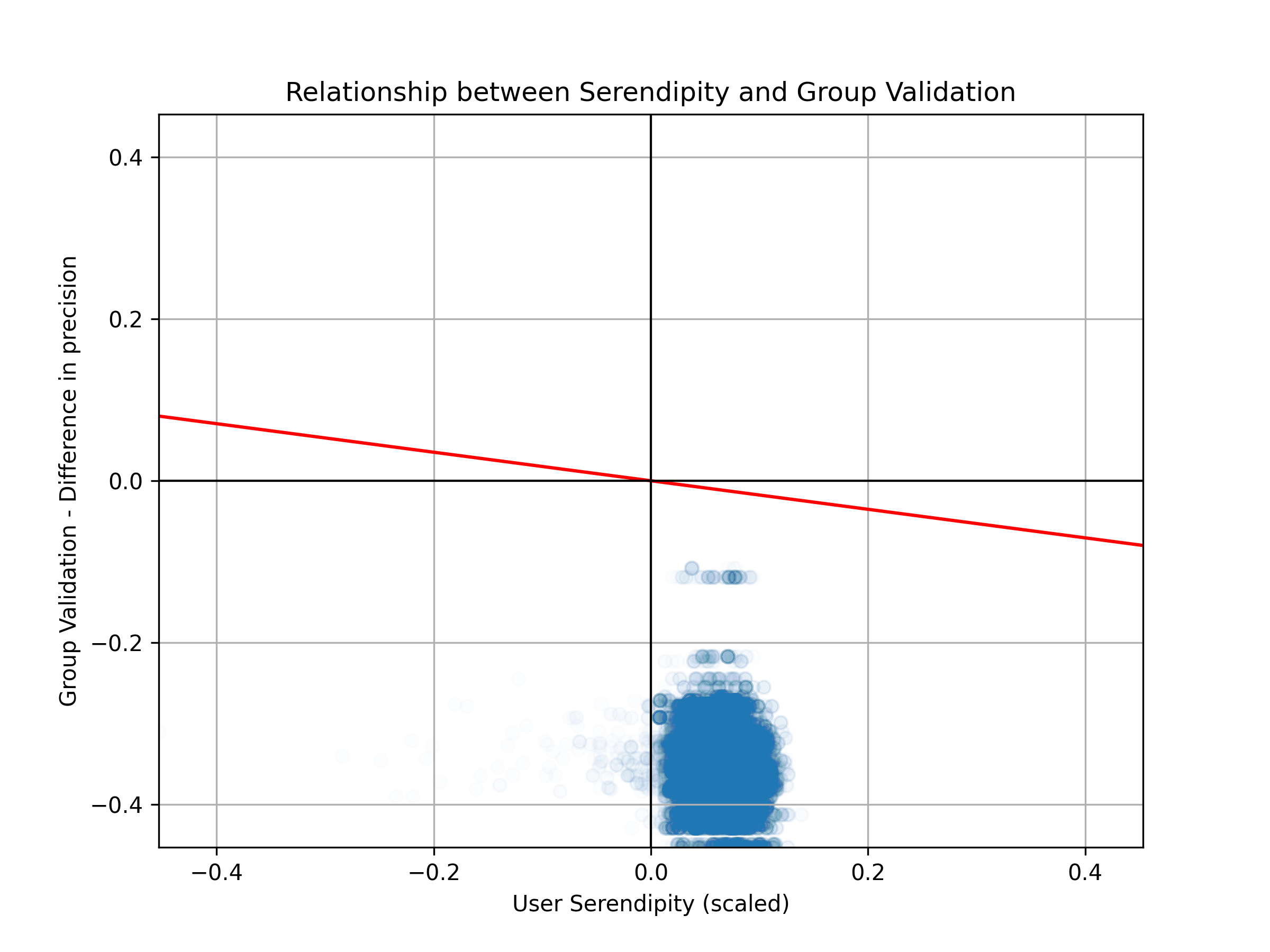} & \includegraphics[width=3cm]{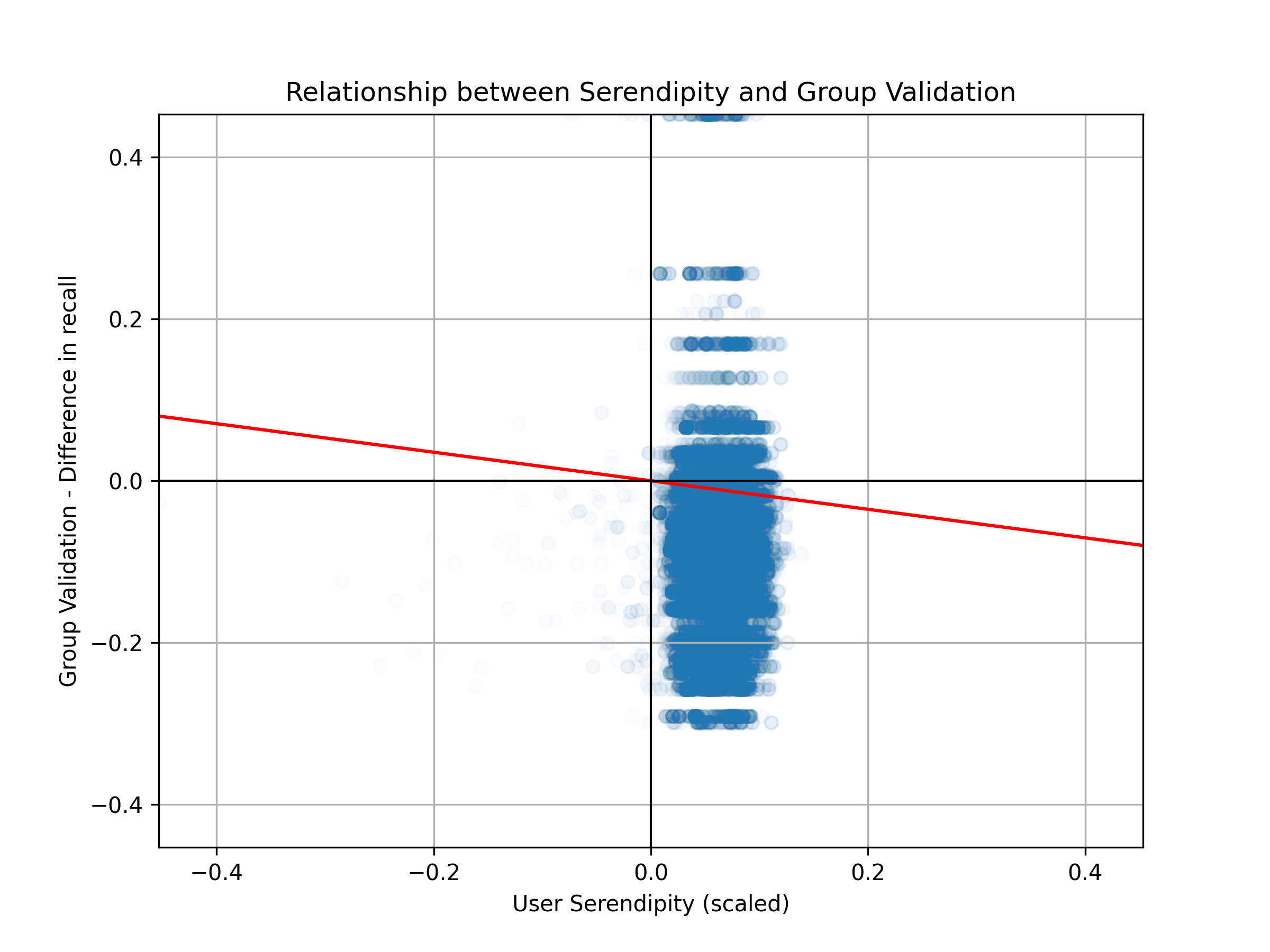} \\
    EL2.2 & \includegraphics[width=3cm]{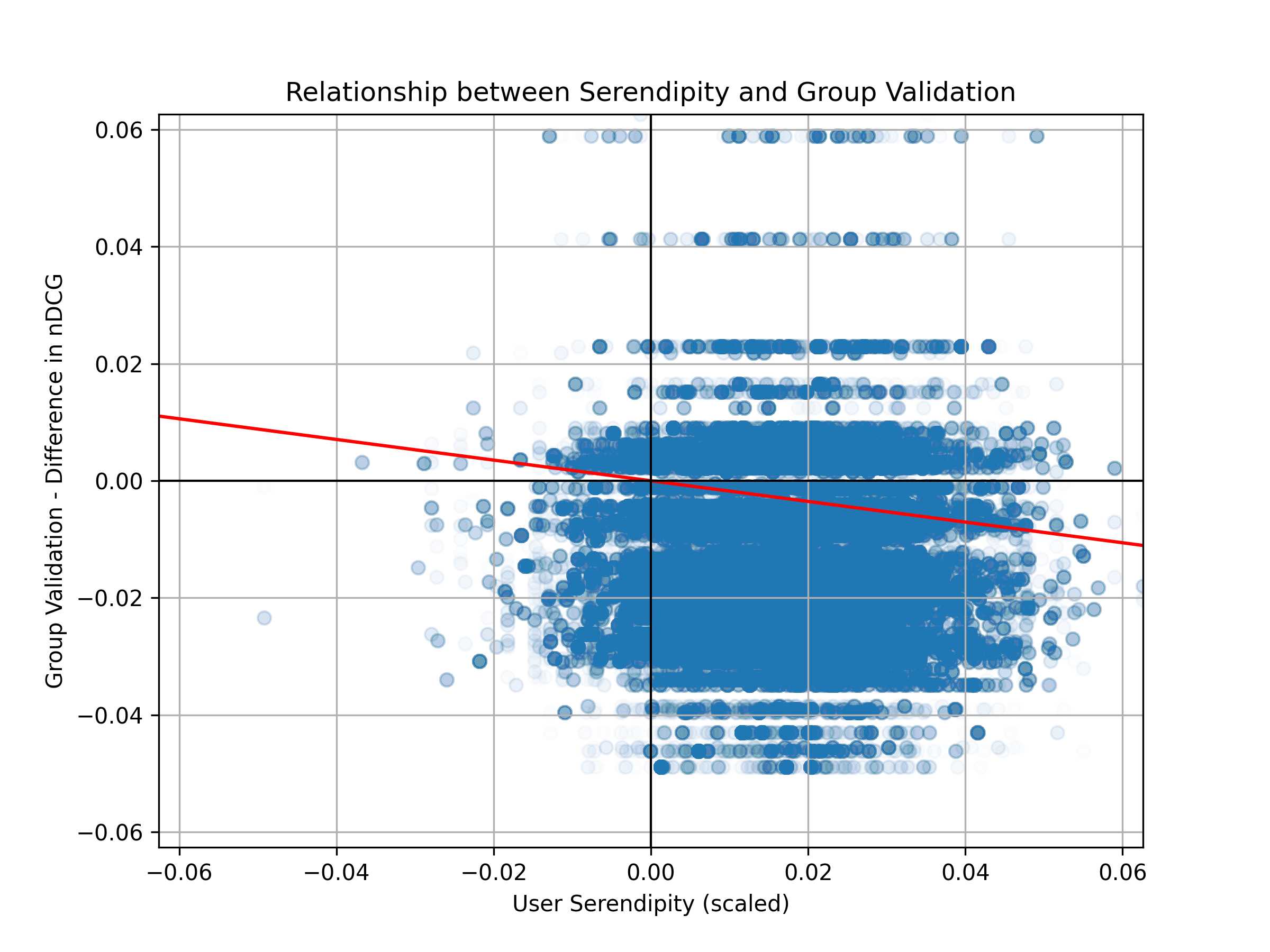} & \includegraphics[width=3cm]{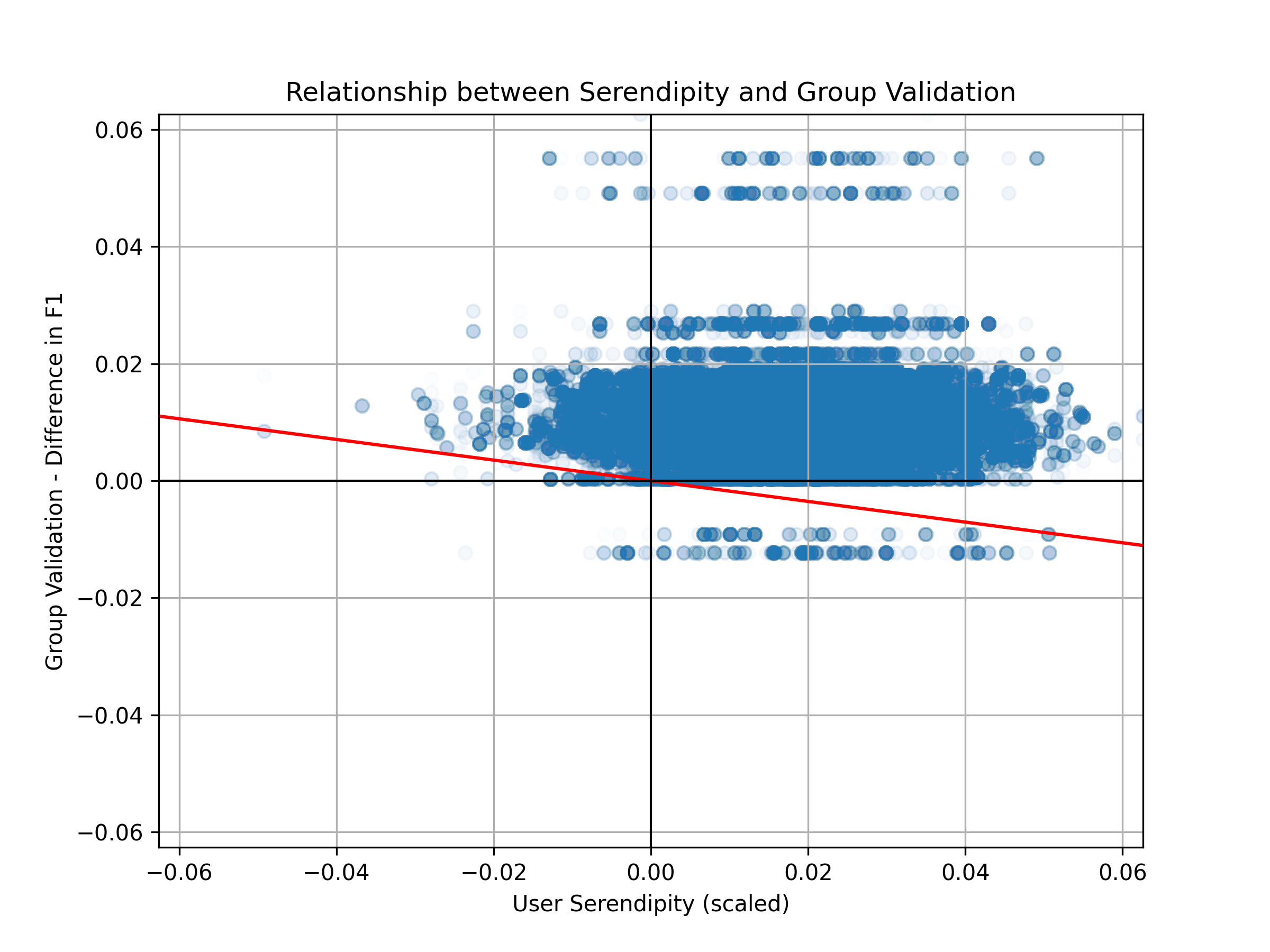}\ & \includegraphics[width=3cm]{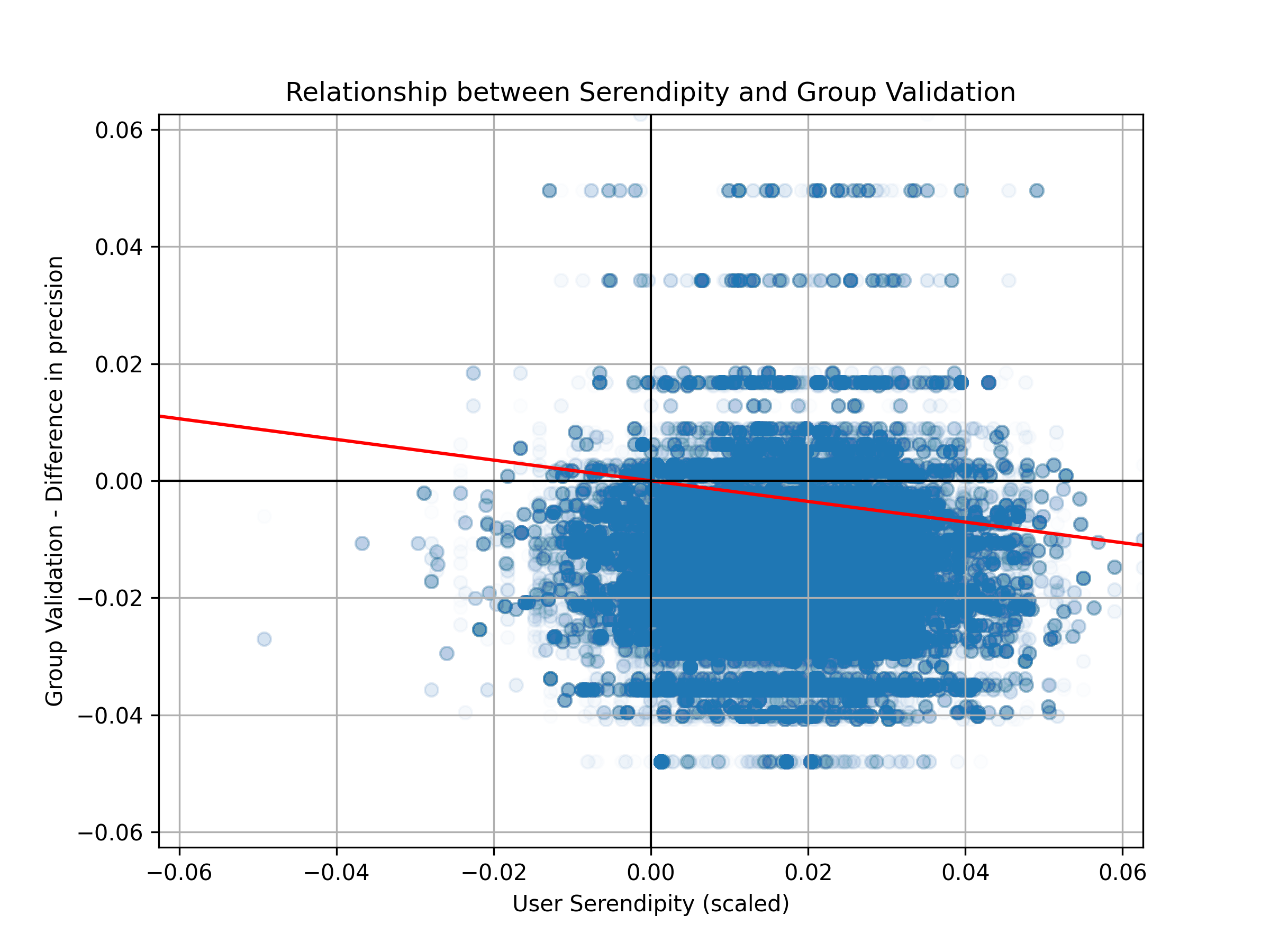} & \includegraphics[width=3cm]{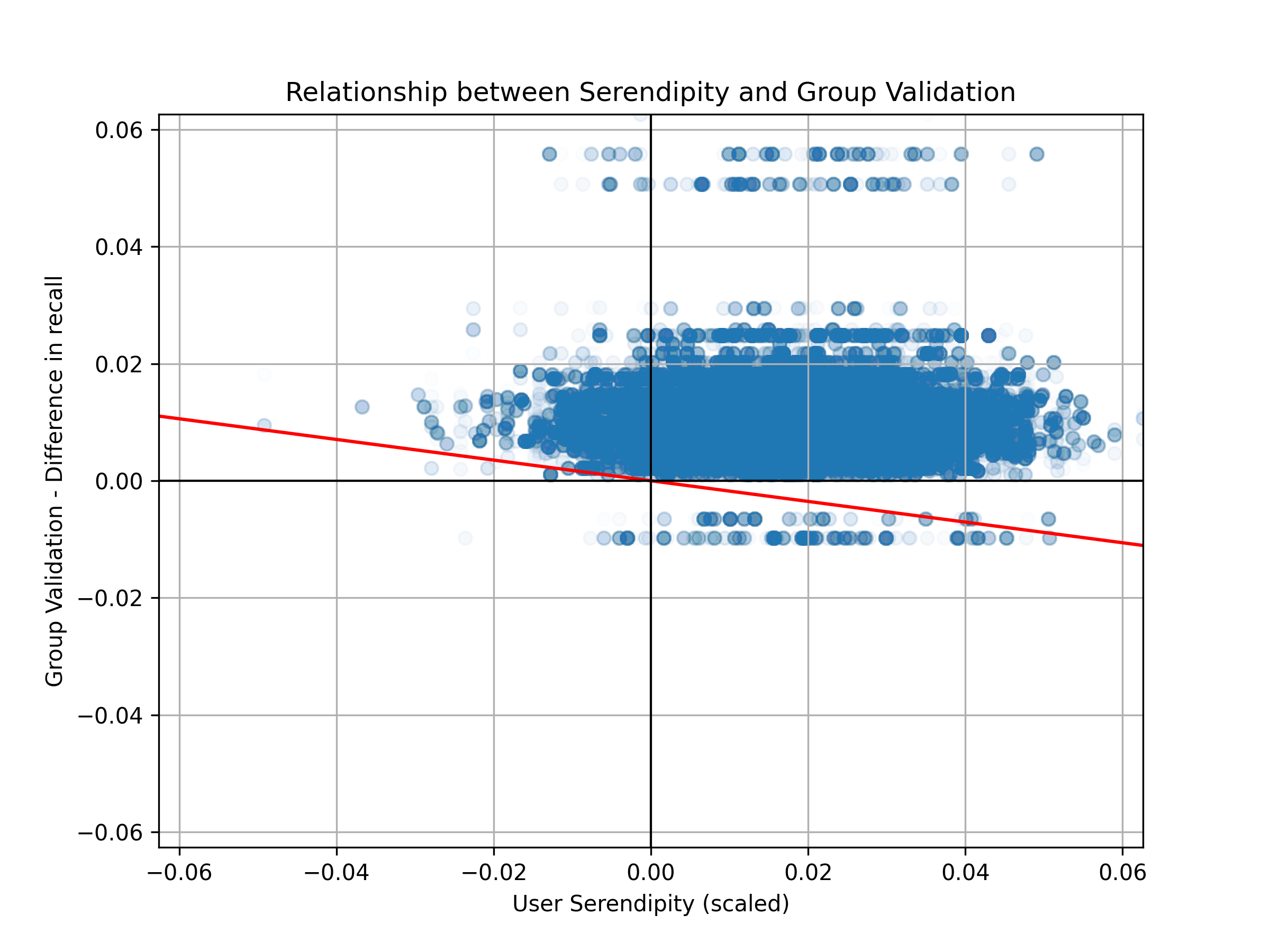} \\
    EL3 & \includegraphics[width=3cm]{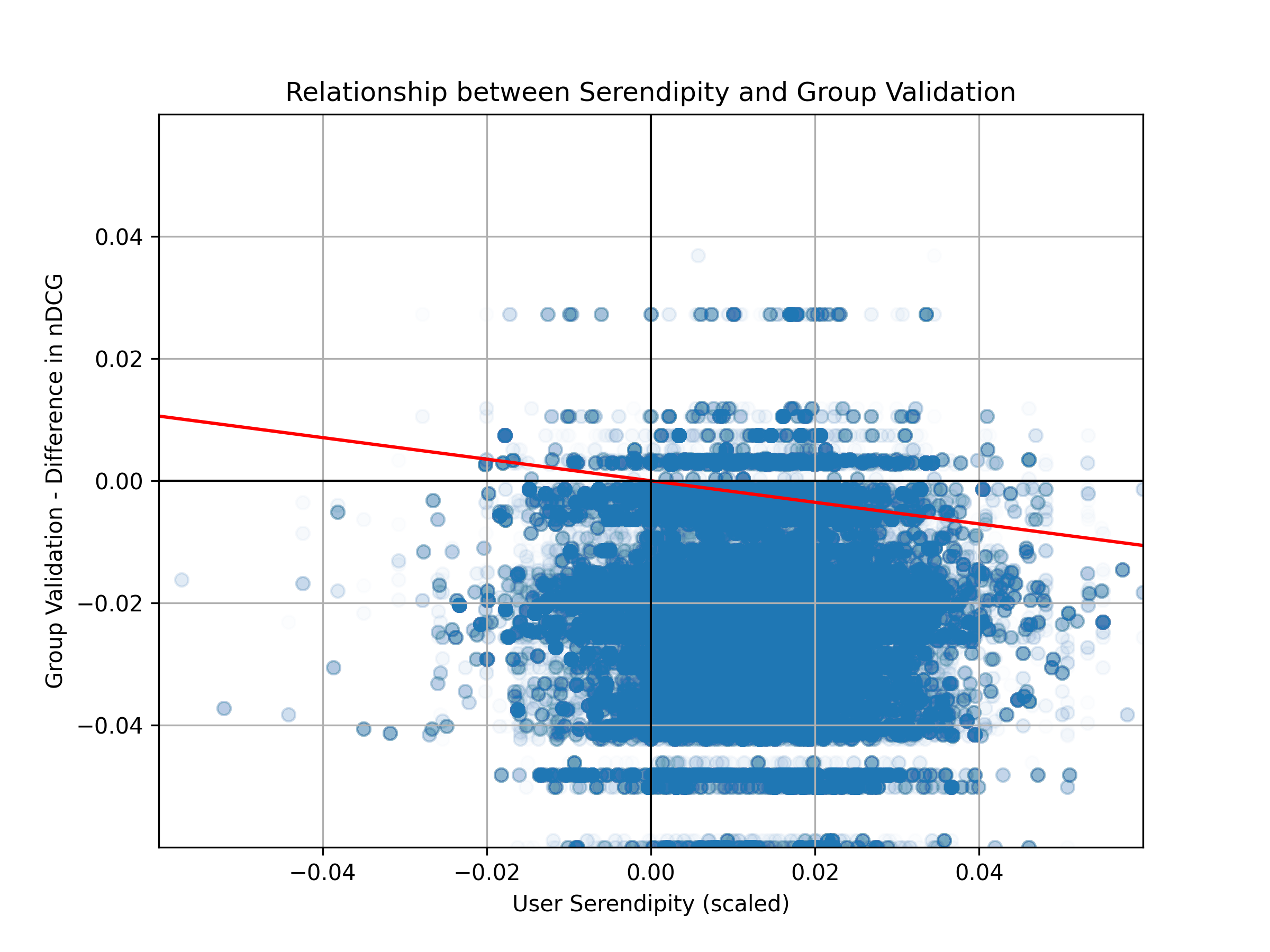} & \includegraphics[width=3cm]{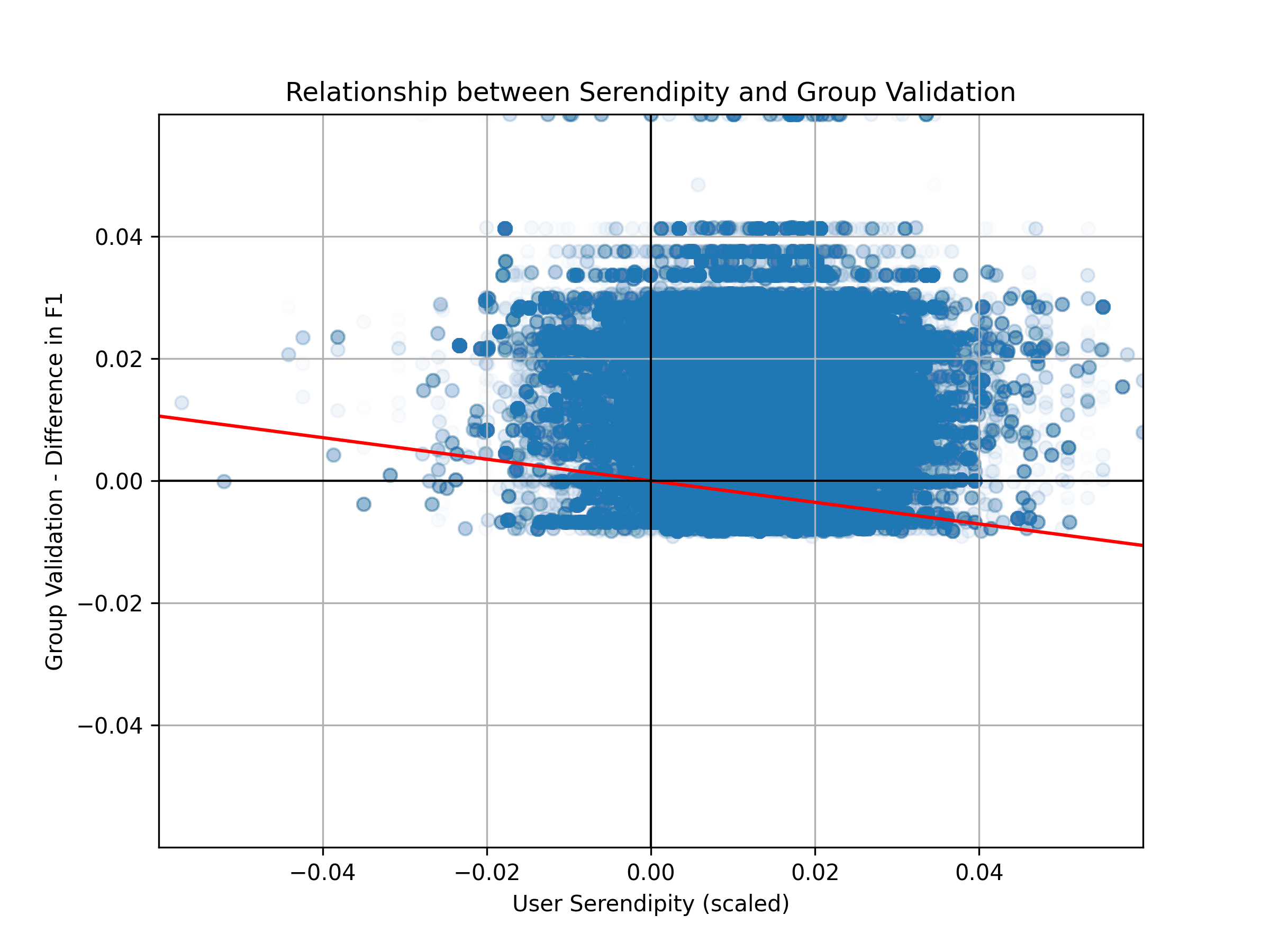}\ & \includegraphics[width=3cm]{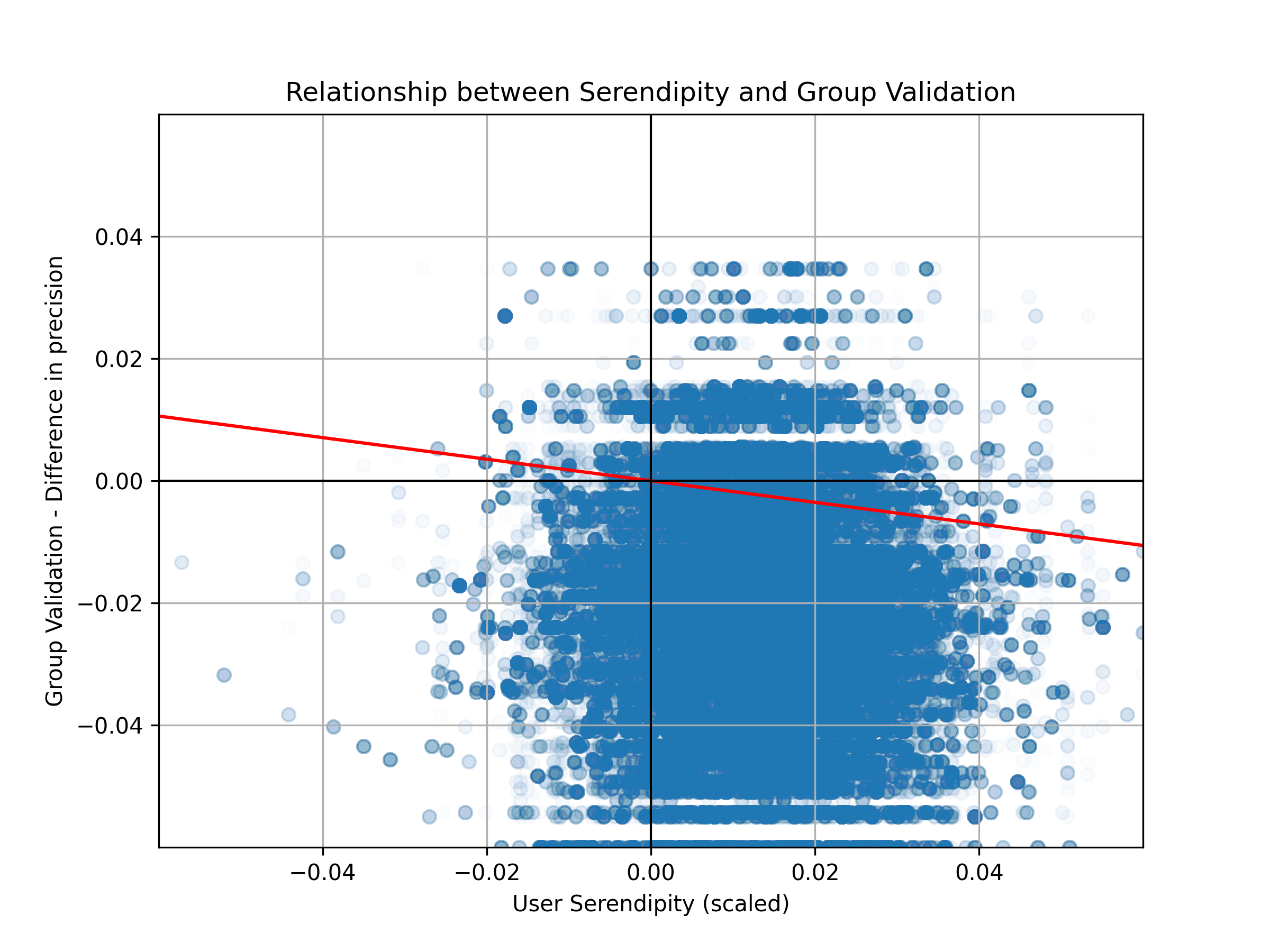} & \includegraphics[width=3cm]{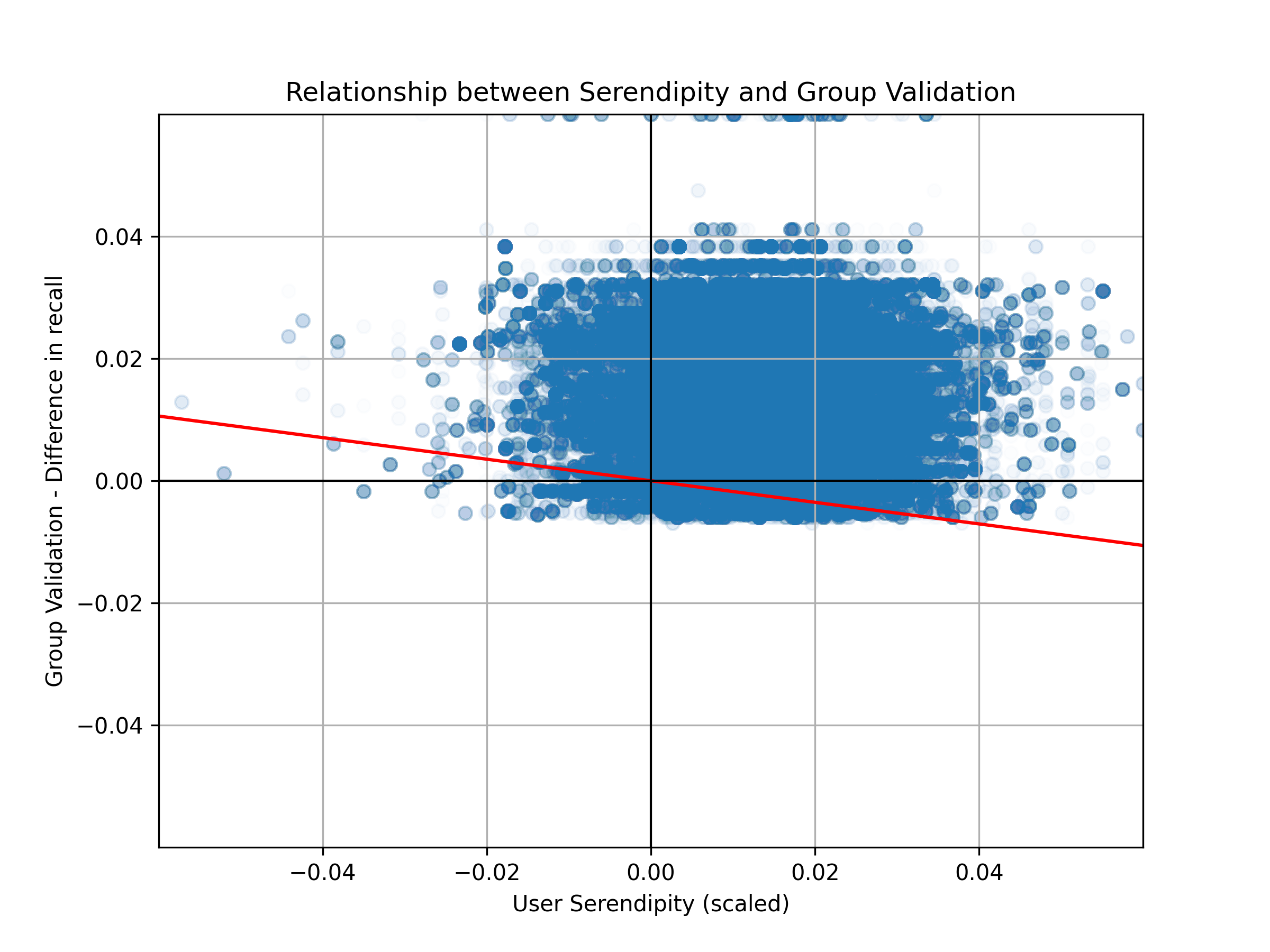} \\
    EL4.1  & \includegraphics[width=3cm]{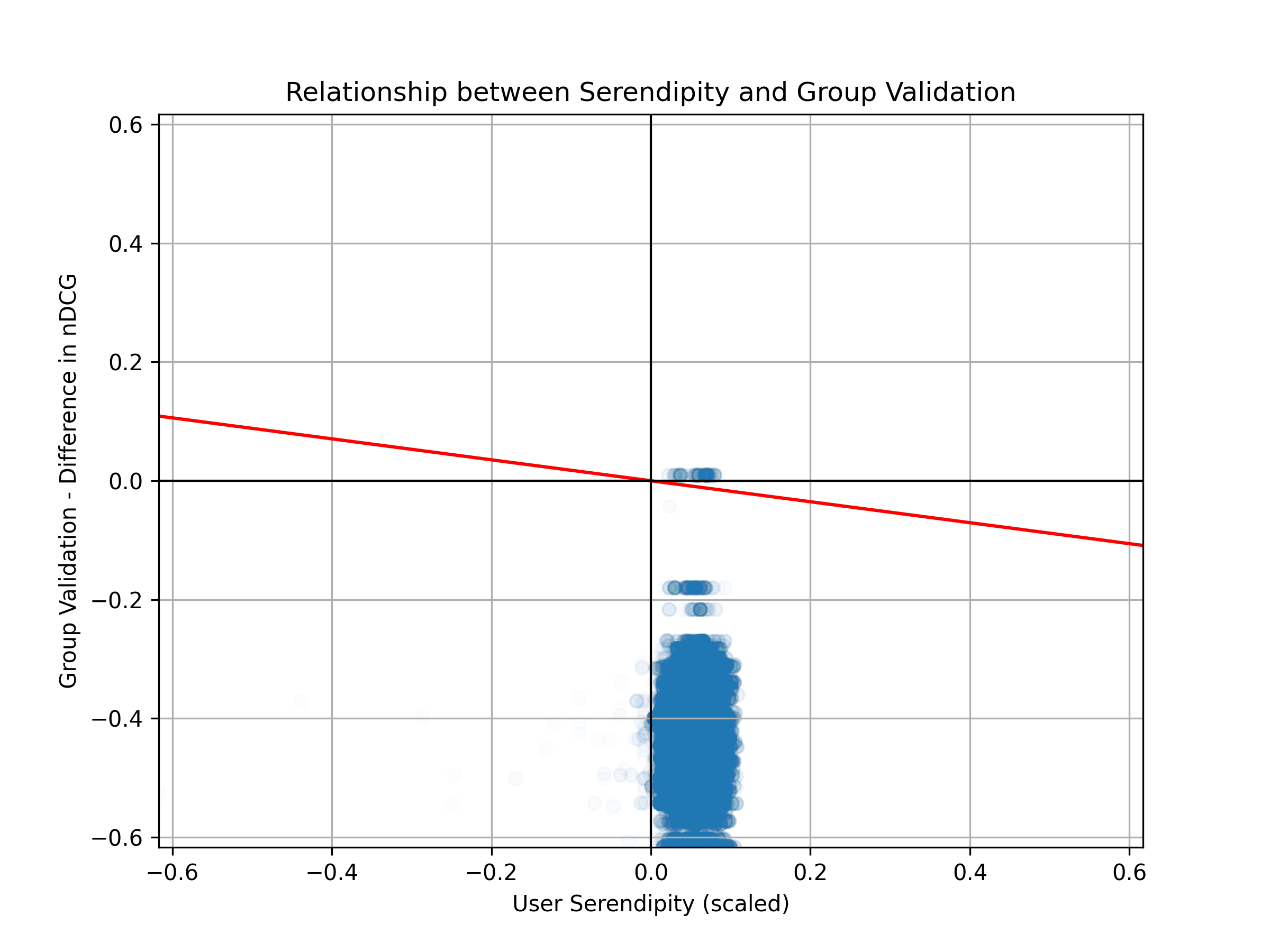} & \includegraphics[width=3cm]{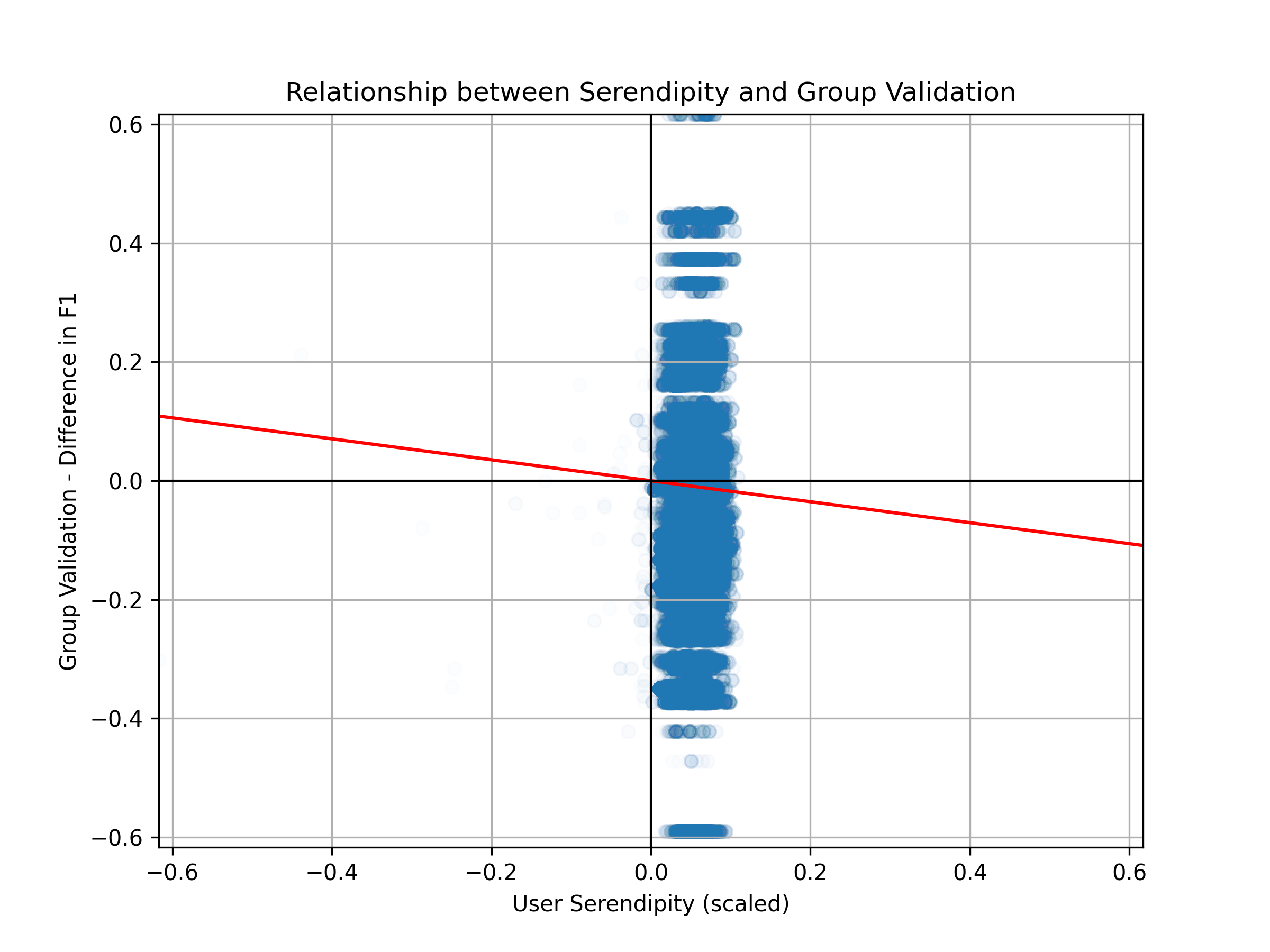}\ & \includegraphics[width=3cm]{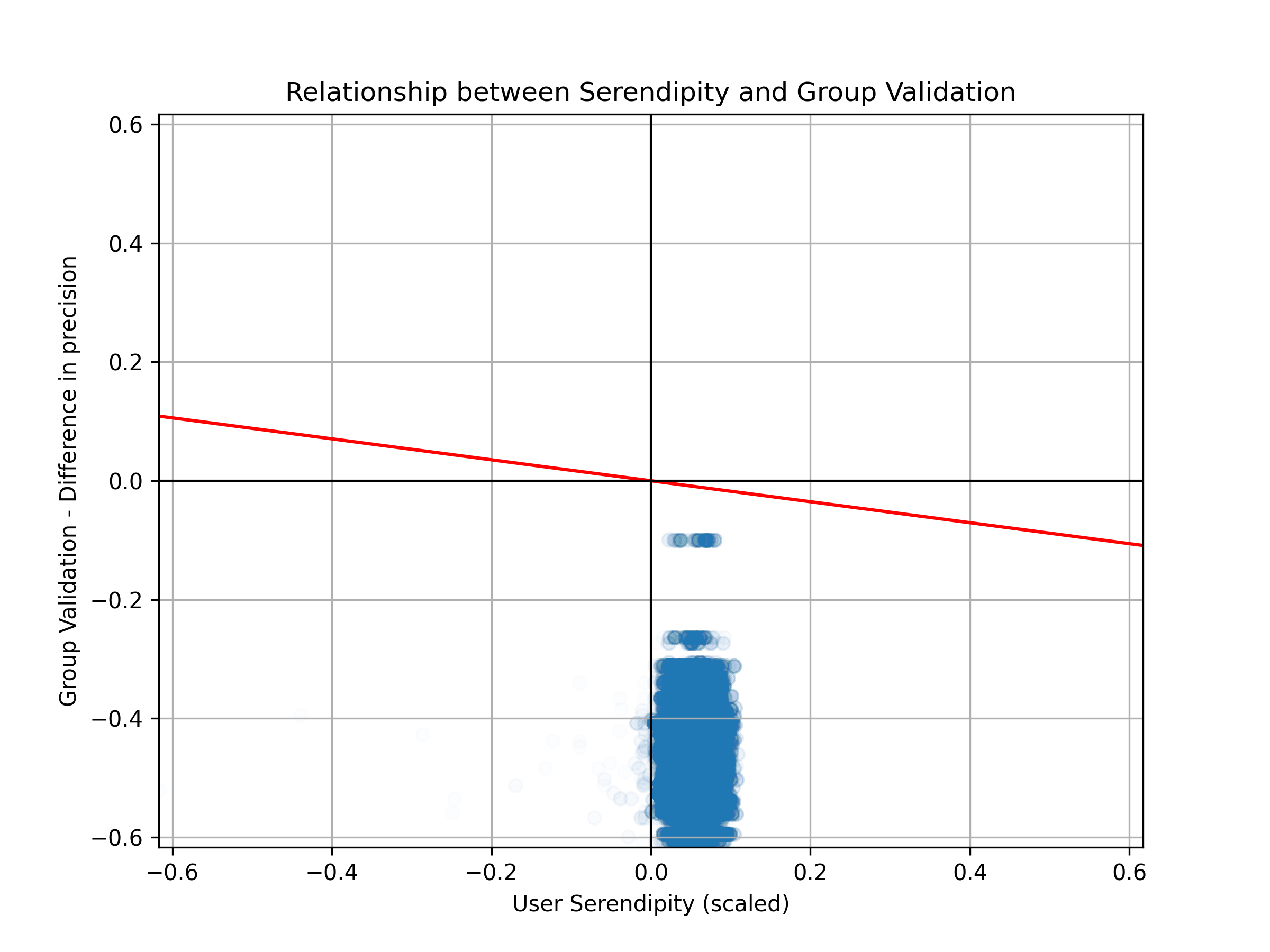} & \includegraphics[width=3cm]{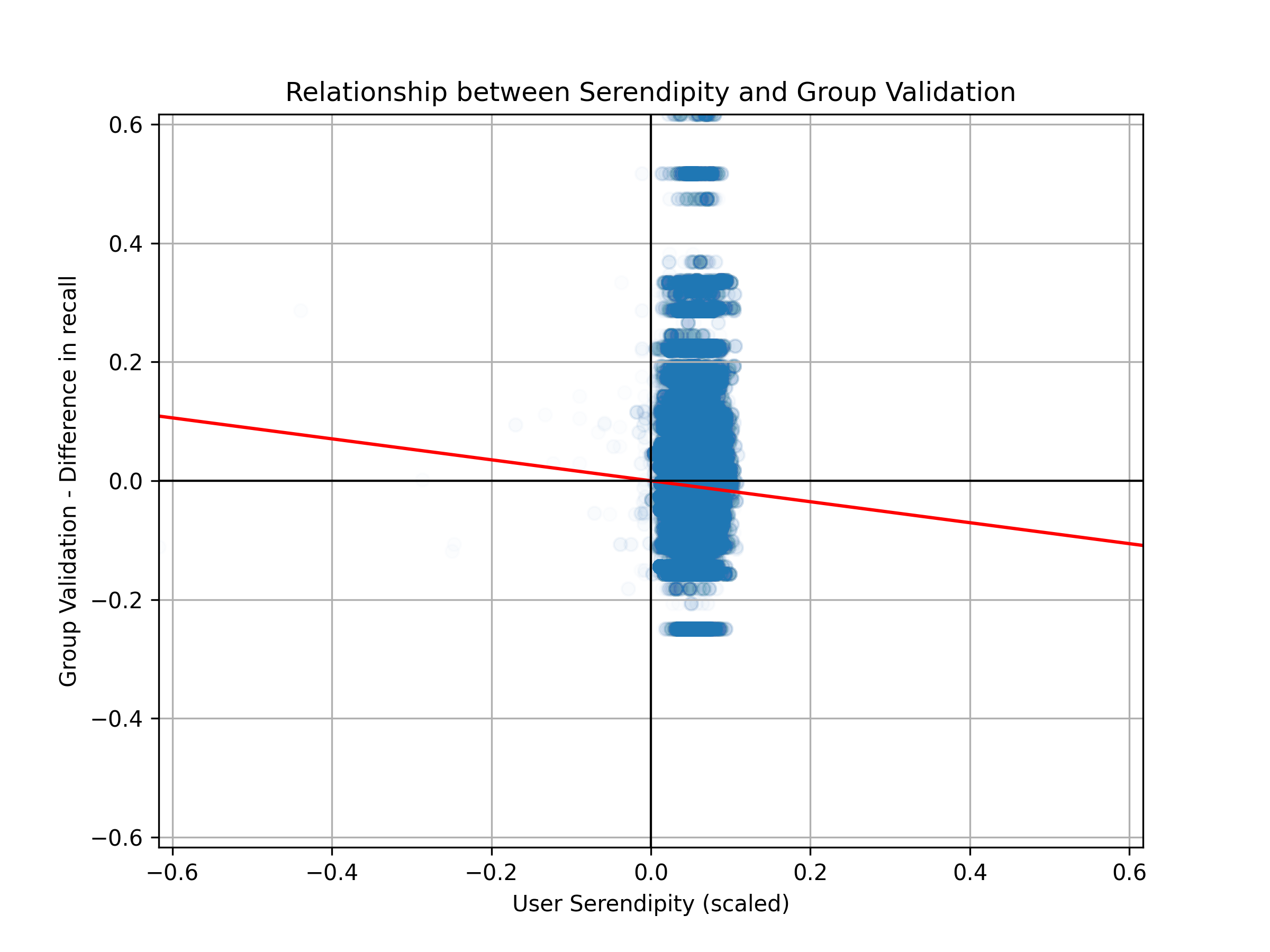} \\
    EL4.2 & \includegraphics[width=3cm]{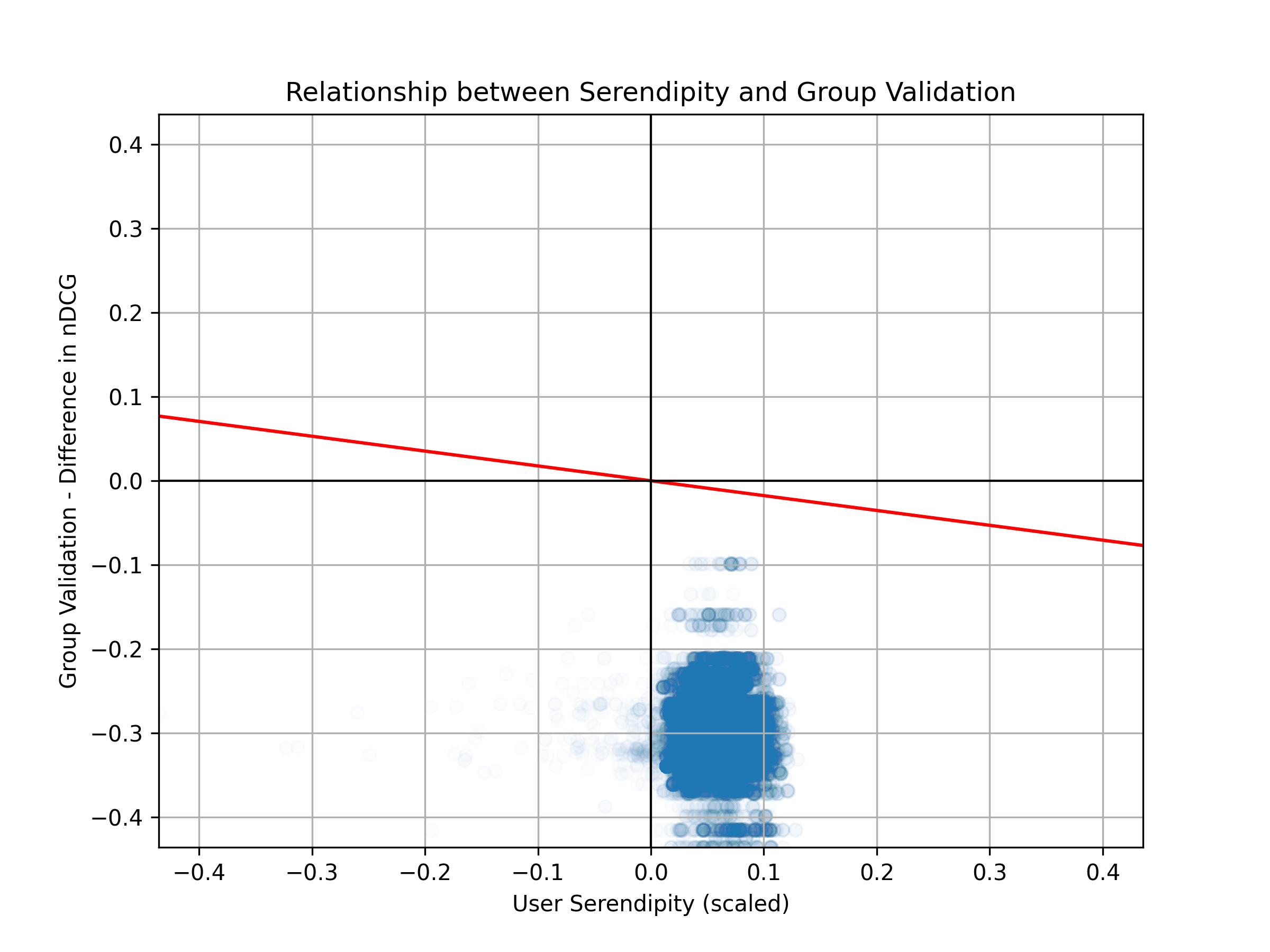} & \includegraphics[width=3cm]{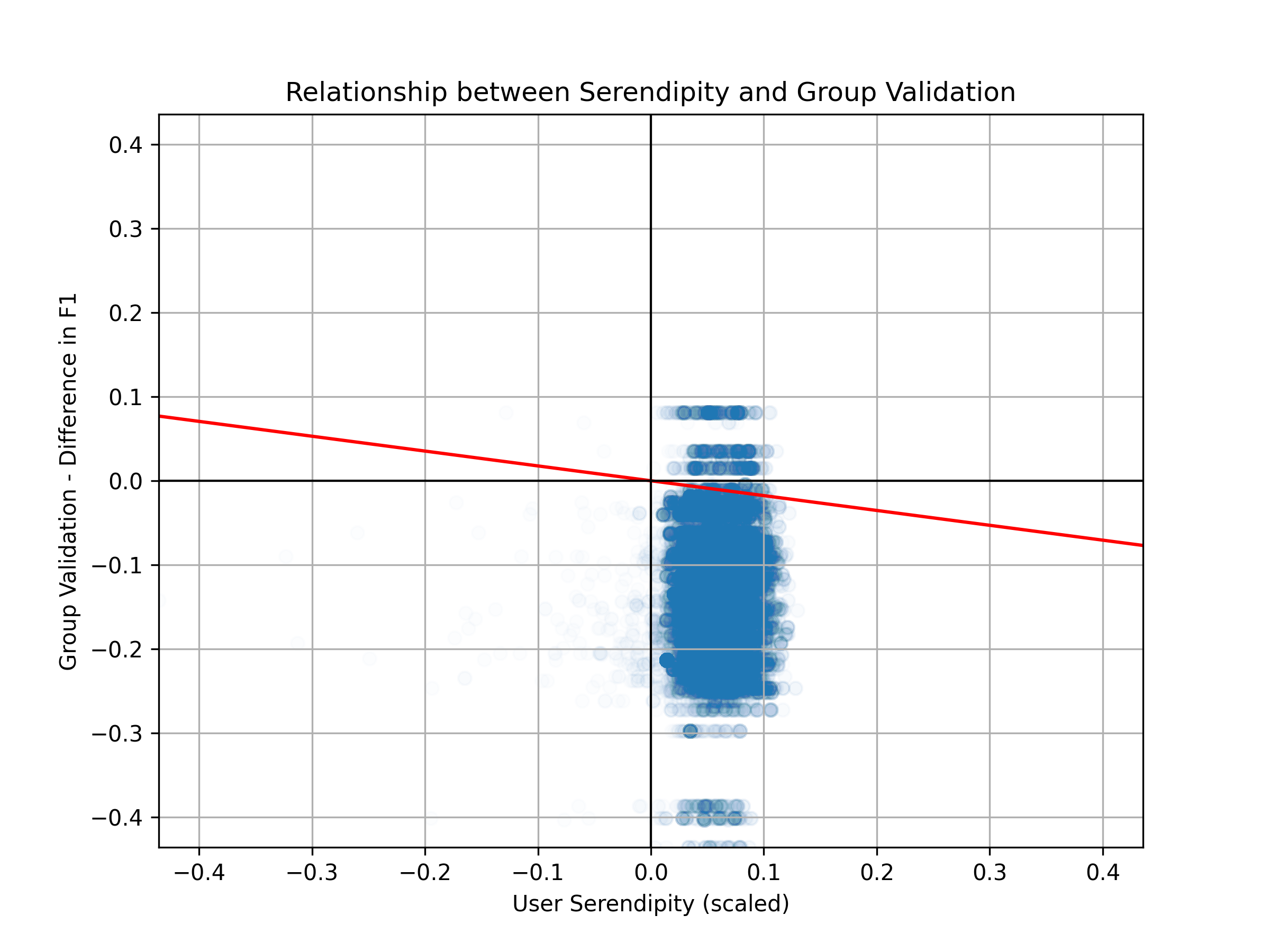}\ & \includegraphics[width=3cm]{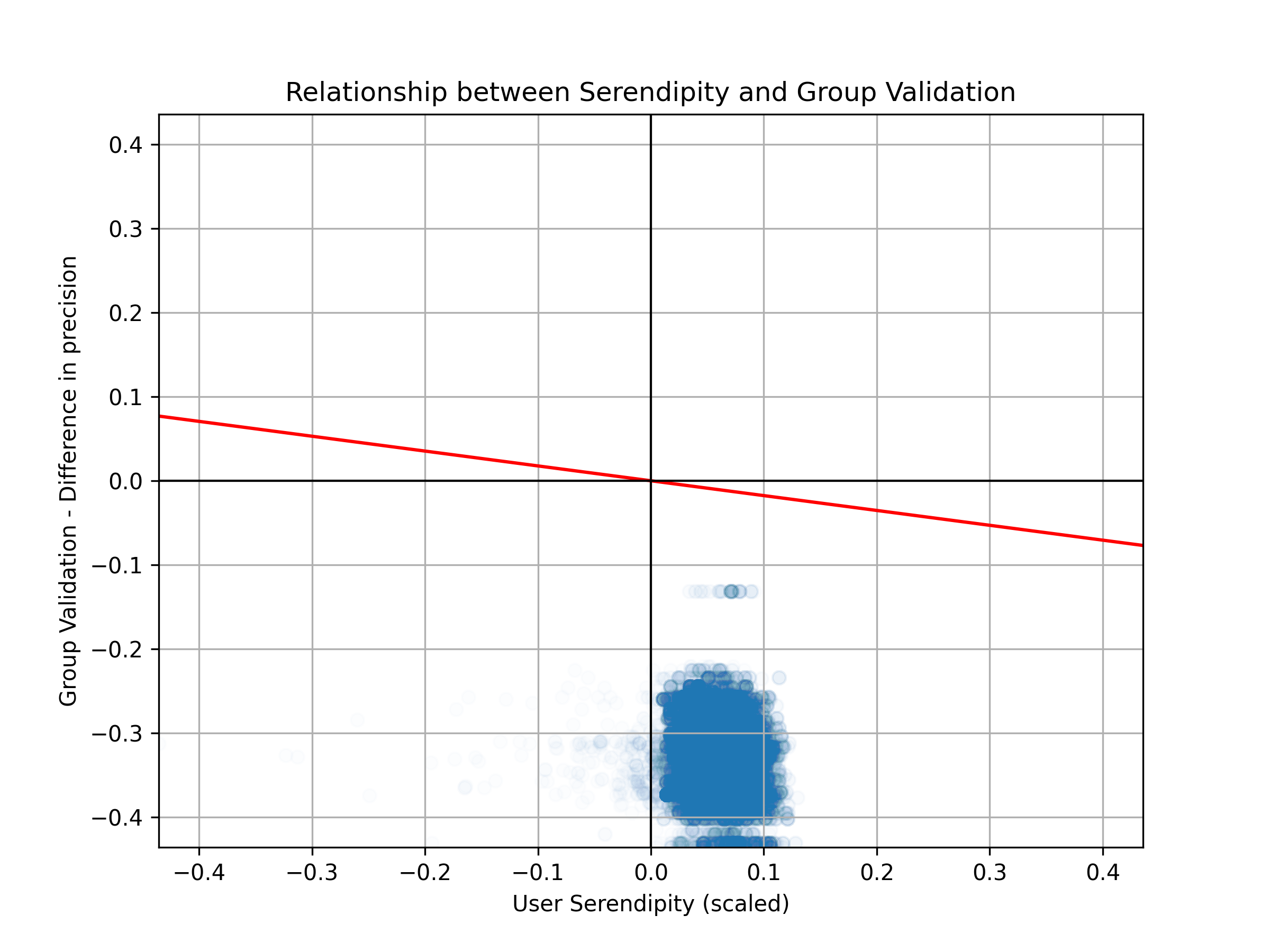} & \includegraphics[width=3cm]{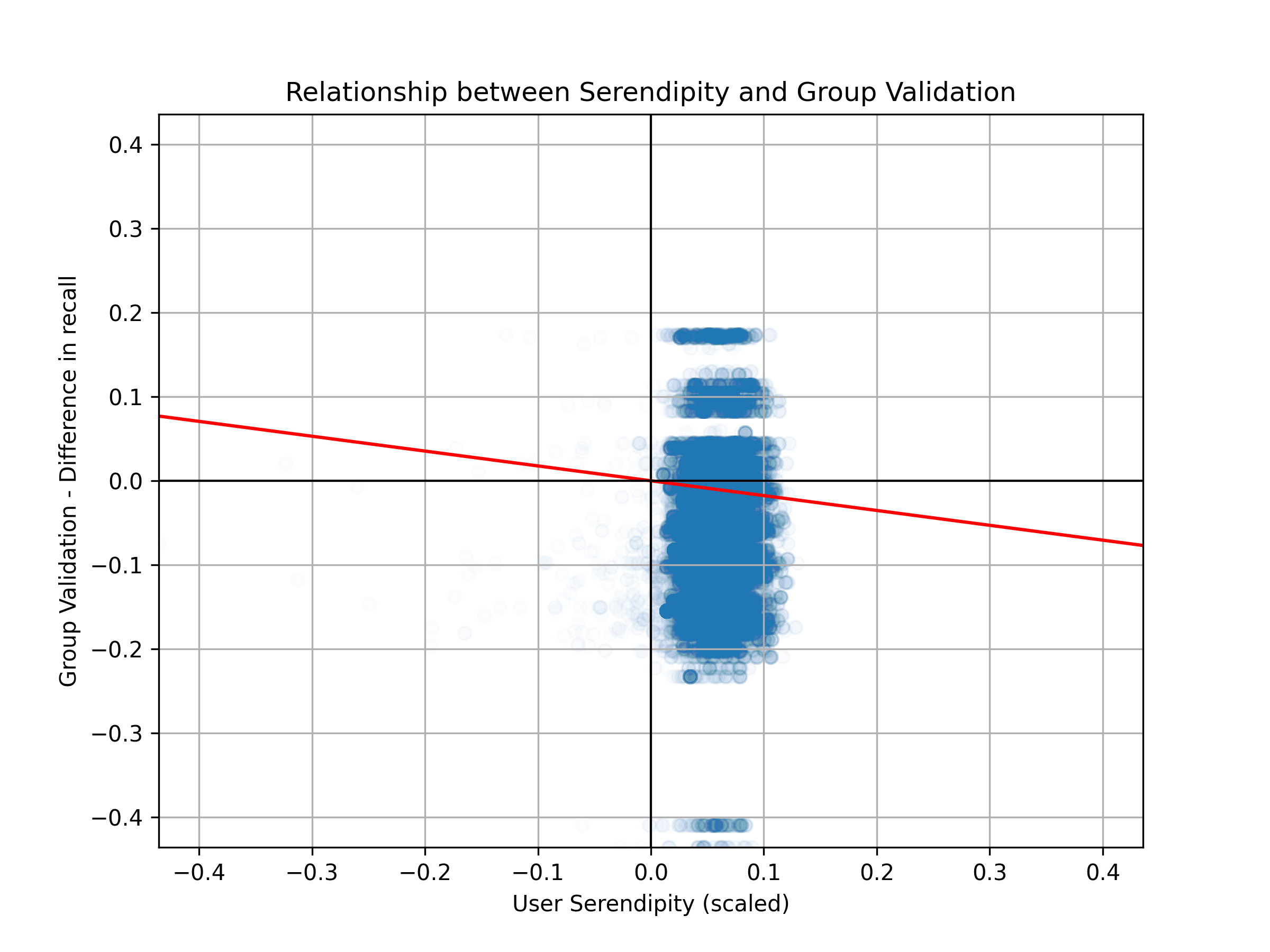} \\
    EL5 & \includegraphics[width=3cm]{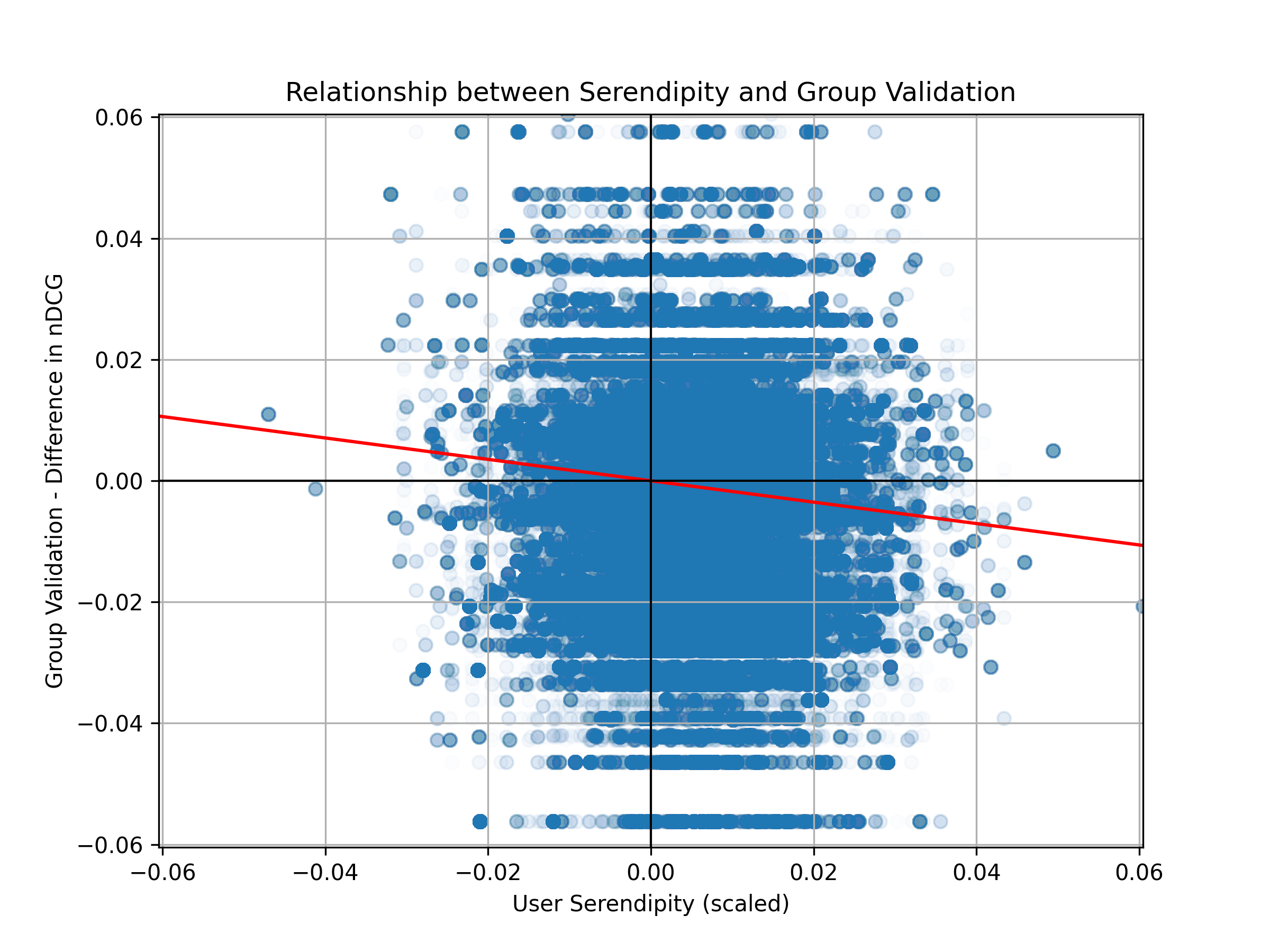} & \includegraphics[width=3cm]{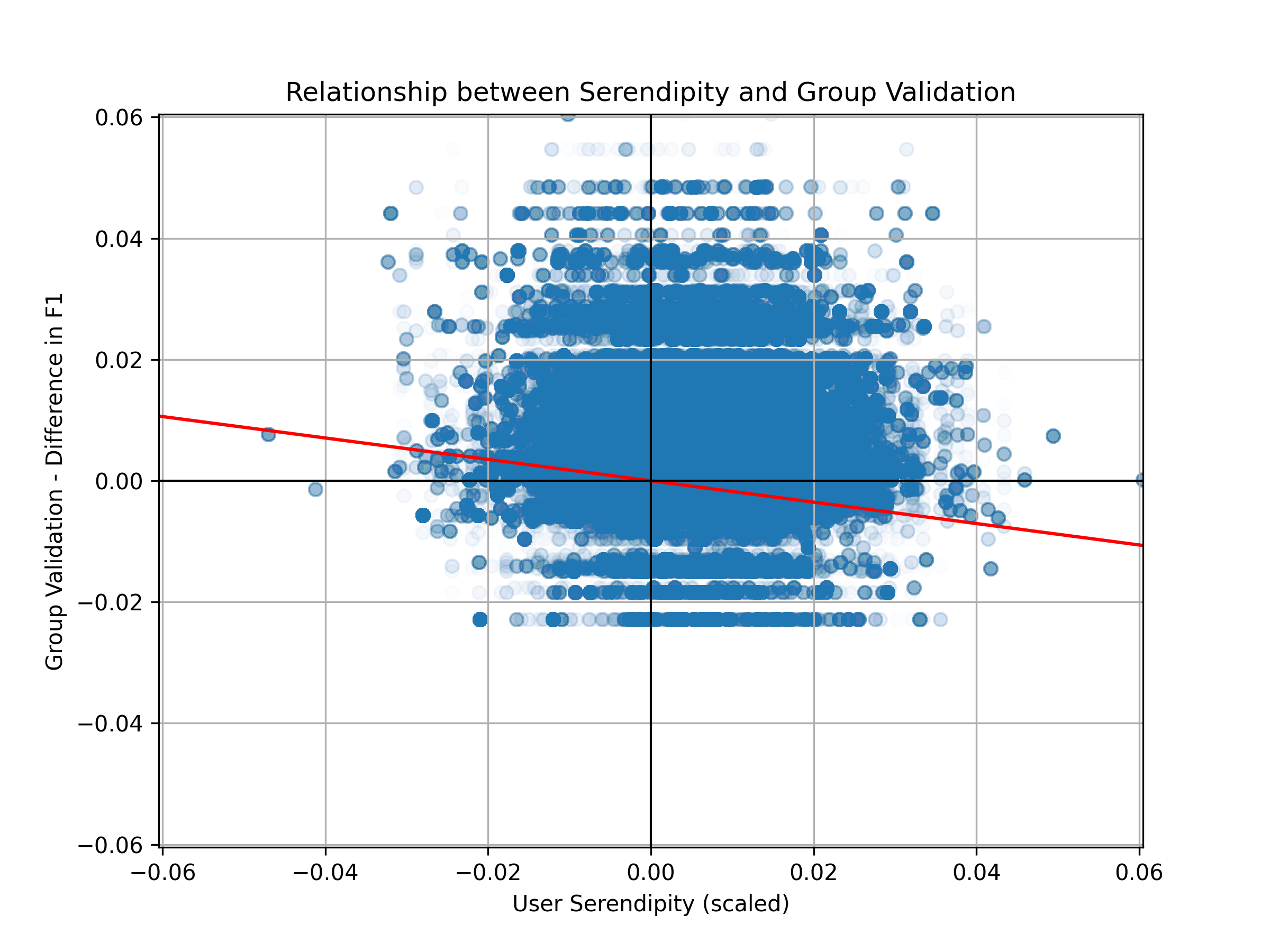}\ & \includegraphics[width=3cm]{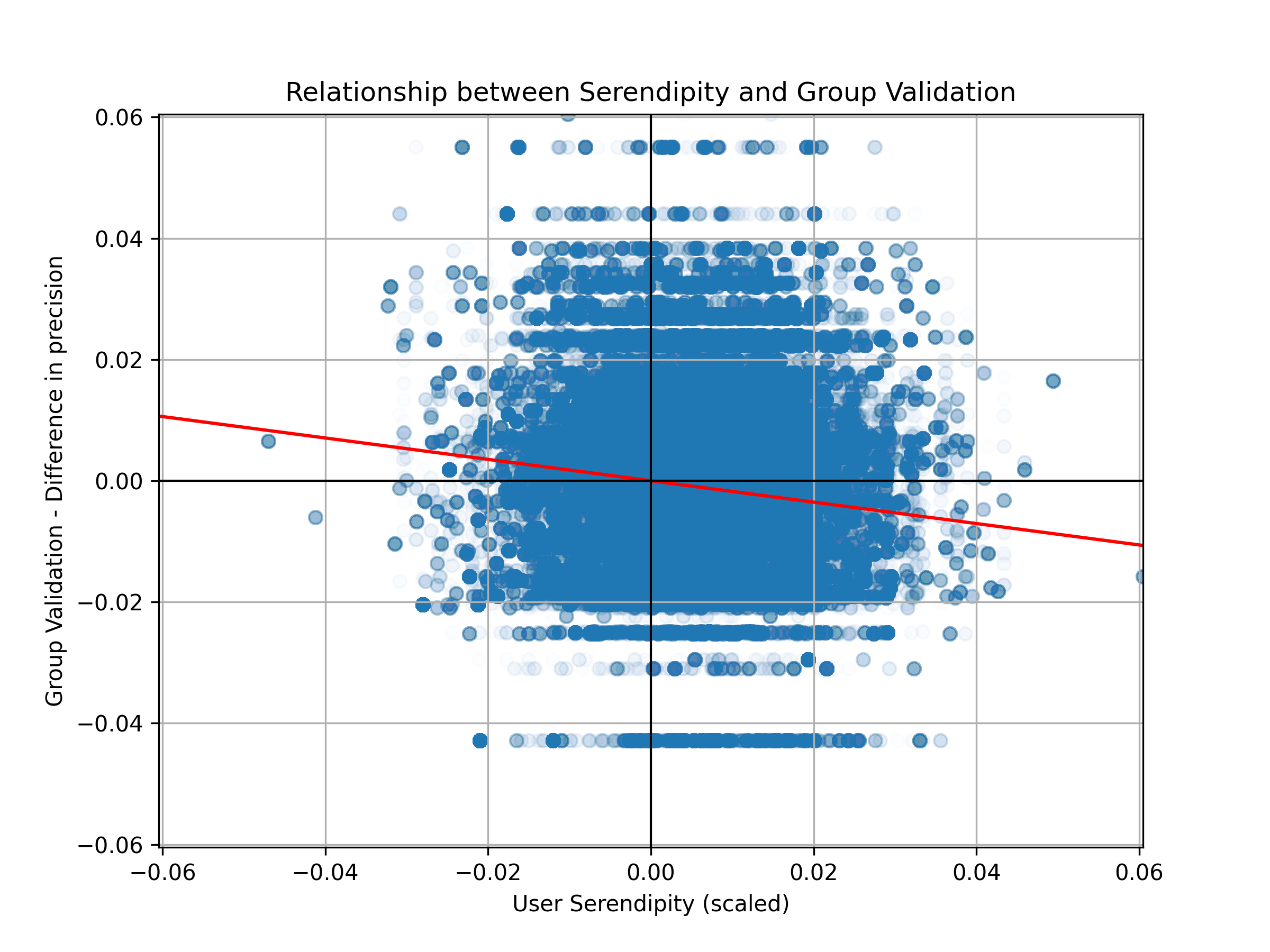} & \includegraphics[width=3cm]{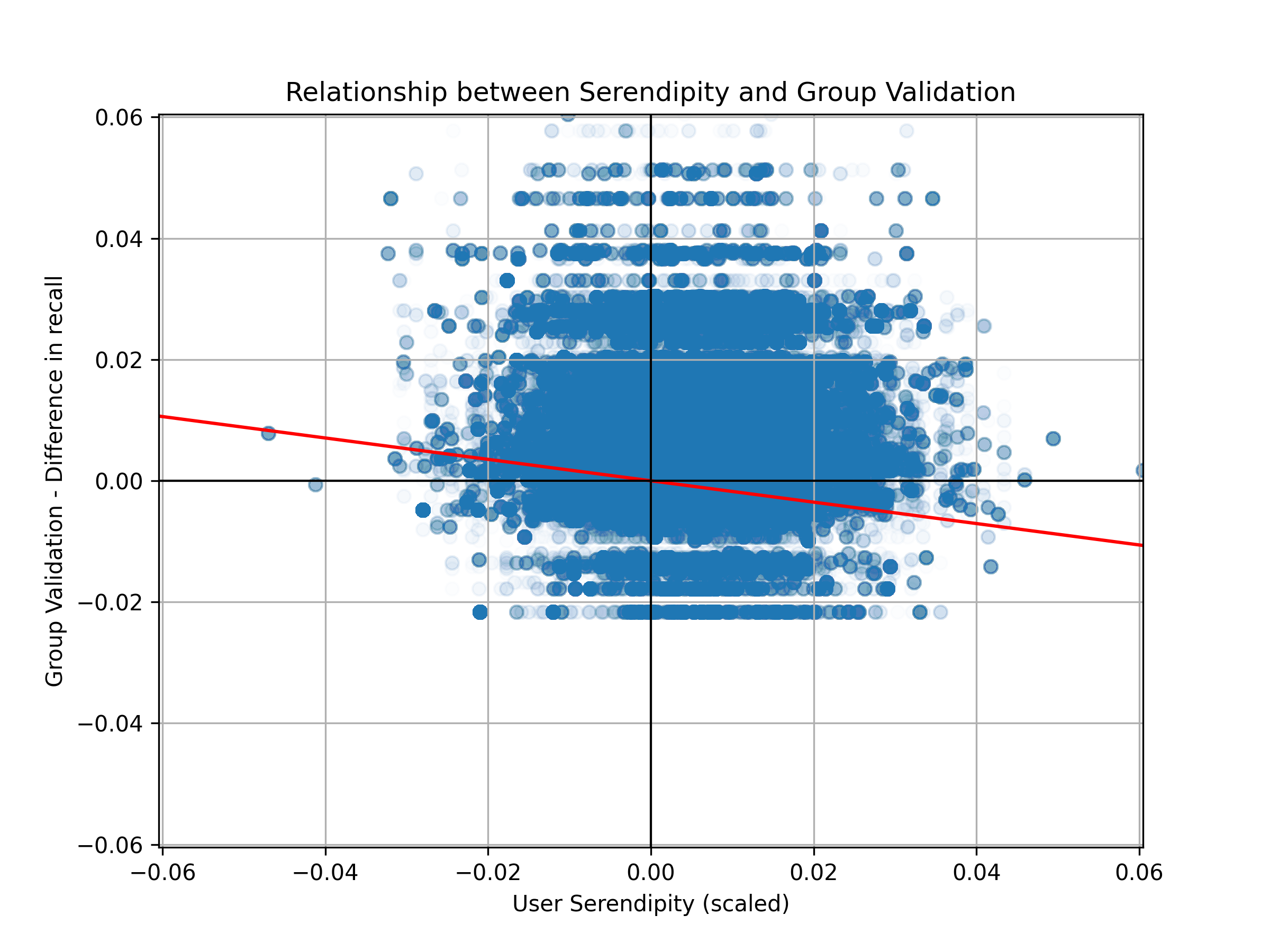} \\
    \bottomrule
    \end{tabular}
}
\end{table}

The proposed framework is modular by definition, enabling flexibility in decision-making based on user needs. Depending on the specific goals of the user, either to maximize recall, improve prediction or F1, the decision can reflect on all layers, including layer 2 and Ensemble learning selection criteria.

We tested the two ensemble learning algorithms (EL3 and EL4.2) that have the best results and chose the one with the better results in this paper.

As this framework is modular, choosing the best Ensemble Learning model to include in layer 2 is relative to the use case and the expectations from the system.
Having selected the four noise filtering algorithms to be used in the first layer, forming the decision board as well as the algorithm that will make up the second layer, we can now move forward to the next stage of building the framework in the following section.


\vspace*{-\baselineskip}
\section{Results and Analysis}
\label{sec:testing_analysis}
With all the necessary layers now in place what's left is to construct the framework itself, in order to establish a comprehensive method for managing noise within the system.
We begin by outlining the environmental configuration taken to test the proposed framework in section \ref{subsec:test_setup} and present the final results and analysis in section \ref{subsec:results}. 
\FloatBarrier
\subsection{Testing Setup}
\label{subsec:test_setup}
The framework was tested on the second subset of the MovieLens dataset,  ml-25m. We employed the four 2D system metrics [nDCG-Serendipity, F1-Serendipity, Precision-Serendipity and Recall-Serendipity respectively], along with global nDCG, Recall, Precision and F1-score. Additionally, we measured the critical groups percentage measured on nDCG.
We compared the framework results to each noise filtering algorithm of layer 1, To ensure that we are comparing noise management algorithms based on the full proposed methods, we applied the same noise mitigation strategy, noise removal, to all algorithms.

The proposed framework consists of the following three key layers:
\begin{itemize}
    \item Layer 1 \\
        - Noise Identification: \\
        - Apply four noise identification algorithms (NF1, NF2, NF3, and NF4).\newline
        - Reach a consensus on the identified noise.
    \item Layer 2 \\
        - Uncertain Dataset Labeling: \newline
        - Employ Ensemble learning methods, specifically semi-supervised Ensemble learning (RESSEL).\newline
        - Label uncertain datasets.
    \item Layer 3 \\
        - Noise Signature Detection:\newline
        - Utilize an obfuscation signature to remove Opt-out users
\end{itemize}

    

\subsection{Framework Results}
\label{subsec:results}
In this section, we present the results of the experiments utilizing the selected noise filtering algorithm, Ensemble learning algorithm as well as the obfuscation signature. Our assessment will focus on the effectiveness of these algorithms in the proposed 2D system, results will be presented as percentages in tables, bar plots, and graphs, showing different aspects of the results.
Note that as the results were not well displayed in the graphs, we've stretched the axis of the data points to be able to visualize their distribution clearly.
\setlength{\textfloatsep}{0pt}
\begin{table}[htbp!]
 \caption{Performance evaluation of the algorithms}

    \begin{tabular}{|l|c|c|c|c|c|}
       \hline
        & NF1 & NF2 & NF3 & NF4 & Framework Output (EL3) \\ \hline
        \% nDCG - Serendipity & 43.44 & 52.6 & 0.22 & 28.89 & 40.28 \\ \hline
        \% F1 - Serendipity & 81.44 & 55.5 & 36.51 & 87.7 & 99.31 \\ \hline
        \% Recall - Serendipity & 82.97 & 55.58 & 65.23 & 89.98 & 99.42 \\ \hline
        \% Precision - Serendipity & 44.56 & 55.1 & 0.2 & 27.8 & 38.05 \\ \hline
        nDCG & 0.042651365 & 0.041987621 & 0.017163298 & 0.040027887 & 0.04327625 \\ \hline
        precision & 0.03847795163 & 0.0393314367 & 0.0089311334 & 0.034903983 & 0.03593974 \\ \hline
        recall & 0.034701350 & 0.030043197 & 0.025065275 & 0.0353888673 & 0.0390435789\\ \hline
        F1 & 0.036492201 & 0.034065538 & 0.013169704 & 0.035144753 & 0.037427421 \\ \hline
    \end{tabular}
    \label{tab:eval_performance1}
\end{table}


\begin{figure}[ht]
    \centering
    \begin{minipage}[b]{0.45\textwidth}
        \centering
        \caption{Improvement of nDCG-Serendipity on the algorithms}
        \label{graph:ndcg-graph}
        \includegraphics[width=\textwidth]{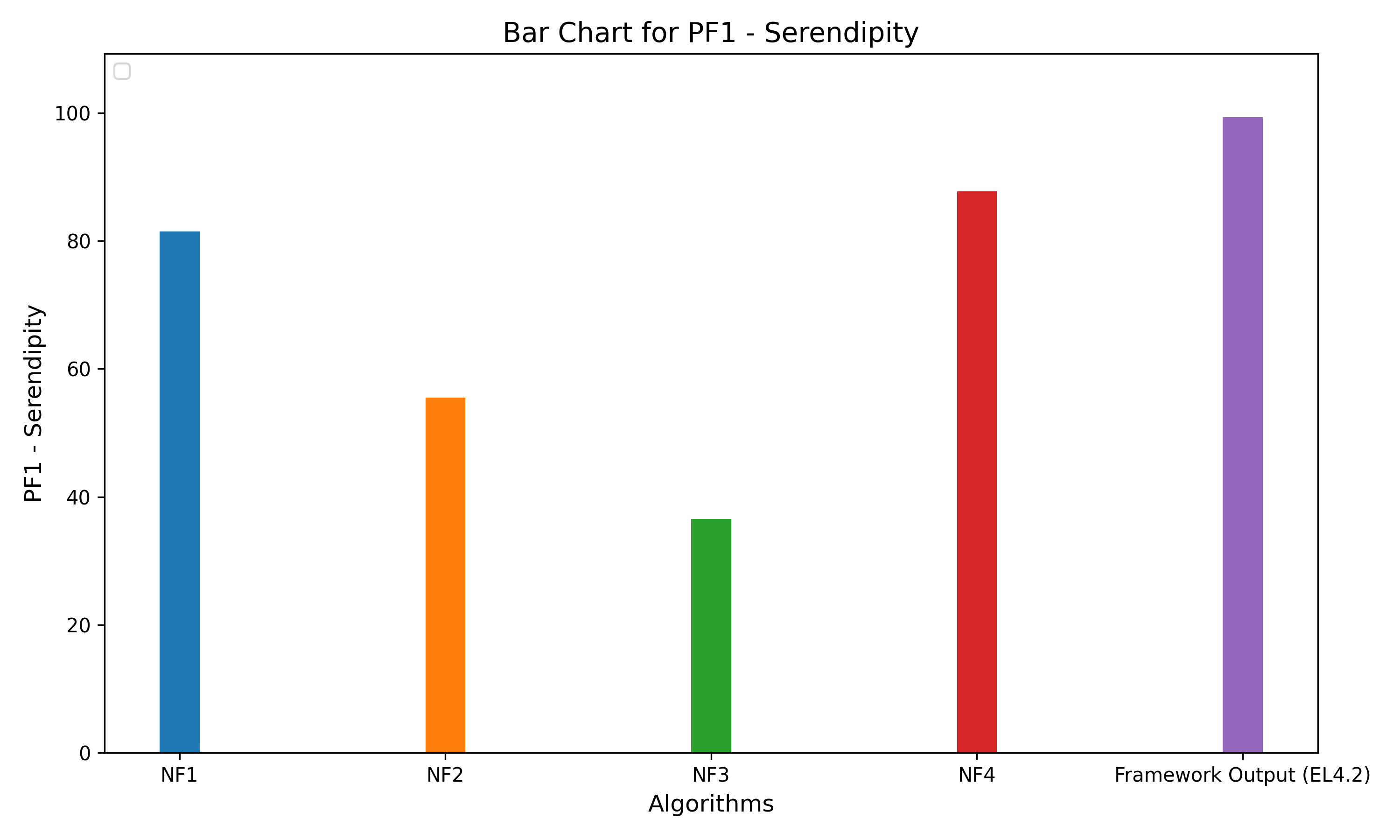}
        
    \end{minipage}
    \hfill
    \begin{minipage}[b]{0.45\textwidth}
        \centering
        \caption{Improvement of F1-Serendipity on the algorithms}
        \label{graph:f1-graph}
        \includegraphics[width=\textwidth]{images/PF1___Serendipity.png}
        
    \end{minipage}
    \hfill
    \begin{minipage}[b]{0.45\textwidth}
        \centering
        \caption{Improvement of Recall-Serendipity on the algorithms}
        \label{graph:recall-graph}
        \includegraphics[width=\textwidth]{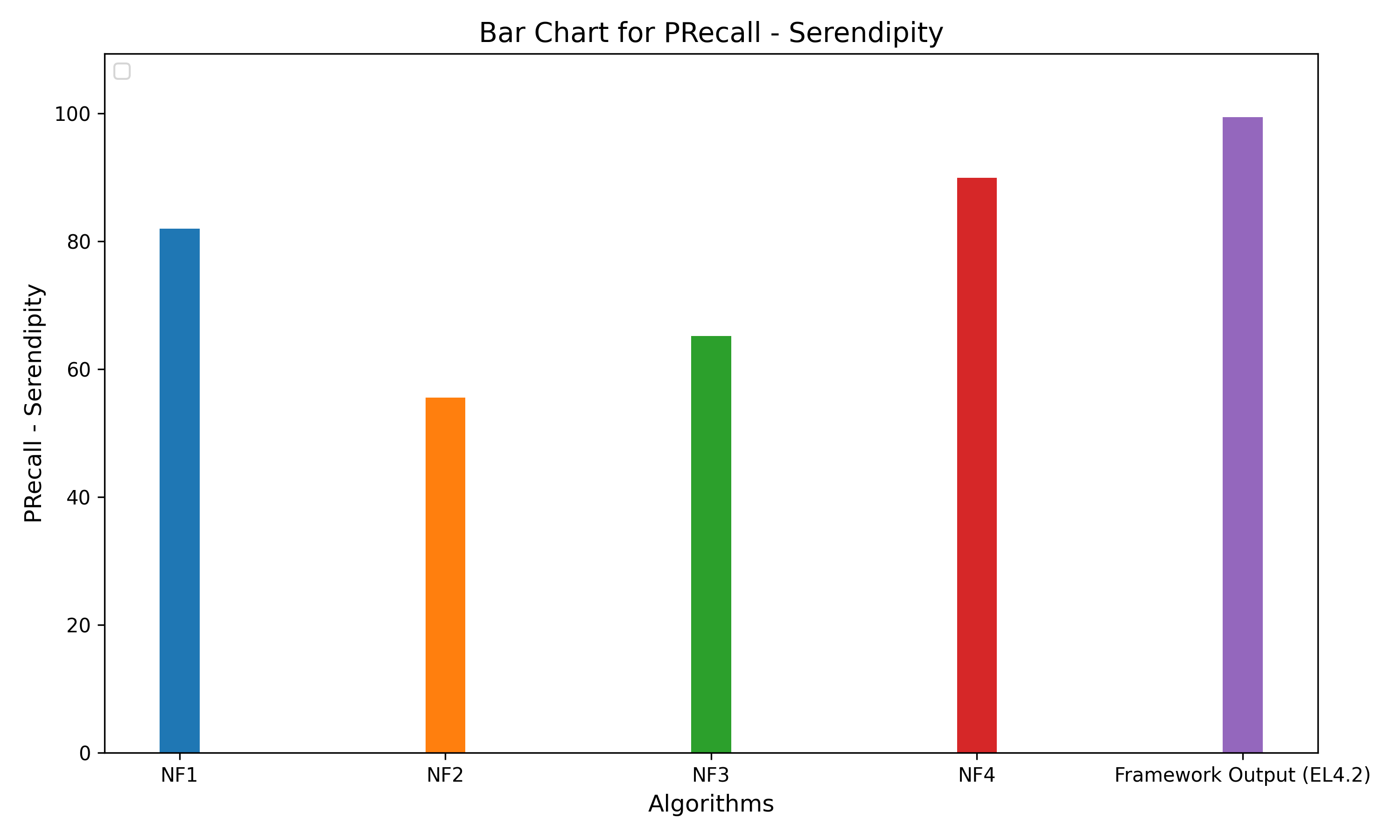}
        
    \end{minipage}
    
\end{figure}
\begin{figure}[ht]
    
    \begin{minipage}[b]{0.45\textwidth}
        \centering
         \caption{Improvement of Precision-Serendipity on the algorithms}
         \label{graph:prec-graph}
        \includegraphics[width=\textwidth]{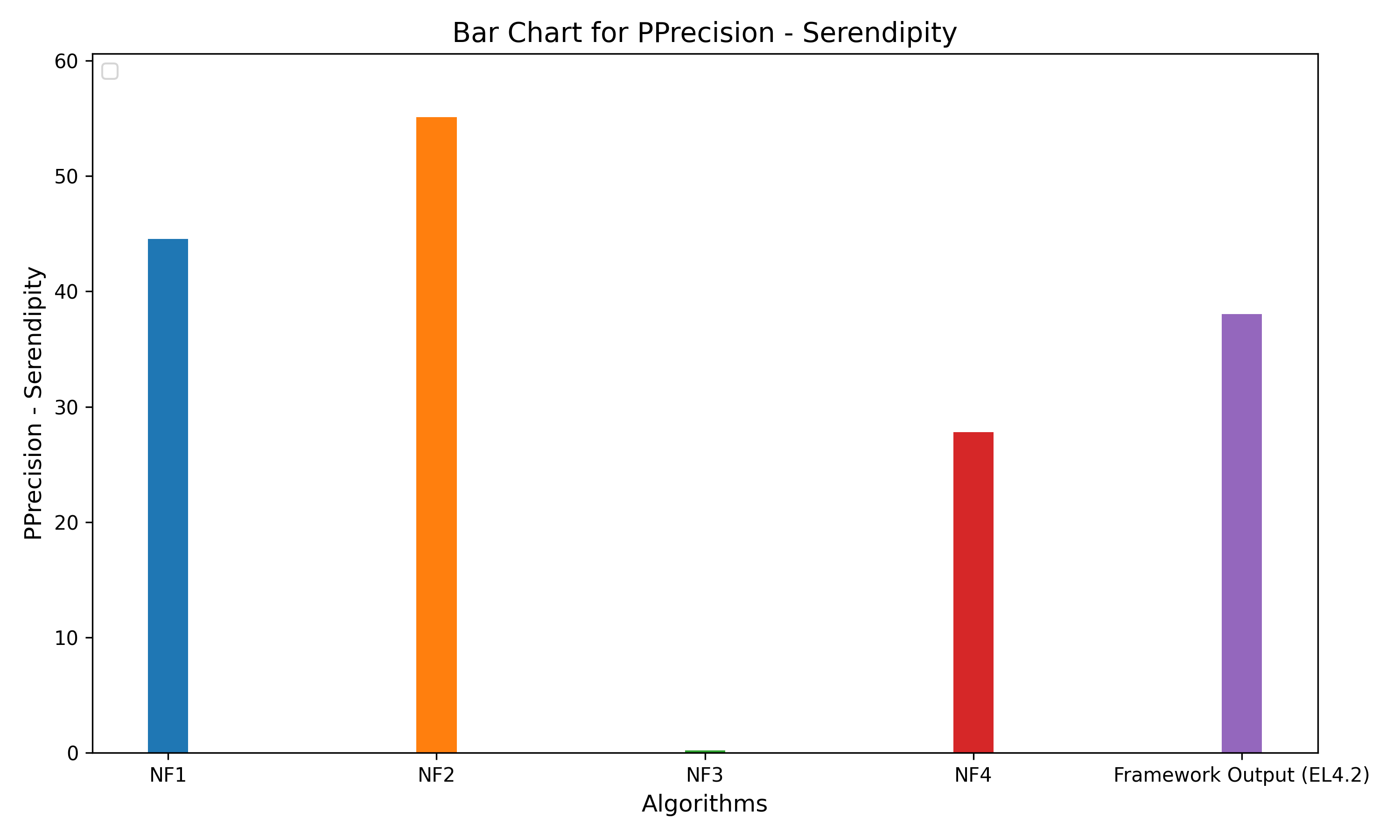}
       
    \end{minipage}
    \hfill
    
         
\end{figure}

The results are displayed in Table \ref{tab:eval_performance1}, it outlines the performance evaluation percentage values, while the figures display the ranking of each algorithm across the evaluation systems in figure \ref{graph:f1-graph}, figure \ref{graph:recall-graph}, figure \ref{graph:prec-graph} and figure \ref{graph:ndcg-graph}.
In addition, we can see the distribution of the data points across each system for the algorithms, in the 2d system in table \ref{graph:propo_nfs}.
\newline
The proposed framework outperformed others on the nDCG - Serndipity and  Precision - Serendipity Systems, while low values were seen on all the algorithms, the proposed algorithm was able to get 40.28\% improvement (percentage of ratings above the line) on  nDCG - Serndipity, while it scored 38.05\% on Precision - Serendipity. NF2 scored second concerning these two systems, with 52.6\% on nDCG-Serendipity and 44.56\% with very close results between the two.



We have also compared other metrics for accuracy beyond the proposed 2D metric. Results on the global metrics showed a difference in performance, the proposed framework was leading in the nDCG and Recall, NF1 got the best scores on Precision and F1. We can see that for global recall the proposed algorithm was able to place first, while all improvements on the other global metrics, was only tagged on the group levels.

The proposed framework placed first on F1-Serendipity and Recall-Serendipity where it got 99.31\% and 99.42\% of data ratings improved. NF4 followed with the second place, then NF1 and NF2.

Based on the table \ref{graph:propo_nfs} containing the graphs demonstrating the various scatter of the data points across the 2D system, a more detailed look can be made on the system, we can first identify that the proposed framework took similar improvement aspects to the chosen Ensemble learning utilized in layer 2. It highlights a good improvement for most of the dataset with F1-Serendipity and Recall-Serendipity, while it maintains a balance for nDCG-Serendipity and Recall-Serendipity. Our proposed frameworks, NF3 depicts very good improvements on Serendipity, which reaches up to an improvement of 0.1 across the data points, while small deteriorations in system performance were observed in Group Validation accuracies for nDCG and precision, NF1, NF2 and NF4 maintain system balance by enhancing performance for some data points while causing a decline for others.

To summarize, the proposed framework demonstrates good capabilities in balancing precision and recall to improve Serendipity in recommendations. On the other hand, being flexible enough to meet the user's needs, whether it is to improve either recall or precision, ranking, or reducing critical groups-all the decisions within each layer of this framework can be modified to meet such needs and ensure a more tailored and user-centric approach to noise management.

\FloatBarrier

\begin{table}[h!]
\centering
\caption{Performance Metrics for ML-25M-Subset Dataset}
\resizebox{\textwidth}{!}{%
\begin{tabular}{@{}l@{}c@{}c@{}c@{}c@{}}
\toprule
\multirow{1}{*}{Implementation} & \multicolumn{4}{c}{ML-25M-subset} \\
\cmidrule(lr){1-5}
  & \% nDCG - Serendipity & \% F1 - Serendipity  & \% Precision - Serendipity  & \% Recall - Serendipity\\
\midrule
Proposed framework & \includegraphics[width=3cm]{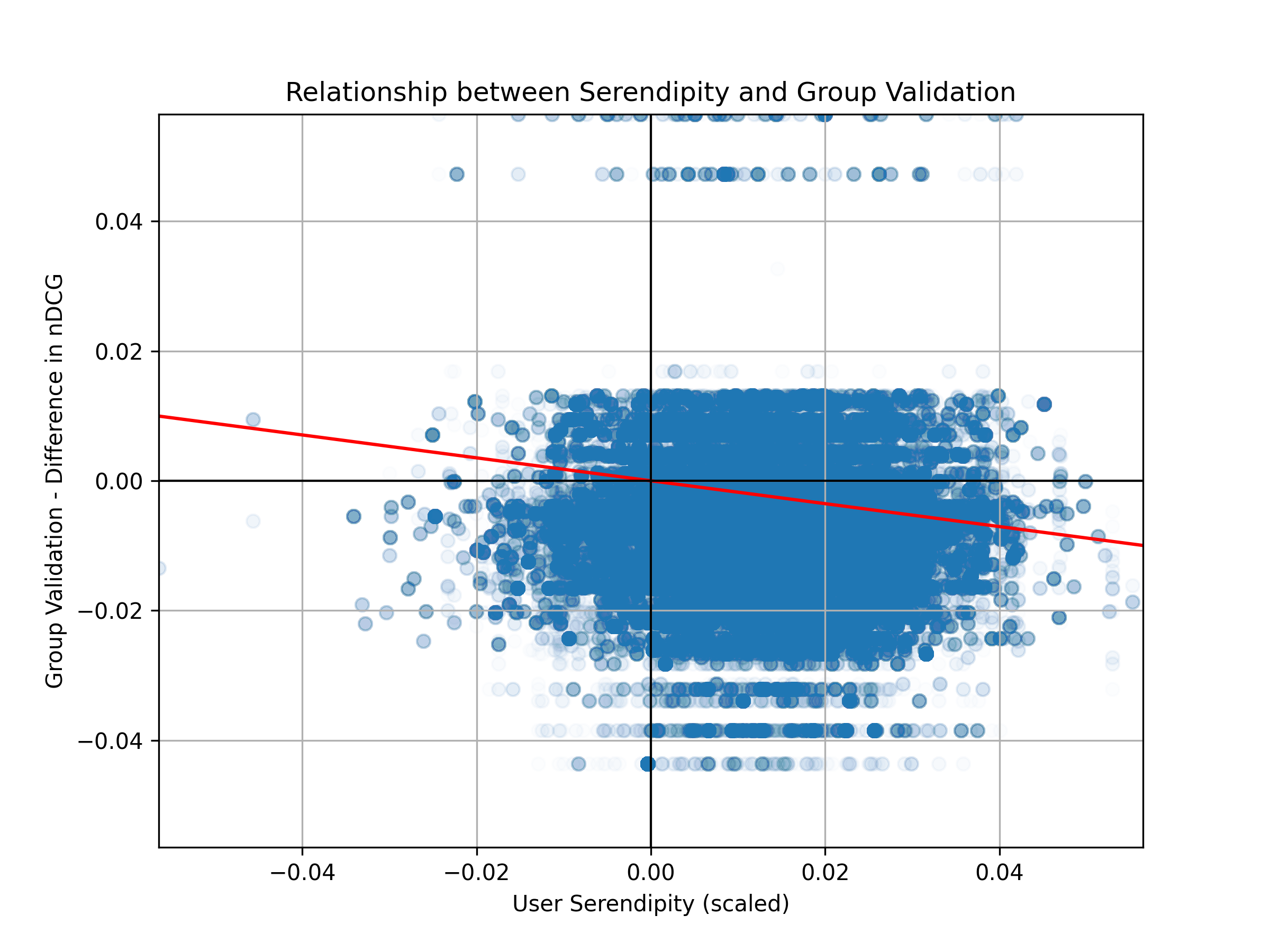} & \includegraphics[width=3cm]{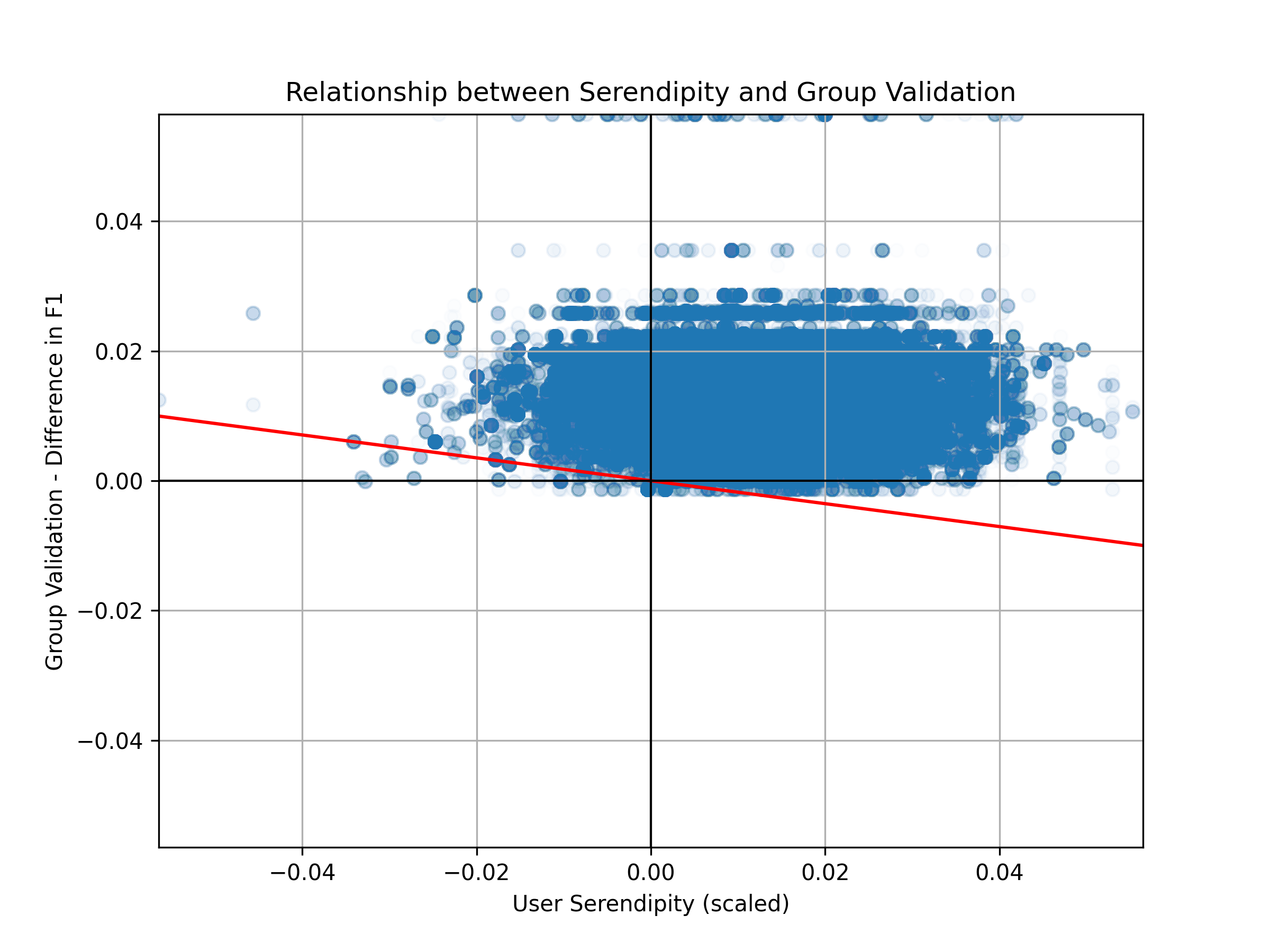} & \includegraphics[width=3cm]{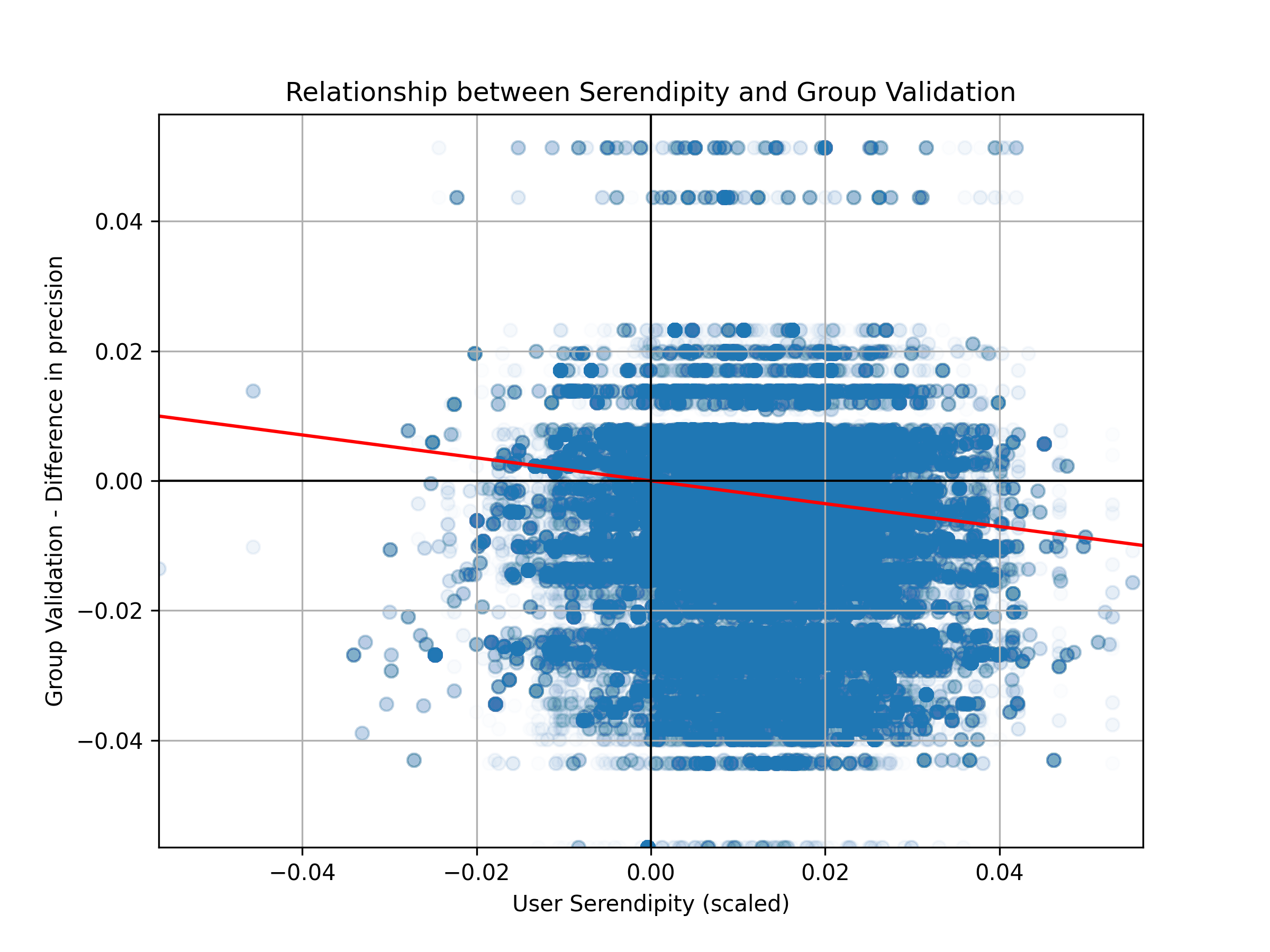} & \includegraphics[width=3cm]{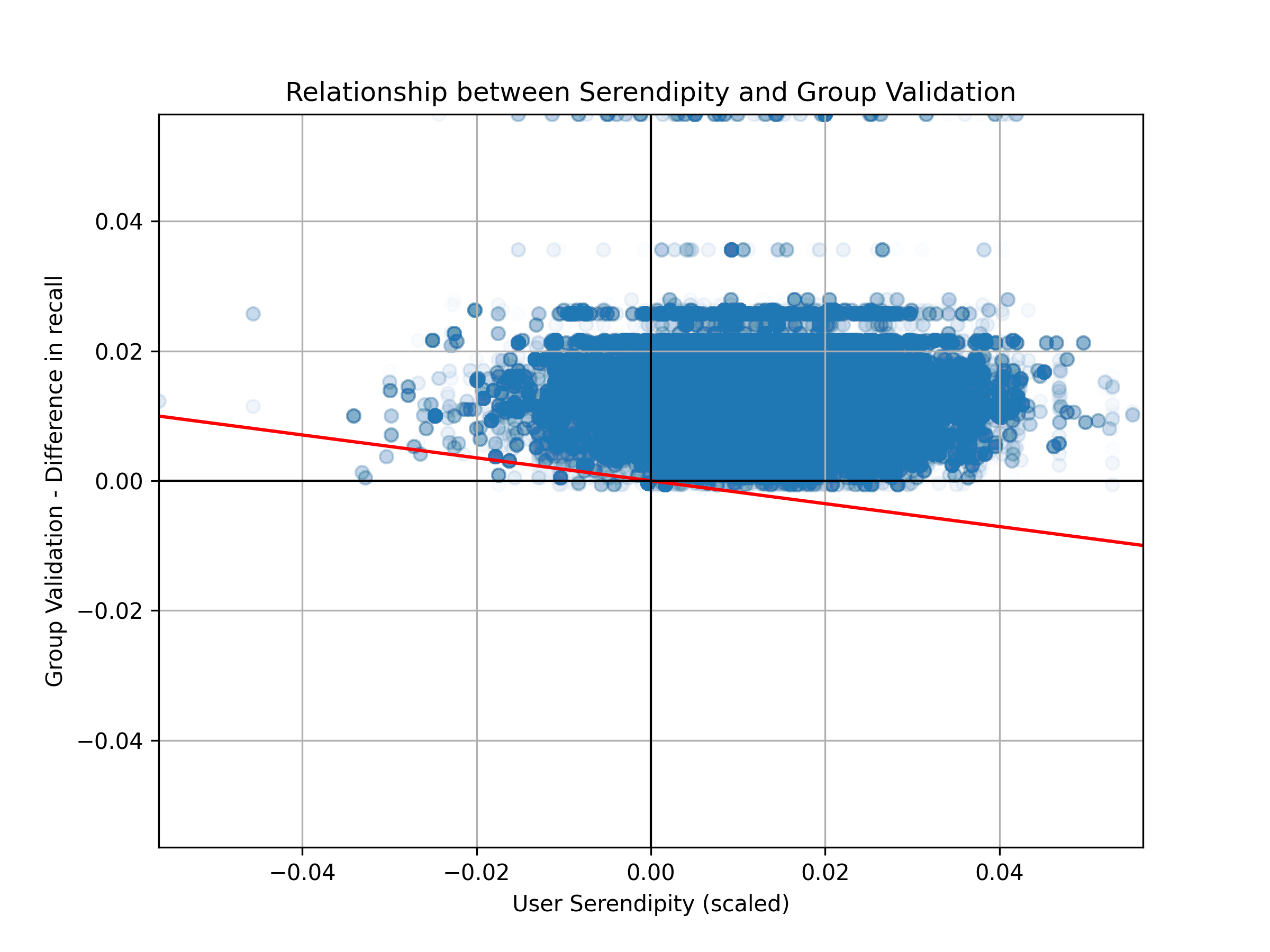} \\
NF1  & \includegraphics[width=3cm]{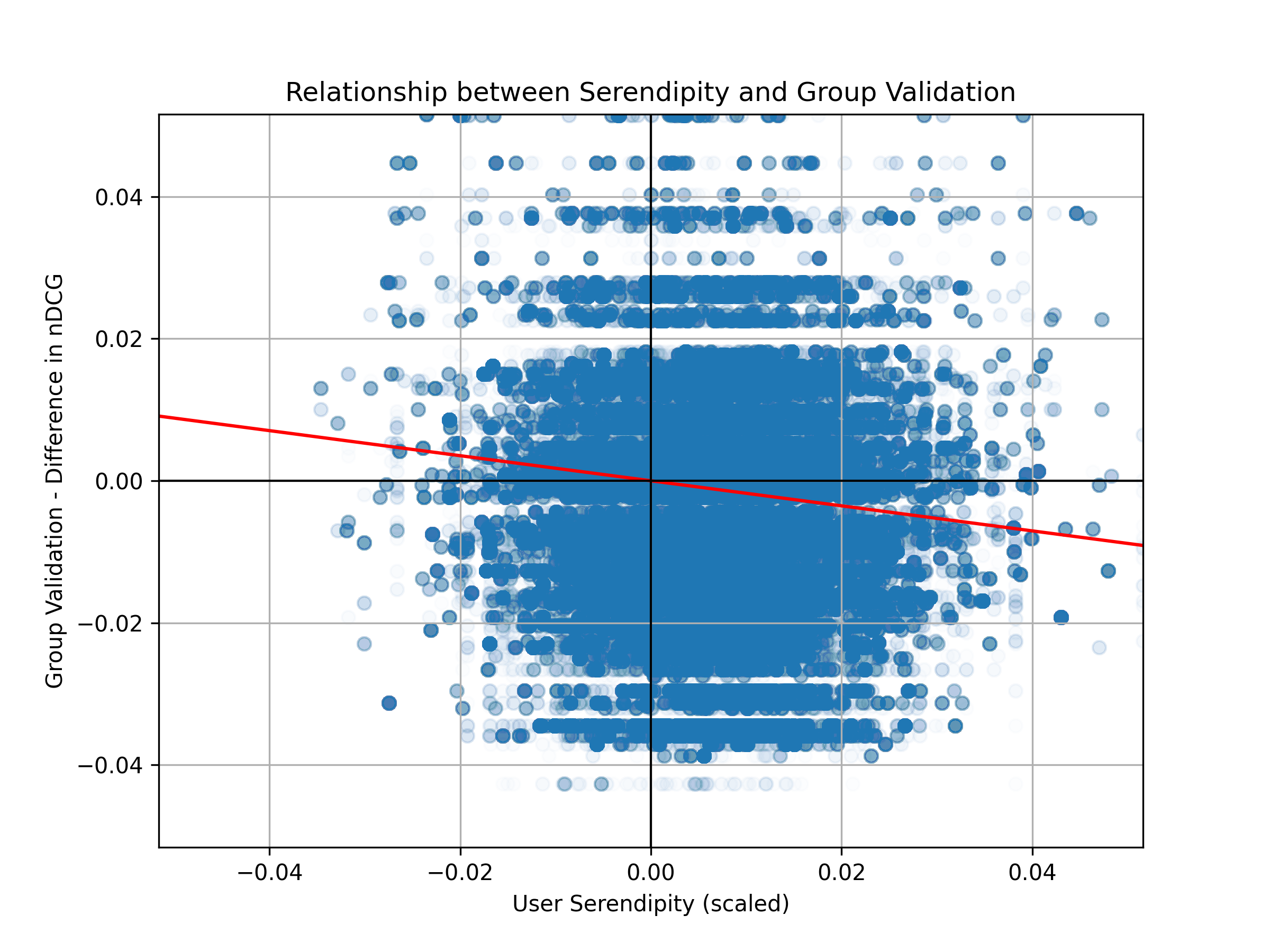} & \includegraphics[width=3cm]{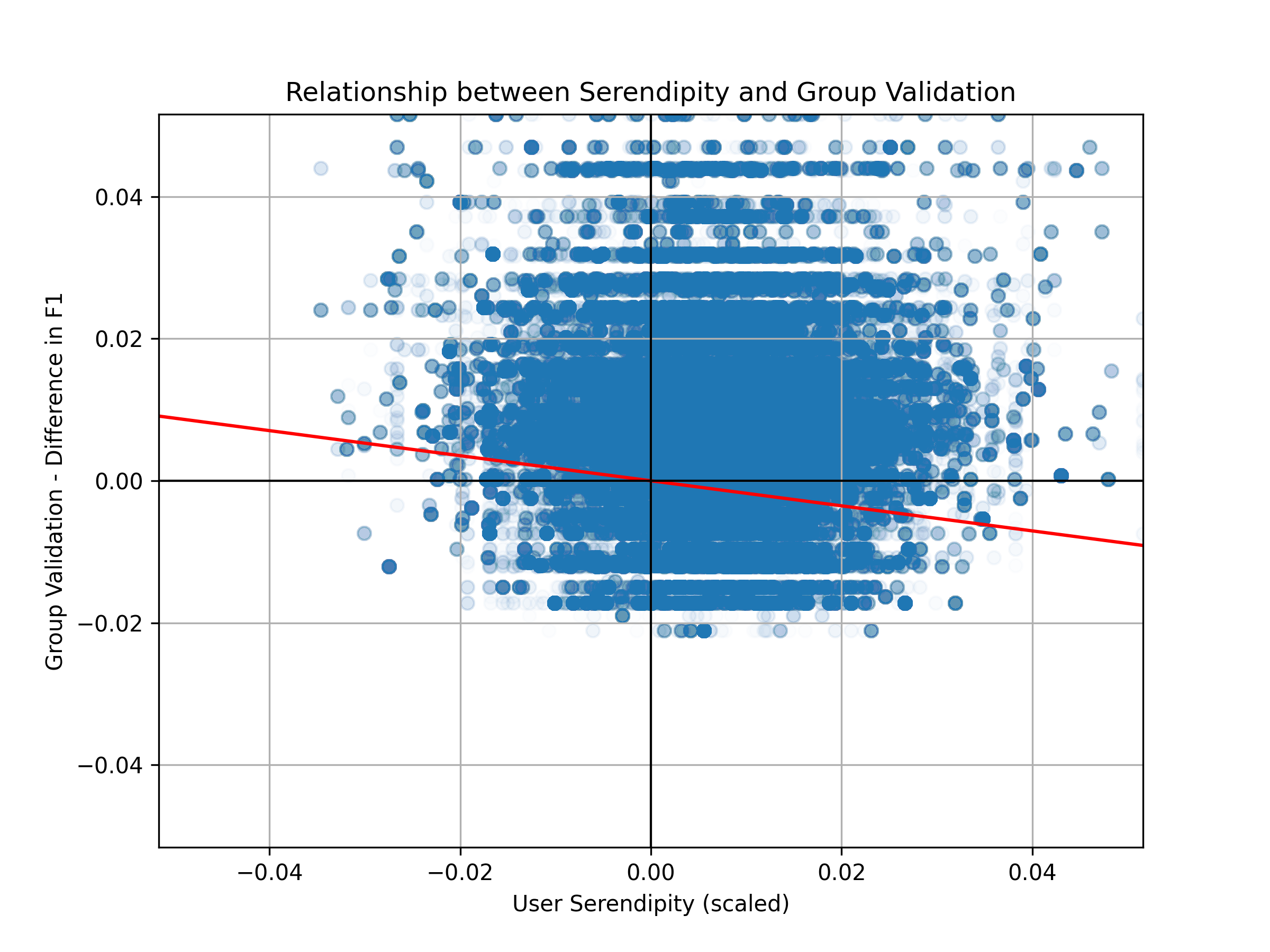} & \includegraphics[width=3cm]{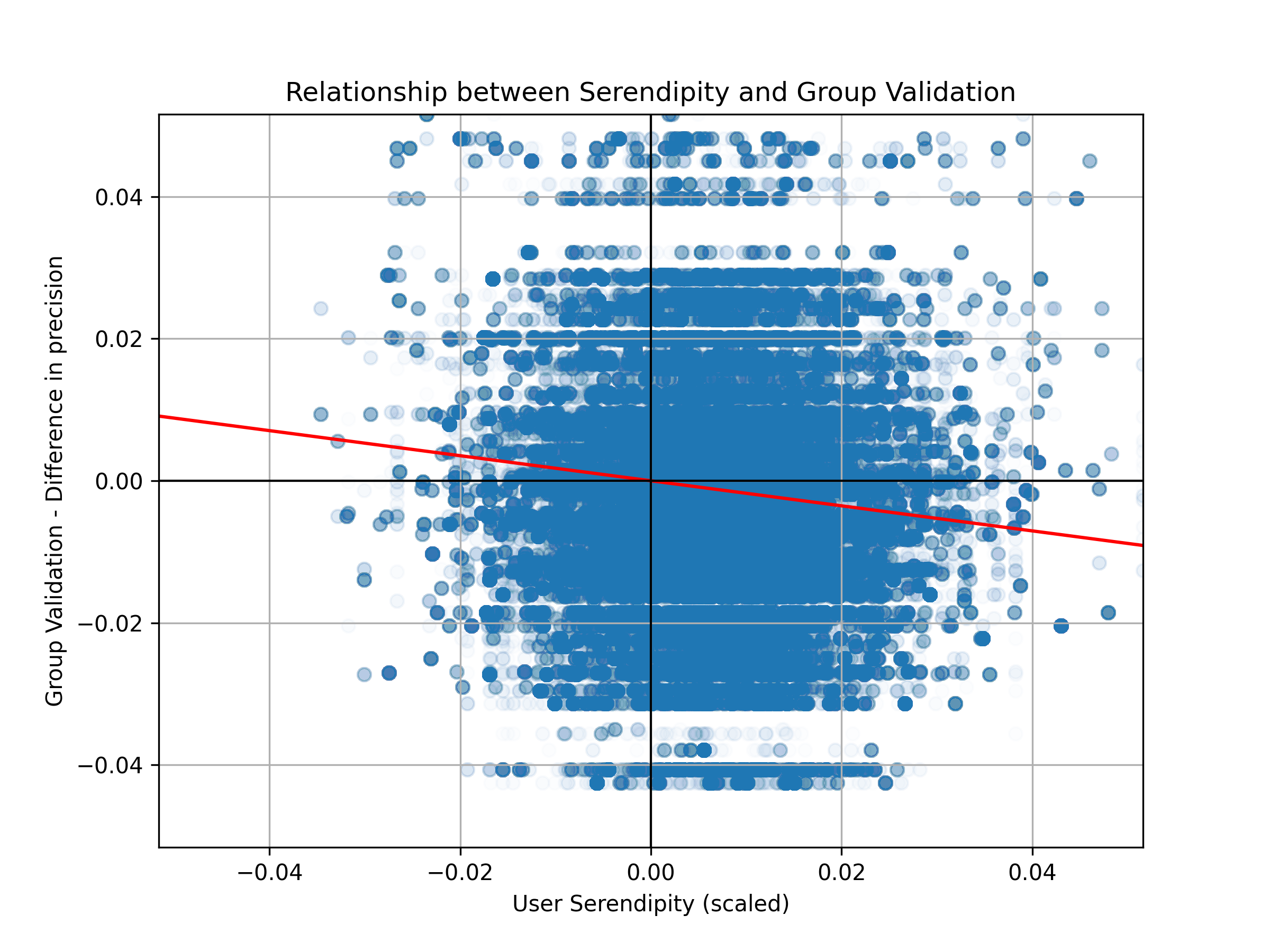} & \includegraphics[width=3cm]{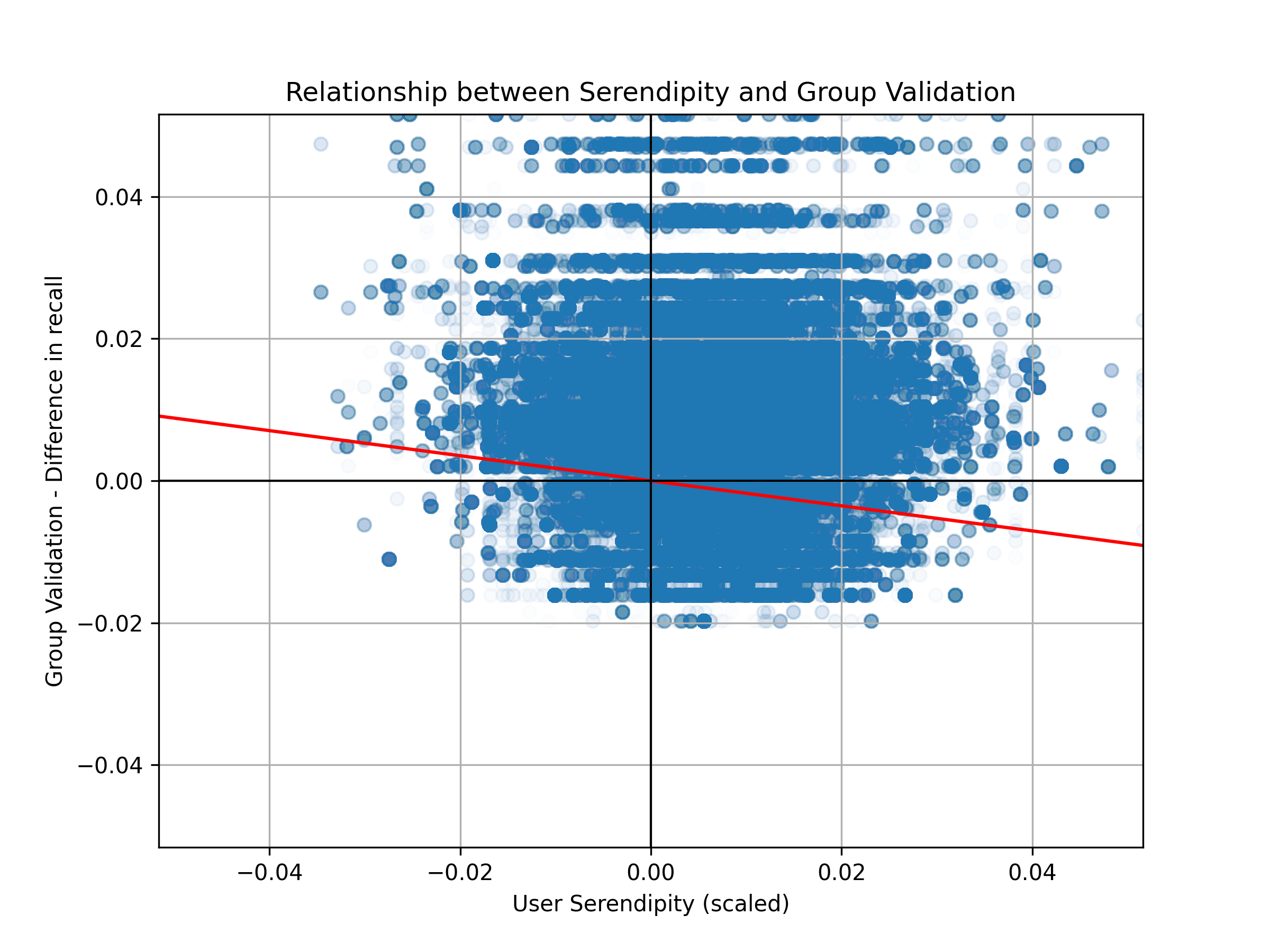} \\
NF2 & \includegraphics[width=3cm]{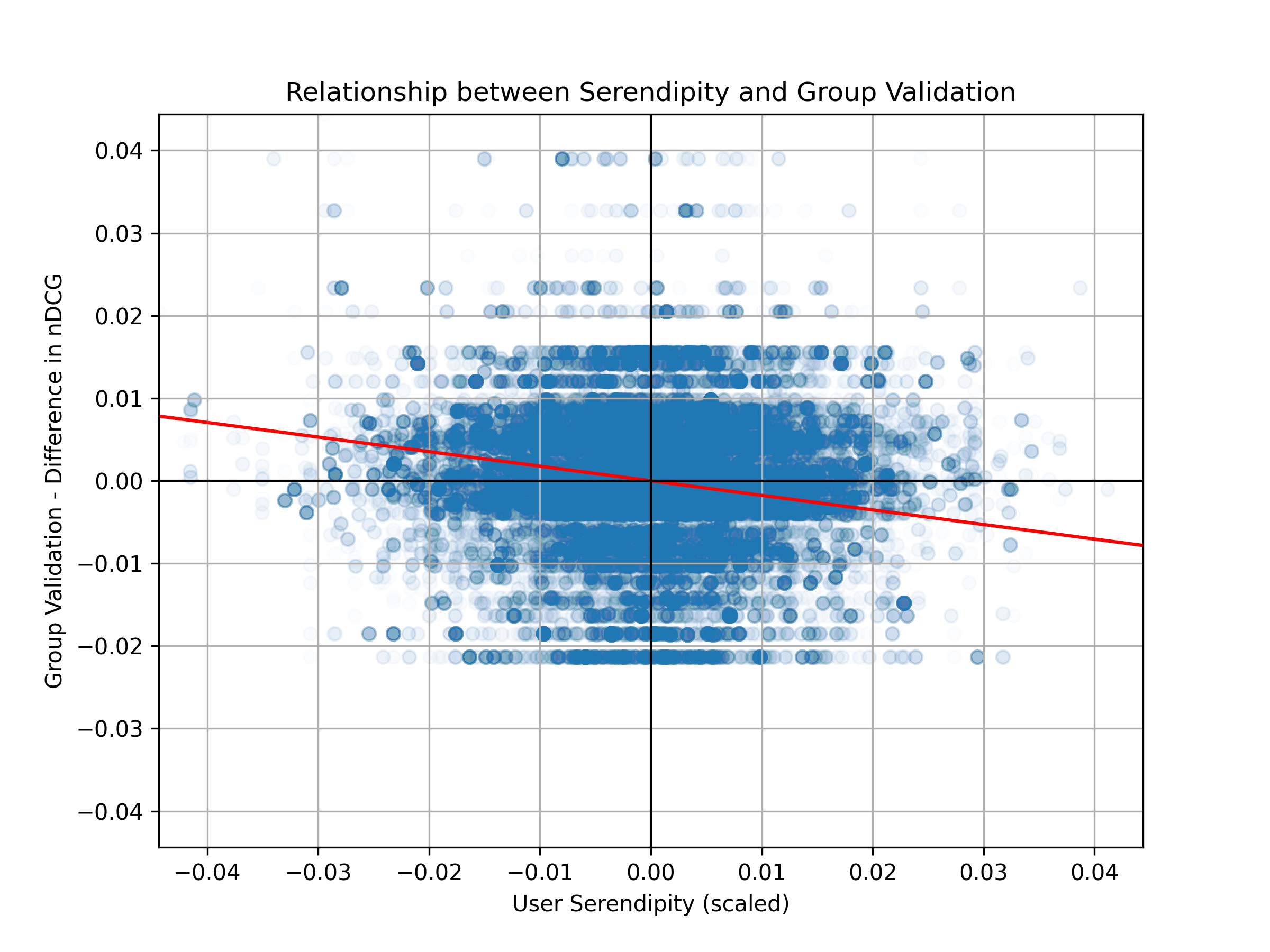} & \includegraphics[width=3cm]{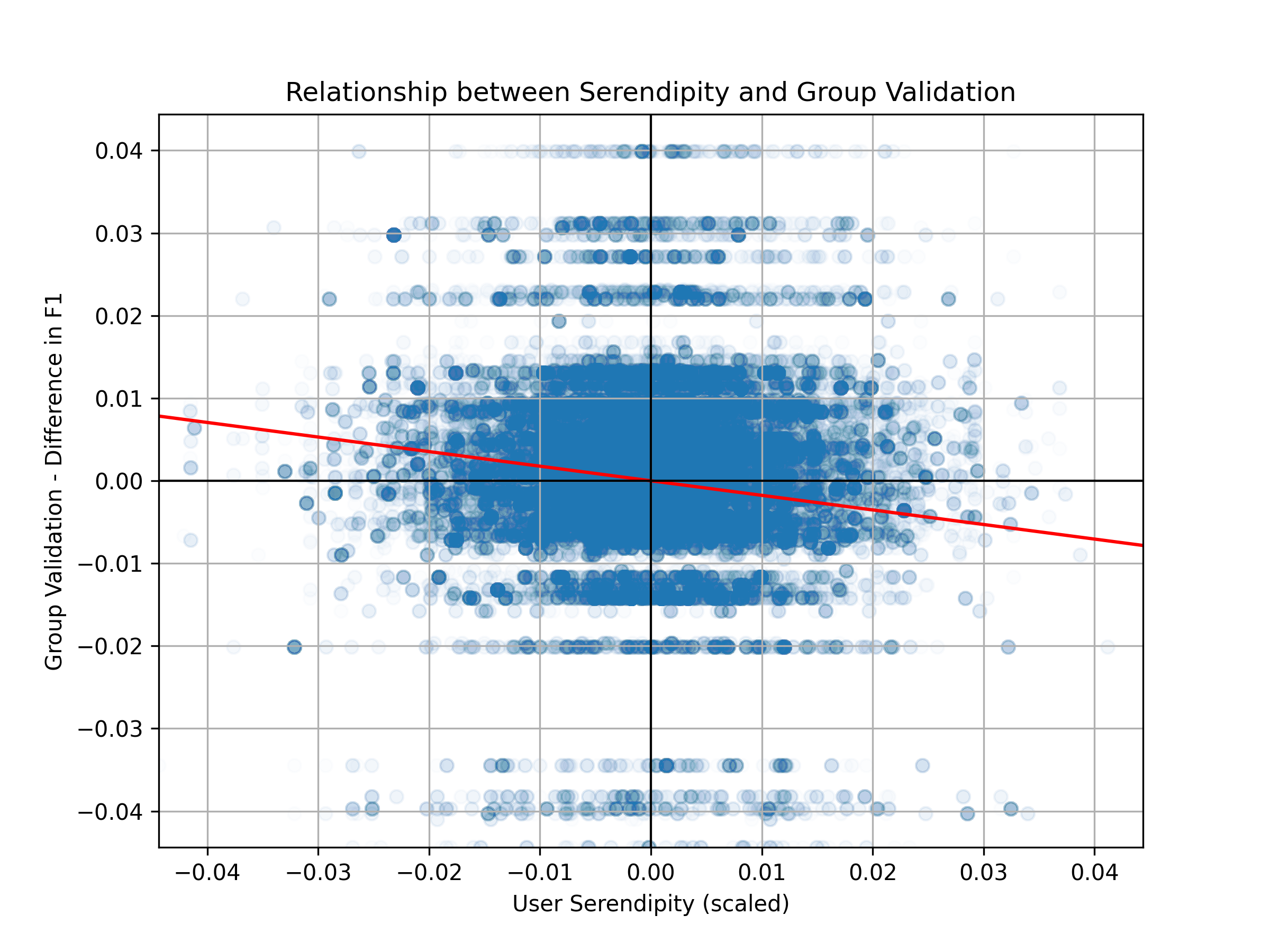} & \includegraphics[width=3cm]{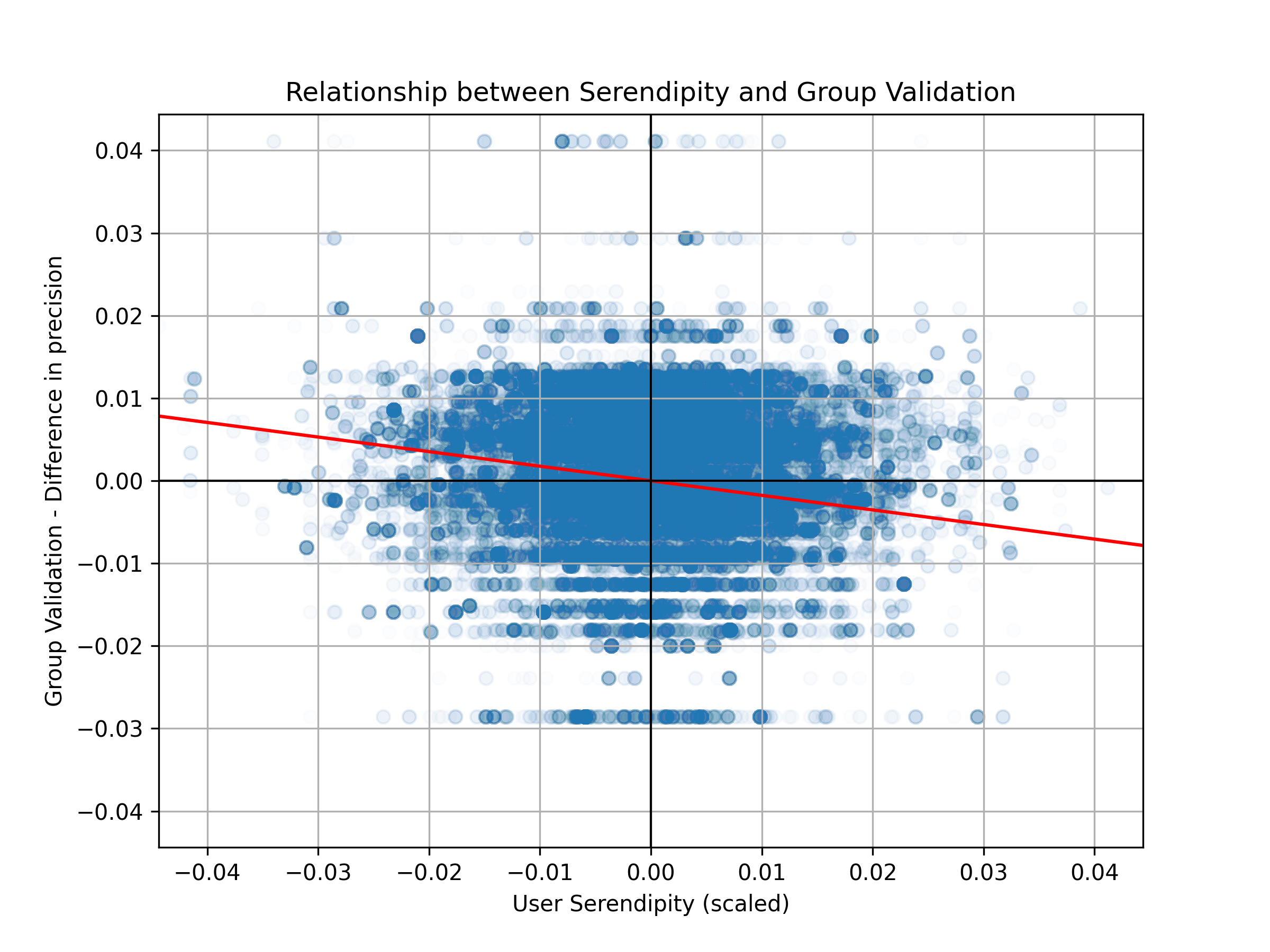} & \includegraphics[width=3cm]{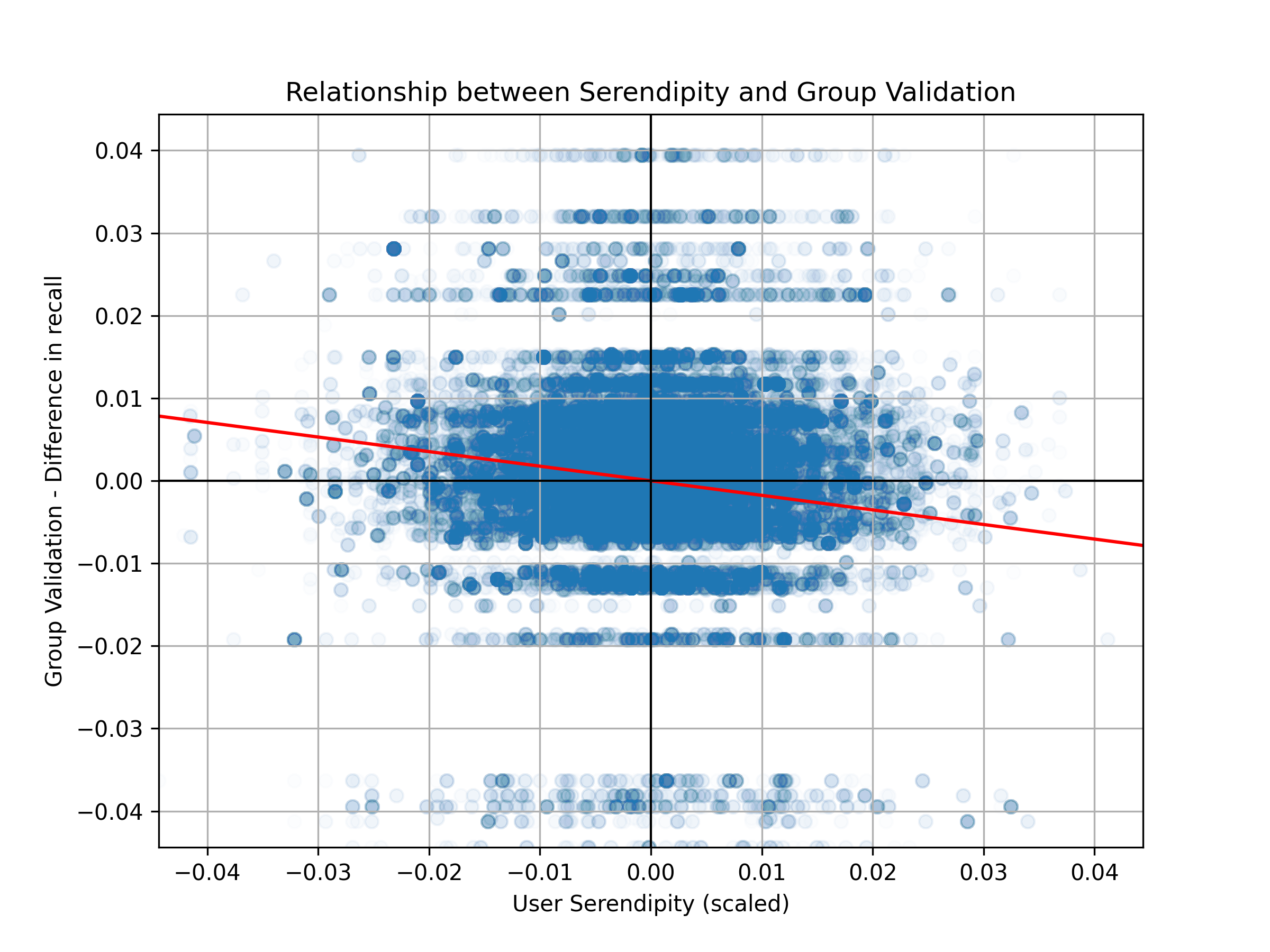} \\
NF3  & \includegraphics[width=3cm]{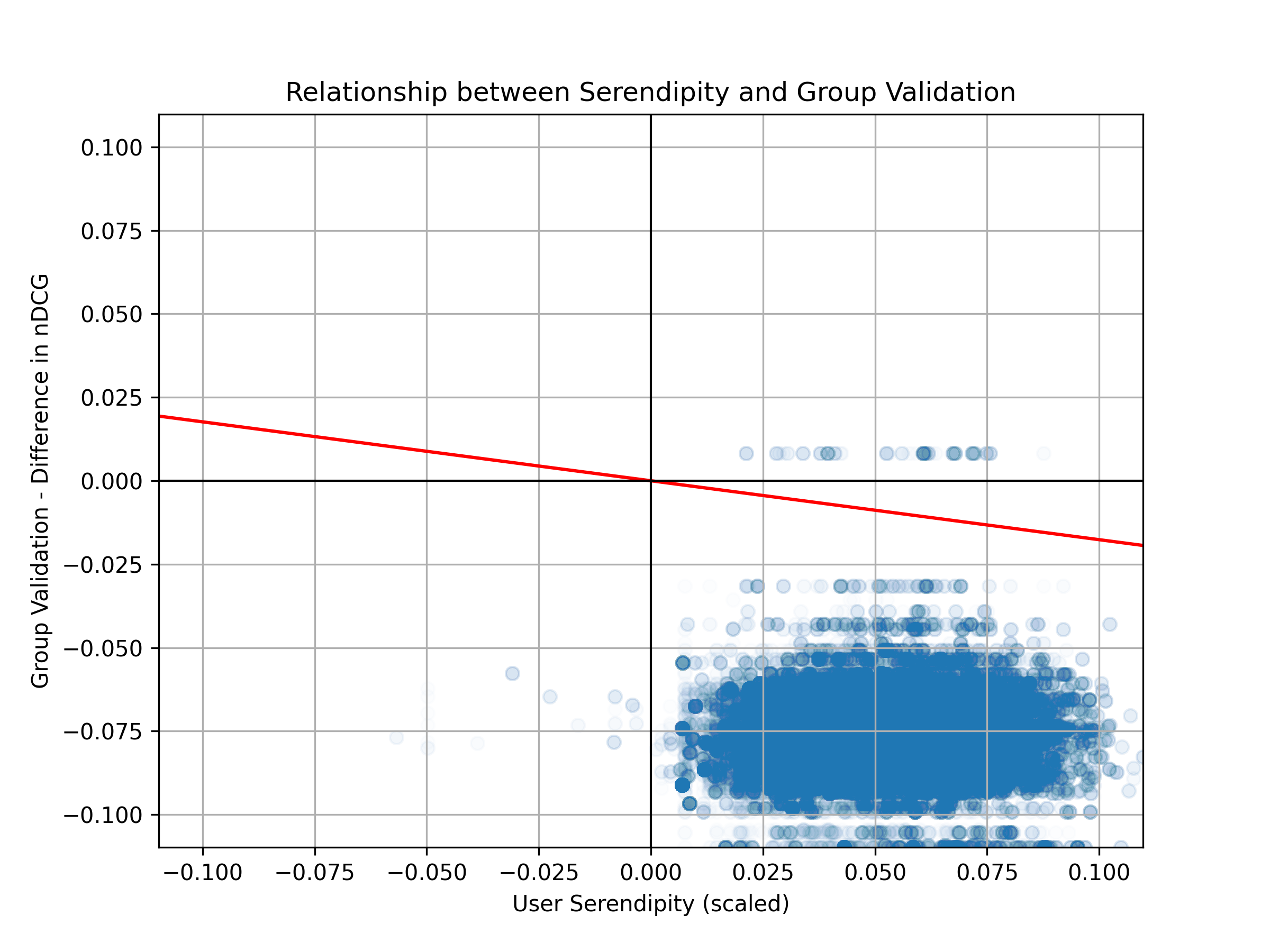} & \includegraphics[width=3cm]{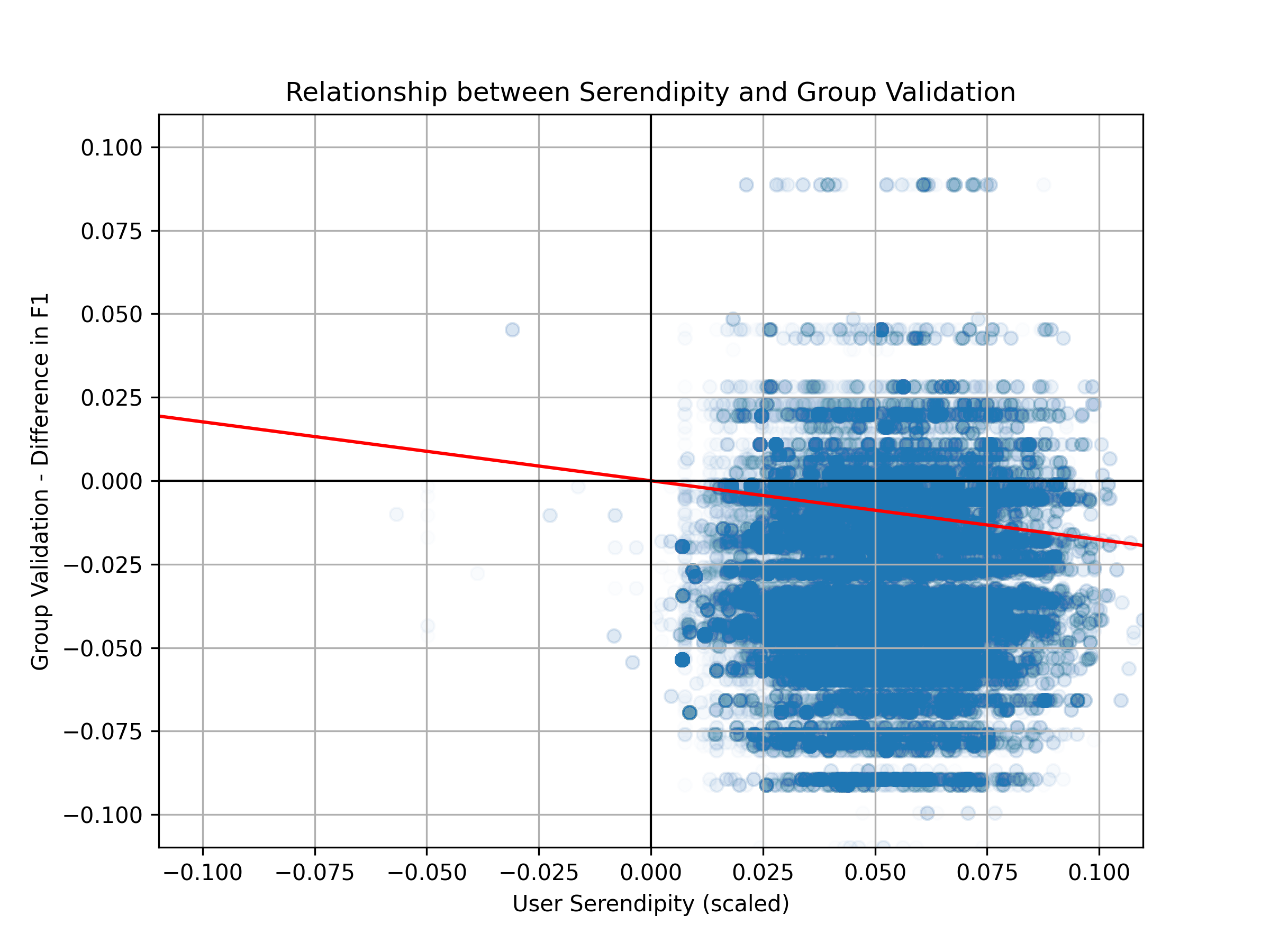} & \includegraphics[width=3cm]{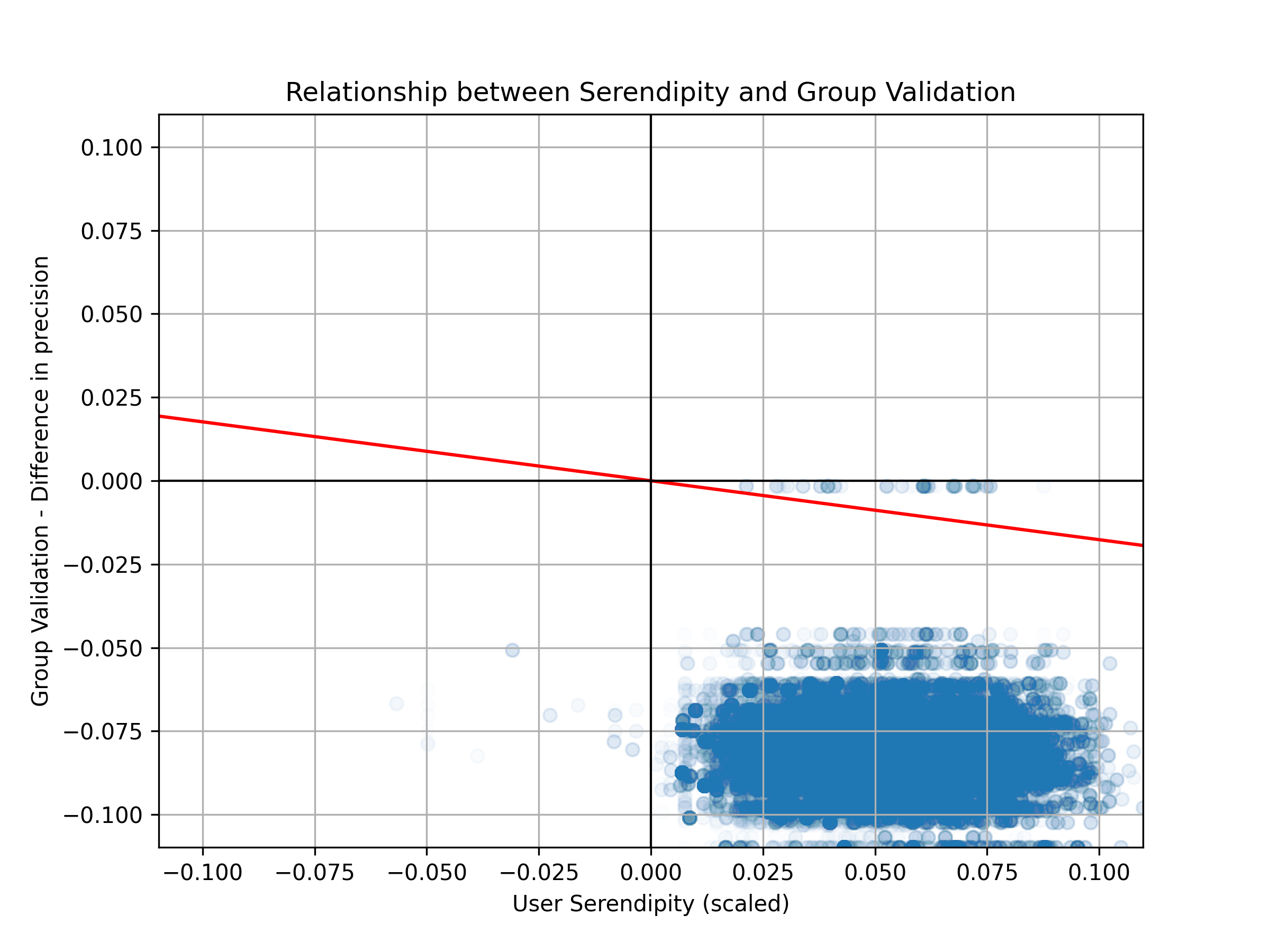} & \includegraphics[width=3cm]{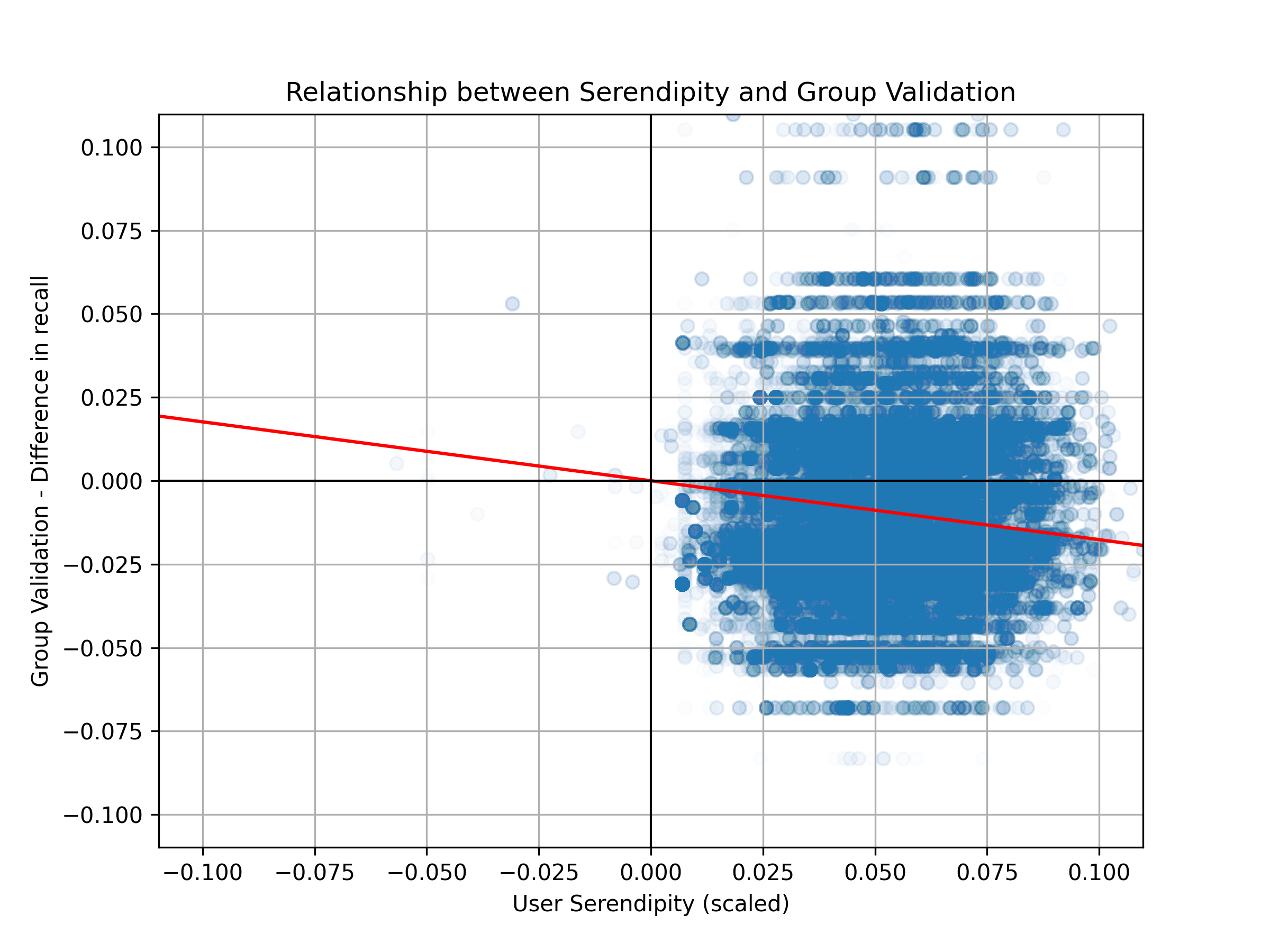} \\
NF4 & \includegraphics[width=3cm]{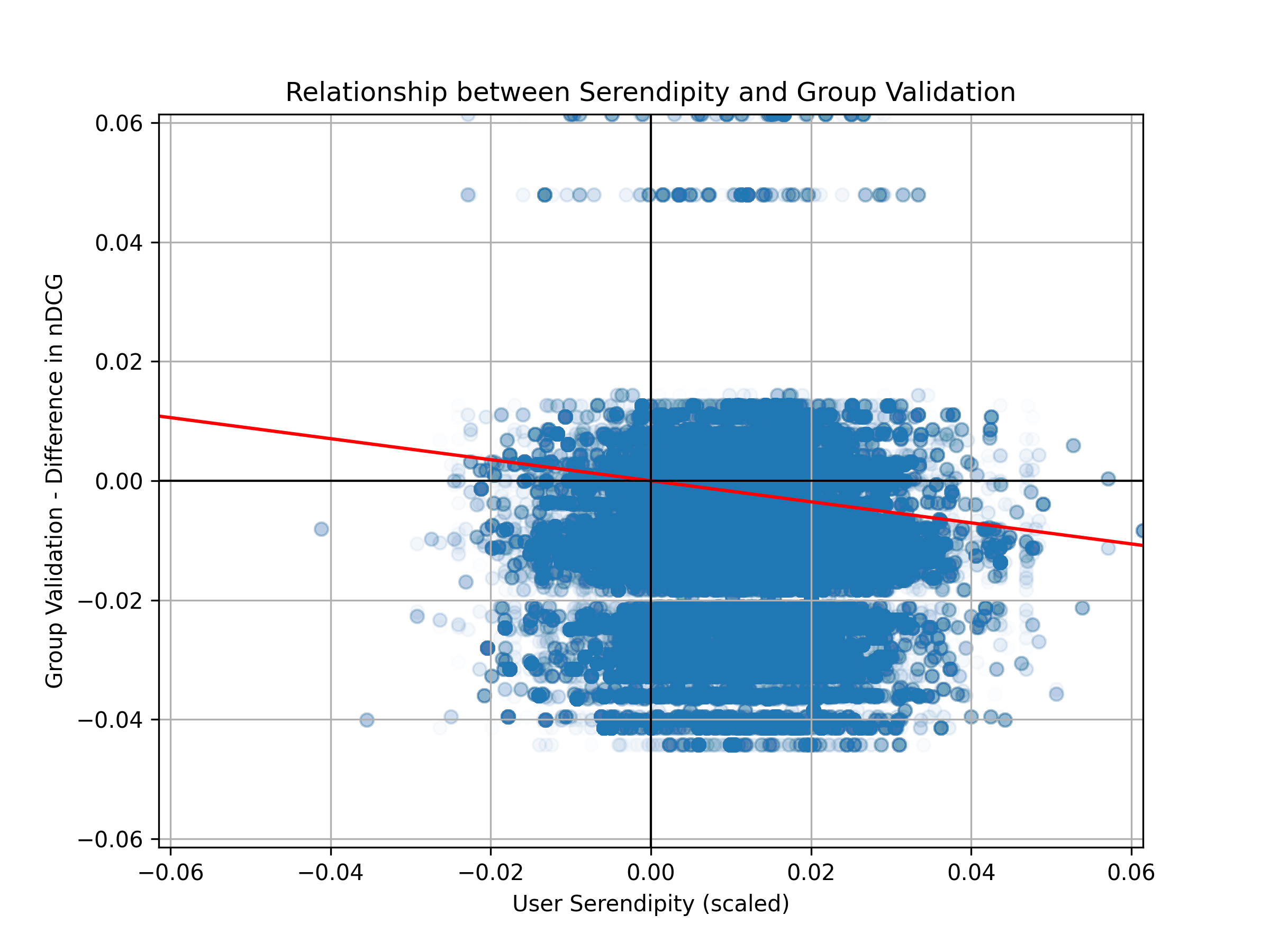} & \includegraphics[width=3cm]{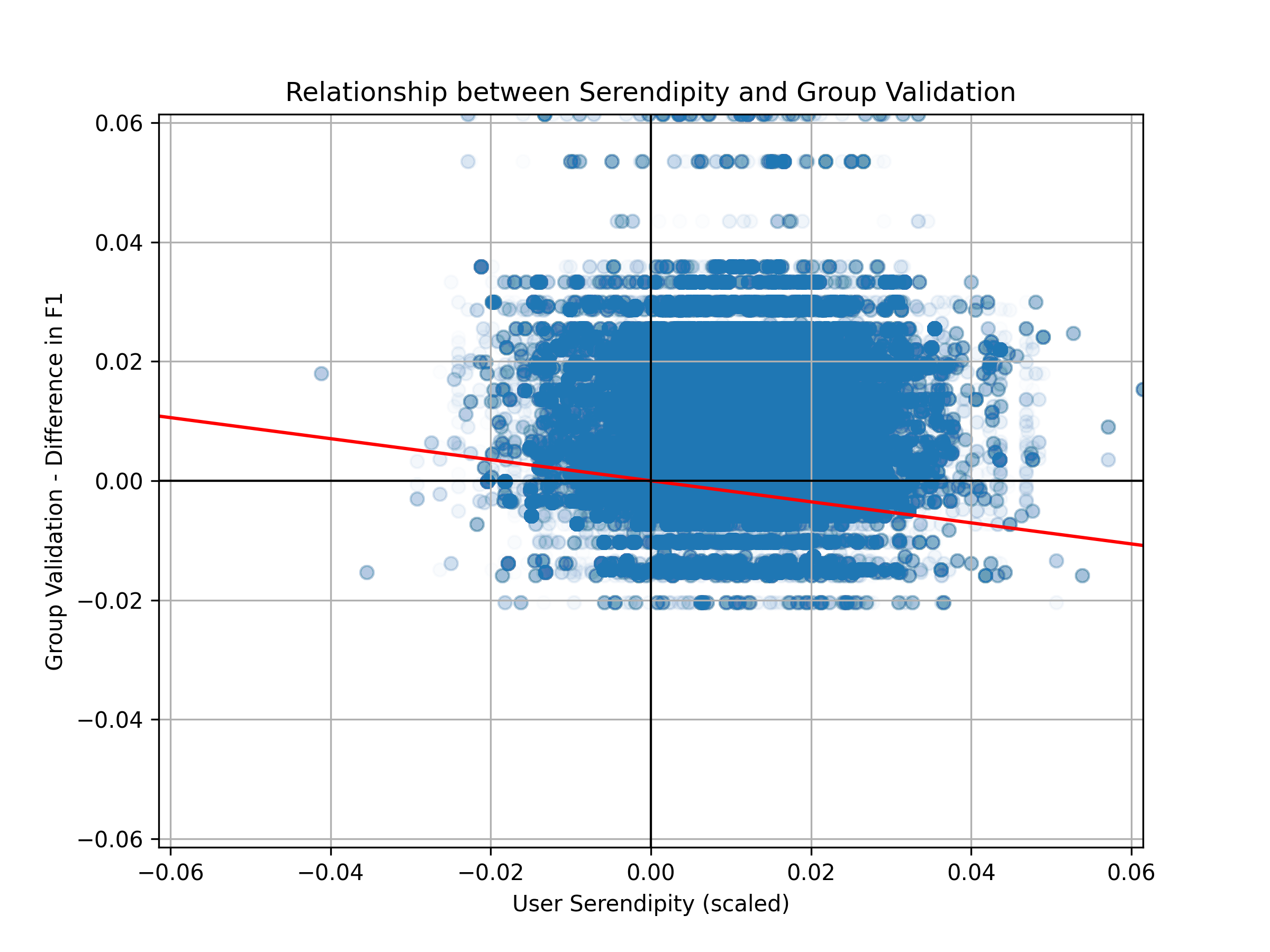} & \includegraphics[width=3cm]{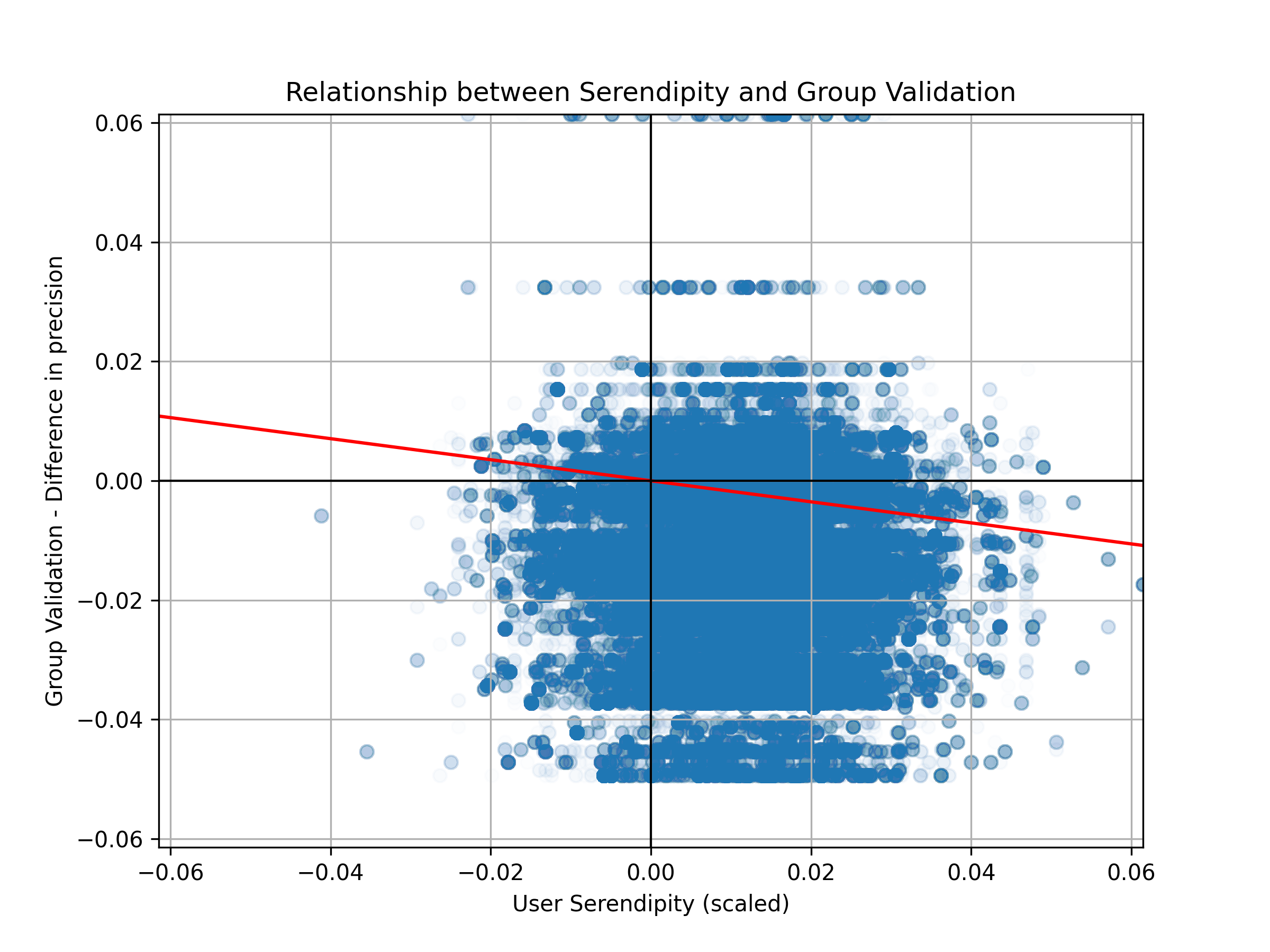} & \includegraphics[width=3cm]{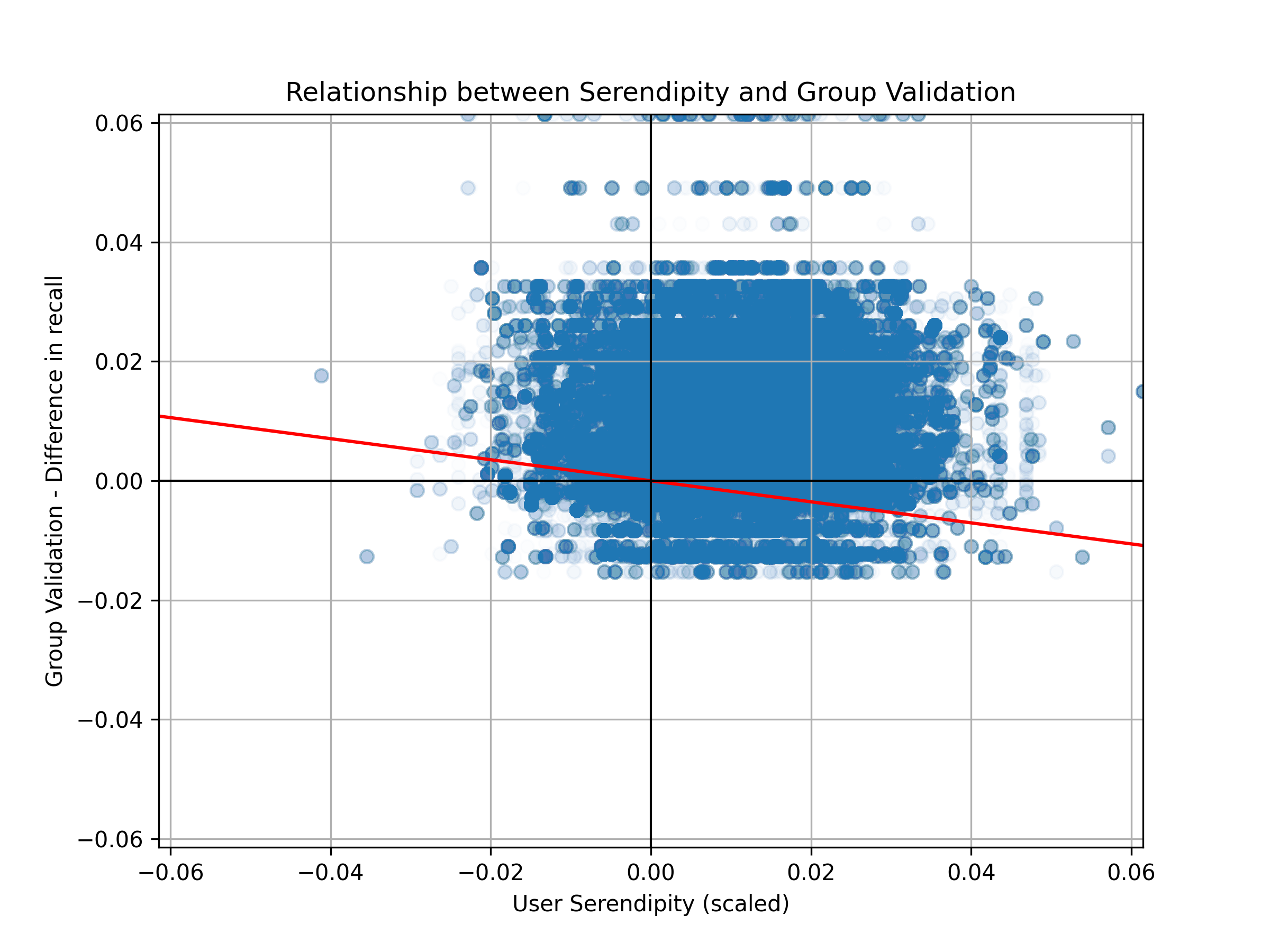}  \\
\bottomrule
\end{tabular}%
}
\label{graph:propo_nfs}
\end{table}

\FloatBarrier

\subsection{Analysis}
\label{subsec:analysis}
The 2D metric system, designed to highlight the importance of both accuracy and Serendipity, proved valuable during testing. By balancing these two metrics, the framework aims to not only improve system performance but also to enhance user satisfaction, which was reflected in the overall results. It was highlighted that improvements over groups and user experience are being discarded by traditional accuracy metric, stating the need for a system that is able to catch these improvements.\\ 

We were also able to see, that while previous noise filtering algorithms had good accuracy results following traditional methods, or, there results are not aligned, and that the non-consensus between these algorithms lets us wonder, if the noise definition is really coherent between these algorithms. This non-consensus between the algorithms, lead to the need of a system guiding towards consensus.

The proposed framework, inspired by IPS in Cyber Security, consisting of three layers, successfully integrates the strengths of multiple algorithms, yielding significant improvements in accuracy and Serendipity. Specifically, the framework achieved an overall accuracy improvement of 40.28\%, 99.31\%, 99.42\% and 38.05\% for nDCG, F1, recall and precision' systems respectively; while making sure that Serendipity is present and enhanced, this demonstrates its ability to maintain a balance between these often conflicting objectives. As seen to make sure that the comparison against known noise filtering algorithms, we utilized the proposed algorithms in their full actions, even if our algorithm does not take into account an action other than noise removal.
\\ 
What sets this framework apart is its flexibility. It adapts to varying needs, allowing users to prioritize different metrics based on their specific goals, whether it's increasing Serendipity or improving precision. Also adaptability in the threshold following this.
\\ 
Future improvements on the framework can be made by adding more 'Signature' based algorithms independent from noise, signatures that do not depend on noise patterns and can be placed as layer 1 to filter out noise before reaching the noise management algorithms, another improvement can be as well as integrating other actions post noise identification as correction, as well as it needs to be tested on real world-scenarios.
\section{Conclusion}
\label{sec:conclusion}

Measuring the accuracy of a Recommender Systems poses significant challenges, particularly when attempting to incorporate factors that influence user experience. One often-overlooked aspect is serendipity, which plays a crucial role in enhancing user satisfaction. We propose a novel method to effectively capture these influential factors, ensuring a balanced relationship between accuracy and serendipity.\\
Our proposed three-layered framework aims to optimize the performance of various algorithms while ensuring versatility to adapt to different system requirements.
Additionally, we raise critical questions regarding user security; as we've seen, some users choose to opt out or delete their profiles to safeguard their identities, prompting us to consider how user profiles can be created in alignment with data security policies, especially as security concerns are increasingly prevalent worldwide. This leads us to wonder if security considerations influence user behavior and preferences. In future work, we will investigate how shortened user profiles may enhance serendipity while emphasizing the importance of security and anonymity in Recommender Systems.
\section{Acknowledgments}
In this paper, the authors acknowledge the assistance of OpenAI's ChatGPT in refining grammar and enhancing paragraphs. However all technical content and analysis are entirely the result of the authors' own work and research.

\bibliography{EnsembleLearning_NF_v1}
\end{document}